\documentclass[reqno]{amsart}  
\usepackage{amssymb}
\usepackage{upref}
\usepackage{hyperref}
\newcommand{\bbI}{{\mathbb{I}}}
\newcommand{\bbN}{{\mathbb{N}}}
\newcommand{\bbR}{{\mathbb{R}}}

\newcommand{\bbZ}{{\mathbb{Z}}}
\newcommand{\bbC}{{\mathbb{C}}}

\newcommand{\calB}{{\mathcal B}}

\newcommand{\calD}{{\mathcal D}}

\newcommand{\calG}{{\mathcal G}}
\newcommand{\calH}{{\mathcal H}}

\newcommand{\frh}{\mathfrak{h}}
\newcommand{\frf}{\mathfrak{f}}
\newcommand{\frg}{\mathfrak{g}}

\newcommand{\dott}{\,\cdot\,}
\newcommand{\hatt}{\widehat}  

\newcommand{\no}{\nonumber}
\newcommand{\lb}{\label}
\newcommand{\f}{\frac}
\newcommand{\ul}{\underline}
\newcommand{\ol}{\overline}

\newcommand{\Oh}{O}

\newcommand{\spec}{\text{\rm{spec}}}

\newcommand{\dom}{\text{\rm{dom}}}

\DeclareMathOperator{\tr}{tr}

\DeclareMathOperator{\sAL}{s-AL}

\DeclareMathOperator{\AL}{AL}

\newcommand{\gam}{\gamma}

\newcommand{\deven}{\delta_{\rm even}}
\newcommand{\dodd}{\delta_{\rm odd}}

\allowdisplaybreaks
\numberwithin{equation}{section}

\newtheorem{theorem}{Theorem}[section]
\newtheorem{lemma}[theorem]{Lemma}

\newtheorem{hypothesis}[theorem]{Hypothesis}

\theoremstyle{definition}
\newtheorem{remark}[theorem]{Remark}

\begin{document}
\title[Conservation Laws and Hamiltonians for  the AL Hierarchy]{Local Conservation Laws and the
Hamiltonian Formalism for the Ablowitz--Ladik Hierarchy}
\author[F.\ Gesztesy]{Fritz Gesztesy}
\address{Department of Mathematics,
University of Missouri,
Columbia, MO 65211, USA}
\email{\href{mailto:fritz@math.missouri.edu}{fritz@math.missouri.edu}}
\urladdr{\href{http://www.math.missouri.edu/personnel/faculty/gesztesyf.html}
{http://www.math.missouri.edu/personnel/faculty/gesztesyf.html}}
\author[H.\ Holden]{Helge Holden}
\address{Department of Mathematical Sciences,
Norwegian University of
Science and Technology, NO--7491 Trondheim, Norway}
\email{\href{mailto:holden@math.ntnu.no}{holden@math.ntnu.no}}
\urladdr{\href{http://www.math.ntnu.no/~holden/}{http://www.math.ntnu.no/\~{}holden/}}

\author[J. Michor]{Johanna Michor}
\address{Department of Mathematics\\ Imperial College\\
180 Queen's Gate\\ London SW7 2BZ\\ and International Erwin Schr\"odinger
Institute for Mathematical Physics, Boltzmanngasse 9\\ 1090 Wien\\ Austria}
\email{\href{mailto:Johanna.Michor@esi.ac.at}{Johanna.Michor@esi.ac.at}}
\urladdr{\href{http://www.mat.univie.ac.at/~jmichor/}{http://www.mat.univie.ac.at/\~{}jmichor/}}

\author[G. Teschl]{Gerald Teschl}
\address{Faculty of Mathematics\\
University of Vienna\\
Nordbergstrasse 15\\ 1090 Wien\\ Austria\\ and International Erwin Schr\"odinger
Institute for Mathematical Physics, Boltzmanngasse 9\\ 1090 Wien\\ Austria}
\email{\href{mailto:Gerald.Teschl@univie.ac.at}{Gerald.Teschl@univie.ac.at}}
\urladdr{\href{http://www.mat.univie.ac.at/~gerald/}{http://www.mat.univie.ac.at/\~{}gerald/}}

\thanks{Research supported in part by the Research Council of Norway,
the US National Science Foundation under Grant No.\ DMS-0405526, and
the Austrian Science Fund (FWF) under Grants No.\ Y330, J2655}
\date{\today}
\subjclass[2000]{Primary 37K10, 37K20, 47B36; Secondary 35Q58, 37K60.}
\keywords{Ablowitz--Ladik hierarchy, Hamiltonian formalism, conservation laws.}

\begin{abstract}
We derive a systematic and recursive approach to local conservation 
laws and the Hamiltonian formalism for the Ablowitz--Ladik (AL) hierarchy. 
Our methods rely on a recursive approach to the AL hierarchy using 
Laurent polynomials and on asymptotic expansions of the Green's function 
of the AL Lax operator, a five-diagonal finite difference operator. 
\end{abstract}

\maketitle

\section{Introduction}\lb{sALh1}

The principal aim of this paper is to provide a systematic and recursive approach to local conservation laws and the Hamiltonian formalism for the Ablowitz--Ladik (AL) hierarchy of integrable differential-difference equations.

Consider sequences $\{\alpha(n,t), \beta(n,t)\}_{n\in\bbZ}\in\ell^1 (\bbZ)$ satisfying some additional  assumptions to be specified later, parametrized by the deformation (time) parameter 
$t\in \bbR$, that are solutions of the Ablowitz--Ladik equations
\begin{equation}
 \begin{pmatrix} -i \alpha_{t}- (1-\alpha\beta) (\alpha^- + \alpha^+) + 2 \alpha  \\   
  -i\beta_{t}+ (1-\alpha\beta) (\beta^- + \beta^+) - 2 \beta \end{pmatrix}=0.
\end{equation}
Here $c^{\pm}$ denote shifts, that is, $c^{\pm}(n)=c(n\pm 1)$, $n\in\bbZ$.  Then clearly
\begin{equation}
\partial_{t}\sum_{n\in\bbZ}\alpha^+(n,t)\beta(n,t) = \partial_{t}\sum_{n\in\bbZ}\alpha(n,t)\beta^+(n,t) =0.
\end{equation}
Indeed, one can show the existence of an infinite sequence $\{\rho_{j,\pm}\}_{j\in\bbN}$ of polynomials of $\alpha,\beta$ and certain shifts thereof, with the property that the lattice sum is time-independent,  
\begin{equation}
\partial_{t}\sum_{n\in\bbZ} \rho_{j,\pm}(n,t)=0, \quad j\in\bbN.
\end{equation}
This result is obtained by deriving local conservation laws of the type
\begin{equation}
\partial_{t} \rho_{j,\pm}+(S^+ -I)J_{j,\pm}=0, \quad j\in\bbN,
\end{equation} 
for certain polynomials $J_{j,\pm}$  of $\alpha,\beta$ and certain shifts thereof. The polynomials $J_{j,\pm}$ will be constructed via an explicit recursion relation. For a detailed discussion of these results we refer to Theorem \ref{tALh5.7} and 
Remarks \ref{rALh5.8} and ref{rALh5.8a}. 

The above analysis extends to the full Ablowitz--Ladik hierarchy as follows. The $\ul p$th equation, $\ul p=(p_-,p_+)\in\bbN_0^2$ (where $\bbN_0=\bbN\cup\{0\}$), in the AL hierarchy is given by
\begin{align}   
\begin{split}
& \AL_{\ul p} (\alpha, \beta) = \begin{pmatrix}-i\alpha_{t_{\ul p}} 
- \alpha(g_{p_+,+} + g_{p_-,-}^-) + f_{p_+ -1,+} - f_{p_- -1,-}^-\\
  -i\beta_{t_{\ul p}}+ \beta(g_{p_+,+}^- + g_{p_-,-}) 
  - h_{p_- -1,-} + h_{p_+ -1,+}^- \end{pmatrix}=0,  \\
& \hspace*{6.44cm} t_{\ul p}\in\bbR, \; \ul p=(p_-,p_+)\in\bbN_0^2,   
\end{split}
\end{align}
where $f_{\ell,\pm}$, $g_{\ell,\pm}$, and $h_{\ell,\pm}$ are carefully designed polynomial expressions of $\alpha,\beta$ and certain shifts thereof. Recursively, they are given by 
\eqref{AL0+}--\eqref{ALh_l-}. On each level in the recursion an arbitrary constant 
$c_{\ell,\pm} \in\bbC$ is introduced. In the homogeneous case, where all these constants 
$c_\ell$, $\ell\in\bbN$, are set equal to zero, a hat \,$\hat {}$\, is added in the notation, that is, $\hat f_{\ell,\pm}, \hat g_{\ell,\pm}, \hat h_{\ell,\pm}$, etc., denote the corresponding homogeneous quantities.  The homogeneous coefficients 
$\hat f_{\ell,\pm}, \hat g_{\ell,\pm}, \hat h_{\ell,\pm}$ can also be  expressed explicitly in terms of appropriate matrix elements of powers of the AL Lax finite difference expression $L$ defined in \eqref{ALLop}, \eqref{ALLrec} and the finite difference expressions $D$ and $E$ in \eqref{ALDE}, as described in Lemma \ref{lALh3.1}.  The conserved densities 
$\rho_{j,\pm}$ are independent of the equation in the hierarchy while the currents 
$J_{\ul p,j,\pm}$ depend on $\ul p$; thus one finds (cf.\ Theorem \ref{tALh5.7})
\begin{equation}
\partial_{t_{\ul p}} \rho_{j,\pm}+(S^+ -I)J_{\ul p,j,\pm}=0, \quad t_{\ul p} \in\bbR, \; 
j\in\bbN, \; \ul p\in\bbN_0^2.  
\lb{ALh1.7}
\end{equation}
For $\alpha, \beta \in \ell^1(\bbZ)$ it then follows  that
\begin{equation}\lb{ALh1.7a}
\frac{d}{d t_{\ul p}}\sum_{n\in\bbZ} \rho_{j,\pm}(n,t_{\ul p}) =0, 
\quad t_{\ul p}\in\bbR, \; j\in\bbN, \; \ul p\in\bbN_0^2.
\end{equation}
By showing that $\rho_{j,\pm}$ equals $\hat g_{j,\pm}$ up to a first-order difference expression (cf.\ Lemma \ref{lALh4.4}), and by investigating the time-dependence of 
$\gamma = 1 - \alpha\beta$, one concludes (cf.\ Remark \ref{rALh5.8}) that
\begin{equation}\lb{ALh1.7b}
\frac{d}{d t_{\ul p}}\sum_{n\in\bbZ} \ln(\gamma(n,t_{\ul p})) =0, \quad 
\frac{d}{d t_{\ul p}}\sum_{n\in\bbZ} \hat g_{j,\pm}(n,t_{\ul p}) =0, \quad  
t_{\ul p} \in\bbR, \; j\in\bbN, \; \ul p\in\bbN_0^2,  
\end{equation}
represent the two infinite sequences of AL conservation laws. 
Our approach to \eqref{ALh1.7} is based on a careful analysis of asymptotic expansions of the Green's function (as the spectral parameter tends to zero and to infinity) for the operator realization $\breve L$ in $\ell^2(\bbZ)$ corresponding to the Lax difference  expression $L$ in \eqref{ALLop}, \eqref{ALLrec}. 

In addition, we provide a detailed study of the Hamiltonian formalism for the AL hierarchy. In particular, the $\ul p$th equation in the AL hierarchy can be written as (cf.\ Theorem \ref{tALh6.5})
\begin{equation}
\AL_{\ul p}(\alpha,\beta)= 
\begin{pmatrix}-i\alpha_{t_{\ul p}} \\ -i \beta_{t_{\ul p}} \end{pmatrix}
+\calD\nabla \calH_{\ul p}=0, \quad  \ul p \in \bbN_0^2,  
\lb{ALh1.12}
\end{equation}
where the Hamiltonians $\calH_{\ul p}$ are given by
\begin{align} \lb{ALh1.13}
\calH_{\ul p} &=
\sum_{\ell=1}^{p_+} c_{p_+-\ell,+} \widehat \calH_{\ell,+} +
\sum_{\ell=1}^{p_-} c_{p_--\ell,-} \widehat \calH_{\ell,-} + c_{\ul{p}} \widehat \calH_0, \quad
\ul p=(p_-,p_+) \in\bbN_0^2,\\
\widehat \calH_0 &= \sum_{n\in\bbZ} \ln(\gam(n)), \qquad
\widehat \calH_{p_\pm,\pm} = \frac{1}{p_\pm} \sum\limits_{n\in\bbZ}\hat g_{p_\pm,\pm}(n), \quad p_\pm \in \bbN.
\end{align}
Here  $\calD=(1 - \alpha\beta)\left(\begin{smallmatrix}0& 1\\ -1& 0 \end{smallmatrix}\right)$. Furthermore, we show that any $\calH_{\ul r}$ is conserved by the Hamiltonian flows in 
\eqref{ALh1.12} (cf.\  Theorem \ref{tALh6.6}), that is,
\begin{equation}
\frac{d \calH_{\ul r}}{dt_{\ul p}} =0, \quad \ul p, \ul r\in \bbN_0^2.  
\end{equation} 
Moreover, for  general sequences $\alpha, \beta$ (i.e., not assuming that they satisfy an equation in the AL hierarchy),  we show in Theorem \ref{tALh6.7} that
\begin{equation}
\{\hatt \calH_{\ul p}, \hatt \calH_{\ul r}\}=0,  \quad  \ul p, \ul r \in \bbN_0^2,     
\end{equation}
for a suitably defined Poisson bracket $\{\dott,\dott\}$ (see \eqref{ALh6.16a}), that is, 
$\calH_{\ul p}$ and $\calH_{\ul r}$ are in involution for all $\ul p, \ul r \in \bbN_0^2$.

The Ablowitz--Ladik hierarchy has been extensively discussed in the completely integrable system literature (cf., e.g.,  
\cite{AblowitzLadik:1975}--\cite{AblowitzLadik2:1976}, \cite{Ablowitz:1977}, 
\cite[Sect.\ 3.2.2]{AblowitzClarkson:1991}, \cite[Ch.\ 3]{AblowitzPrinariTrubatch:2004},
\cite{ChiuLadik:1977}, \cite{Common:1992}, 
\cite{MillerErcolaniKricheverLevermore:1995}, \cite{Nenciu:2005}, \cite{Nenciu:2005a}, 
\cite{Nenciu:2006}, \cite{Schilling:1989}, \cite{Suris:2003}, \cite{Vekslerchik:1998}, 
\cite{Vekslerchik:2002} and the references cited therein) and in recent years especially due to its close connections with the theory of orthogonal polynomials, a field that underwent a remarkable resurgency in recent years (cf.\ \cite{Simon:2005}, \cite{Simon:2006}, \cite{Simon:2007} and the  quoted therein). Rather than repeating some of the AL hierarchy history and its relevance to the theory of orthogonal polynomials at this place, we refer to the detailed introductions of 
\cite{GesztesyHoldenMichorTeschl:2006}, \cite{GesztesyHoldenMichorTeschl:2007}, 
\cite{GesztesyHoldenMichorTeschl:2007a} and the extensive bibliography listed therein. Here we just mention references intimately connected with the topics discussed in this paper: Infinitely many conservation laws are discussed, for instance, by 
Ablowitz and Ladik \cite{AblowitzLadik:1976}, Ablowitz, Prinari, and Trubatch 
\cite[Ch.\ 3]{AblowitzPrinariTrubatch:2004}, Ding, Sun, and Xu 
\cite{DingSunXu:2006}, Zhang and Chen \cite{ZhangChen:2002}, and 
Zhang, Ning, Bi, and Chen \cite{ZhangNingBiChen:2006}; the bi-Hamiltonian structure of the AL hierarchy is considered by Ercolani and Lozano \cite{ErcolaniLozano:2006}, 
Hydon \cite{Hydon:2005}, and Lozano \cite{Lozano:2004}, multi-Hamiltonian structures for the defocusing AL hierarchy were studied by Gekhtman and Nenciu 
\cite{GekhtmanNenciu:2007}, Zeng and Rauch-Wojciechowski 
\cite{ZengRauchWojciechowski:1995}, and Zhang and Chen \cite{ZhangChen:2002a}; Poisson brackets for orthogonal polynomials on the unit circle relevant to the case of the defocusing AL hierarchy (where $\beta = \ol \alpha$) have been studied by Cantero and Simon \cite{CanteroSimon:2006}, Killip and Nenciu \cite{KillipNenciu:2006}, and Nenciu \cite{Nenciu:2007a}; Lenard recursions and Hamiltonian structures were discussed in Geng and Dai \cite{GengDai:2007} and Geng, Dai, and Zhu \cite{GengDaiZhu:2007}.  

Next we briefly describe the structure of this paper: Section \ref{sALh2} recalls the recursive construction of the AL hierarchy as discussed in detail in 
\cite{GesztesyHoldenMichorTeschl:2006} (see also 
\cite{GesztesyHoldenMichorTeschl:2007}, \cite{GesztesyHoldenMichorTeschl:2007a}). In Section \ref{sALh3} we introduce the Lax pair for the AL hierarchy and prove its equivalence with the corresponding zero-curvature formulation. These results are new. In Section \ref{sALh4} we discuss the Green's function of the Lax operator $\breve L$ and study its asymptotic expansions as the spectral parameter tends to zero and infinity. As a direct consequence of these asymptotic expansions, local conservation laws are then derived in Section \ref{sALh5}. Our final Section \ref{sALh6} then introduces the basics of variational derivatives and provides a detailed derivation of the Hamiltonian formalism for the AL hierarchy.

Finally, we emphasize that our recursive and systematic approach to local conservation laws of the Ablowitz--Ladik hierarchy appears to be new. Moreover, our treatment of Poisson brackets and variational derivatives, and their connections with the diagonal Green's function of the underlying Lax operator, now puts the AL  hierarchy on precisely the same level as the Toda and KdV hierarchy with respect to this particular aspect of the Hamiltonian formalism (cf.\ \cite[Ch.\ 1]{GesztesyHolden:2003}, 
\cite[Ch.\ 1]{GesztesyHolden:2005}).

\section{The Ablowitz--Ladik hierarchy in a nutshell}   \label{sALh2}

In this section we summarize the construction of the Ablowitz--Ladik hierarchy employing a Laurent polynomial recursion formalism and derive the associated sequence of Ablowitz--Ladik zero-curvature pairs. Moreover, we discuss the Burchnall--Chaundy Laurent polynomial in connection with the stationary Ablo\-witz--Ladik hierarchy and the underlying hyperelliptic curve. For a detailed treatment of this material we refer to 
\cite{GesztesyHolden:2005}, \cite{GesztesyHoldenMichorTeschl:2006}.

We denote by $\bbC^{\bbZ}$ the set of complex-valued sequences indexed by $\bbZ$. 

Throughout this section we suppose the following hypothesis. 

\begin{hypothesis} \lb{hAL2.1} 
In the stationary case we assume that $\alpha, \beta$ satisfy
\begin{equation}
\alpha, \beta\in \bbC^{\bbZ}, \quad  \alpha(n)\beta(n)\notin \{0,1\}, \; n\in\bbZ. \lb{AL2.01}
\end{equation}
In the time-dependent case we assume that $\alpha, \beta$ satisfy
\begin{align}
\begin{split}
& \alpha(\dott,t), \beta(\dott,t) \in \bbC^{\bbZ}, \; t\in\bbR, \quad 
\alpha(n,\dott), \beta(n,\dott)\in C^1(\bbR), \; n\in\bbZ,   \lb{AL2.01a}  \\
& \alpha(n,t)\beta(n,t)\notin \{0,1\}, \; (n,t)\in\bbZ\times \bbR.
\end{split}
\end{align}
\end{hypothesis}

We denote by $S^\pm$ the shift operators acting on complex-valued sequences 
$f=\{f(n)\}_{n\in\bbZ} \in\bbC^{\bbZ}$ according to
\begin{equation}
(S^\pm f)(n)=f(n\pm1), \quad n\in\bbZ. \lb{AL2.02}
\end{equation}
Moreover, we will frequently use the notation
\begin{equation}
f^\pm = S^{\pm} f, \quad f\in\bbC^{\bbZ}. 
\end{equation} 

To construct the  Ablowitz--Ladik hierarchy one typically introduces appropriate zero-curvature pairs of  $2\times2$ matrices, denoted by $U(z)$ and $V_{\ul p}(z)$, $\ul p\in\bbN_0^2$  (with $z\in\bbC\setminus\{0\}$ a certain spectral parameter to be discussed later), defined recursively in the following. We take the shortest route to the construction of $V_{\ul p}$ and hence to that of the Ablowitz--Ladik hierarchy by starting from the recursion relation \eqref{AL0+}--\eqref{ALh_l-} below.   

Define sequences $\{f_{\ell,\pm}\}_{\ell\in \bbN_0}$, $\{g_{\ell,\pm}\}_{\ell\in \bbN_0}$, and $\{h_{\ell,\pm}\}_{\ell\in \bbN_0}$ recursively by
\begin{align} \label{AL0+}
g_{0,+} &= \tfrac12 c_{0,+}, \quad f_{0,+} = - c_{0,+}\alpha^+, 
\quad h_{0,+} = c_{0,+}\beta, \\ \label{ALg_l+}
g_{\ell+1,+} - g_{\ell+1,+}^- &= \alpha h_{\ell,+}^- + \beta f_{\ell,+}, \quad \ell\in \bbN_0,\\ \label{ALf_l+}
f_{\ell+1,+}^- &= f_{\ell,+} - \alpha (g_{\ell+1,+} + g_{\ell+1,+}^-), \quad \ell\in \bbN_0, \\  \label{ALh_l+}
h_{\ell+1,+} &= h_{\ell,+}^- + \beta (g_{\ell+1,+} + g_{\ell+1,+}^-), \quad \ell\in \bbN_0,  
\end{align}
and
\begin{align} \label{AL0-}
g_{0,-} &= \tfrac12 c_{0,-}, \quad f_{0,-} = c_{0,-}\alpha, 
\quad h_{0,-} = - c_{0,-}\beta^+, \\ \label{ALg_l-}
g_{\ell+1,-} - g_{\ell+1,-}^- &= \alpha h_{\ell,-} + \beta f_{\ell,-}^-, \quad \ell\in \bbN_0,\\ \label{ALf_l-}
f_{\ell+1,-} &= f_{\ell,-}^- + \alpha (g_{\ell+1,-} + g_{\ell+1,-}^-), \quad \ell\in \bbN_0, \\ \label{ALh_l-}
h_{\ell+1,-}^- &= h_{\ell,-} - \beta (g_{\ell+1,-} + g_{\ell+1,-}^-), \quad \ell\in \bbN_0.
\end{align}
Here $c_{0,\pm}\in\bbC$ are given constants. For later use we also introduce
\begin{equation}\lb{ALminus}
f_{-1,\pm}= h_{-1,\pm}=0.
\end{equation}

\begin{remark}\lb{rAL2.2}
The sequences $\{f_{\ell,+}\}_{\ell\in \bbN_0}$, 
$\{g_{\ell,+}\}_{\ell\in \bbN_0}$, and
$\{h_{\ell,+}\}_{\ell\in \bbN_0}$ can be computed recursively as follows: 
Assume that $f_{\ell,+}$,
$g_{\ell,+}$, and $h_{\ell,+}$ are known.  Equation \eqref{ALg_l+} is a 
first-order difference equation in $g_{\ell+1,+}$ that can be solved directly
and yields a local lattice function that is determined up to a new constant denoted
by $c_{\ell+1,+}\in\bbC$. Relations \eqref{ALf_l+} and \eqref{ALh_l+}
then determine $f_{\ell+1,+}$ and $h_{\ell+1,+}$, etc.  The sequences 
$\{f_{\ell,-}\}_{\ell\in \bbN_0}$, $\{g_{\ell,-}\}_{\ell\in \bbN_0}$, and 
$\{h_{\ell,-}\}_{\ell\in \bbN_0}$ are determined similarly.
\end{remark}

Upon setting 
\begin{equation}
\gamma = 1 - \alpha \beta, \lb{ALgamma}
\end{equation}
one explicitly obtains 
\begin{align}
\begin{split}
f_{0,+} &= c_{0,+}(-\alpha^+), \quad 
f_{1,+} = c_{0,+}\big(- \gamma^+ \alpha^{++} + (\alpha^+)^2 \beta\big) 
+ c_{1,+} (-\alpha^+), \\
g_{0,+} &= \tfrac{1}{2}c_{0,+},  \quad 
 g_{1,+} = c_{0,+}(-\alpha^+ \beta) + \tfrac{1}{2}c_{1,+}, \\
h_{0,+} &= c_{0,+}\beta, \quad 
 h_{1,+} = c_{0,+}\big(\gamma \beta^- - \alpha^+ \beta^2\big) 
+ c_{1,+} \beta, \\
f_{0,-} &= c_{0,-}\alpha, \quad  
f_{1,-} = c_{0,-}\big(\gamma \alpha^- - \alpha^2 \beta^+\big) + c_{1,-} \alpha, \\
g_{0,-} &= \tfrac{1}{2}c_{0,-}, \quad  
g_{1,-} = c_{0,-}(-\alpha \beta^+) + \tfrac{1}{2}c_{1,-}, \\
h_{0,-} &= c_{0,-}(-\beta^+), \quad  
h_{1,-} = c_{0,-}\big(- \gamma^+ \beta^{++} 
+ \alpha (\beta^+)^2 \big) + c_{1,-} (- \beta^+), \, \text{ etc.}
\end{split}
\end{align}
Here $\{c_{\ell,\pm}\}_{\ell \in \bbN}$ denote summation constants
which naturally arise when solving the difference equations for 
$g_{\ell, \pm}$ in \eqref{ALg_l+}, \eqref{ALg_l-}.  
Subsequently, it will also be useful to work with the corresponding homogeneous coefficients $\hat f_{\ell, \pm}$,
$\hat g_{\ell, \pm}$, and $\hat h_{\ell, \pm}$, defined by the vanishing of all summation constants $c_{k,\pm}$ for $k=1,\dots,\ell$, and choosing $c_{0,\pm}=1$,
\begin{align}
& \hat f_{0,+}=-\alpha^+, \quad \hat f_{0,-}=\alpha, \quad 
 \hat f_{\ell,\pm}=f_{\ell,\pm}|_{c_{0,\pm}=1, \, c_{j,\pm}=0, j=1,\dots,\ell},  \quad \ell\in\bbN, 
 \lb{AL2.04a} \\
& \hat g_{0,\pm}=\tfrac12, \quad 
\hat g_{\ell,\pm}=g_{\ell,\pm}|_{c_{0,\pm}=1, \, c_{j,\pm}=0, j=1,\dots,\ell}, 
\quad \ell\in\bbN,  \lb{AL2.04b} \\
& \hat h_{0,+}=\beta, \quad \hat h_{0,-}=-\beta^+,  \quad 
\hat h_{\ell,\pm}=h_{\ell,\pm}|_{c_{0,\pm}=1, \, c_{j,\pm}=0, j=1,\dots,\ell}, 
\quad \ell\in\bbN.  \lb{AL2.04c}
\end{align}
By induction one infers that
\begin{equation} \label{ALhat f}
f_{\ell, \pm} = \sum_{k=0}^\ell c_{\ell-k, \pm} \hat f_{k, \pm}, \quad
g_{\ell, \pm} = \sum_{k=0}^\ell c_{\ell-k, \pm} \hat g_{k, \pm}, \quad 
h_{\ell, \pm} = \sum_{k=0}^\ell c_{\ell-k, \pm} \hat h_{k, \pm}.  
\end{equation} 
In a slight abuse of notation we will occasionally stress the dependence of $f_{\ell,\pm}$, 
$g_{\ell,\pm}$, and $h_{\ell,\pm}$ on $\alpha, \beta$ by writing  
$f_{\ell,\pm}(\alpha,\beta)$, $g_{\ell,\pm}(\alpha,\beta)$, and 
$h_{\ell,\pm}(\alpha,\beta)$. 

One can show (cf.\ \cite{GesztesyHoldenMichorTeschl:2006}) that all homogeneous elements 
$\hat f_{\ell,\pm}$, $\hat g_{\ell,\pm}$, and $\hat h_{\ell,\pm}$, $\ell\in\bbN_0$, are polynomials in $\alpha, \beta$, and some of their shifts. 

\begin{remark}\lb{rAL2.4}
As an efficient tool to distinguish between nonhomogeneous and homogeneous 
quantities $f_{\ell,\pm}$, $g_{\ell,\pm}$, $h_{\ell,\pm}$, and $\hat f_{\ell,\pm}$, 
$\hat g_{\ell,\pm}$, $\hat h_{\ell,\pm}$, respectively, we now introduce the notion of degree as follows. Denote  
\begin{align}
f^{(r)}=S^{(r)}f, \quad f=\{f(n)\}_{n\in\bbZ}\in\bbC^{\bbZ}, \quad
   S^{(r)}&=\begin{cases}(S^+)^r, &\text{$r\ge 0$},\\
(S^-)^{-r}, &\text{$r< 0$},\end{cases} \quad
r\in \bbZ,    \lb{AL2.1AA}
\end{align}
and define
\begin{equation}
\deg \big(\alpha^{(r)}\big)=r, \quad \deg \big(\beta^{(r)}\big)=-r, \quad r\in\bbZ.  
\lb{3.2.15aa}
\end{equation}
This implies
\begin{align}
\begin{split}
\deg\big(\hat f_{\ell,+}^{(r)}\big)&= \ell+1+r, \quad \deg\big(\hat f_{\ell,-}^{(r)}\big)
= -\ell+r, \quad \deg\big(\hat g_{\ell, \pm}^{(r)}\big)= \pm\ell, \\
\deg\big(\hat h_{\ell,+}^{(r)}\big)&= \ell-r, \quad \deg\big(\hat h_{\ell,-}^{(r)}\big)
= -\ell-1-r, \quad \ell\in\bbN_0, \; r\in\bbZ. 
\end{split}
\end{align}
\end{remark}

Alternatively the homogeneous coefficients can be computed directly via the following
nonlinear recursion relations:

\begin{lemma} \label{lALB.1}
The homogeneous quantities $\hat f_{\ell,\pm}$, $\hat g_{\ell,\pm}$, $\hat h_{\ell,\pm}$
are uniquely defined by the following recursion relations:
\begin{align}
\begin{split} \lb{ALB1.9}
\hat g_{0,+} &= \frac{1}{2}, \quad \hat f_{0,+} = -\alpha^+, \quad \hat h_{0,+} = \beta, \\
\hat g_{l+1,+} &= \sum_{k=0}^l \hat f_{l-k,+} \hat h_{k,+}
- \sum_{k=1}^l \hat g_{l+1-k,+} \hat g_{k,+}, \\
\hat f_{l+1,+}^- &= \hat f_{l,+} - \alpha (\hat g_{l+1,+} + \hat g_{l+1,+}^-),\\
\hat h_{l+1,+} &= \hat h_{l,+}^- + \beta (\hat g_{l+1,+} + \hat g_{l+1,+}^-), 
\end{split}
\end{align}
and
\begin{align}
\begin{split}\lb{ALB1.9a}
\hat g_{0,-} &= \frac{1}{2}, \quad \hat f_{0,-} = \alpha, \quad \hat h_{0,-} = -\beta^+, \\
\hat g_{l+1,-} &= \sum_{k=0}^l \hat f_{l-k,-} \hat h_{k,-}
- \sum_{k=1}^l \hat g_{l+1-k,-} \hat g_{k,-}, \\
\hat f_{l+1,-} &= \hat f_{l,-}^- + \alpha (\hat g_{l+1,-} + \hat g_{l+1,-}^-),\\
\hat h_{l+1,-}^- &= \hat h_{l,-} - \beta (\hat g_{l+1,-} + \hat g_{l+1,-}^-). 
\end{split}
\end{align}
\end{lemma}

We also note the following useful result (cf.\ \cite{GesztesyHoldenMichorTeschl:2006}):  Assuming \eqref{AL2.01}, we find
\begin{align} \lb{ALadd}
\begin{split}
g_{\ell,+} - g_{\ell,+}^- &= \alpha h_{\ell,+} + \beta f_{\ell,+}^-, \quad \ell\in\bbN_0, \\
g_{\ell,-} - g_{\ell,-}^- &= \alpha h_{\ell,-}^- + \beta f_{\ell,-}, \quad \ell\in\bbN_0. 
\end{split}
\end{align}
Moreover, we record the following symmetries, 
\begin{equation}\lb{ALsym}
\hat f_{\ell,\pm}(c_{0,\pm},\alpha,\beta)=\hat h_{\ell,\mp}(c_{0,\mp},\beta, \alpha), \quad 
\hat g_{\ell,\pm}(c_{0,\pm},\alpha,\beta)=\hat g_{\ell,\mp}(c_{0,\mp},\beta, \alpha), \quad \ell\in \bbN_0.
\end{equation}

Next we define the $2\times 2$ zero-curvature matrices
\begin{equation}
U(z) = \begin{pmatrix} z & \alpha  \\ z \beta & 1\\ \end{pmatrix}    \lb{AL2.03}
\end{equation}
and 
\begin{equation} \lb{AL_v}
V_{\ul p}(z) = i  \begin{pmatrix}
G_{\ul p}^-(z) & - F_{\ul p}^-(z)     \\
H_{\ul p}^-(z) & - K_{\ul p}^-(z)  \\
\end{pmatrix},  \quad \ul p \in \bbN_0^2,
\end{equation}
for appropriate Laurent polynomials $F_{\ul p}(z)$, $G_{\ul p}(z)$, $H_{\ul p}(z)$, and $K_{\ul p}(z)$ in the  
spectral parameter $z\in \bbC\setminus\{0\}$ to be determined shortly. By postulating the stationary zero-curvature relation,  
\begin{equation}   \lb{ALstatzc}
0=U V_{\ul p} - V_{\ul p}^+ U,
\end{equation}
one concludes that \eqref{ALstatzc} is equivalent to the following relations 
 \begin{align} \label{AL1,1}
z (G_{\ul p}^- - G_{\ul p}) + z \beta F_{\ul p} + \alpha H_{\ul p}^- &= 0,\\ \label{AL2,2}
z \beta F_{\ul p}^- + \alpha H_{\ul p} - K_{\ul p} + K_{\ul p}^- &= 0,\\
\label{AL1,2}
 - F_{\ul p} + z F_{\ul p}^- + \alpha (G_{\ul p} + K_{\ul p}^-) &= 0,\\ \label{AL2,1}
 z \beta (G_{\ul p}^- + K_{\ul p}) - z H_{\ul p} + H_{\ul p}^- &= 0.
\end{align}
In order to make the connection between the zero-curvature formalism and the recursion relations 
\eqref{AL0+}--\eqref{ALh_l-}, we now define Laurent polynomials $F_{\ul p}$, $G_{\ul p}$, $H_{\ul p}$, and $K_{\ul p}$, $\ul p=(p_-,p_+)\in\bbN_0^2$, by\footnote{In this paper, a sum is interpreted as zero whenever the upper limit in the sum is strictly less than its lower limit.}
\begin{align}
F_{\ul p}(z) &= \sum_{\ell=1}^{p_-} f_{p_- -\ell,-} z^{-\ell} + \sum_{\ell=0}^{p_+ -1} f_{p_+-1-\ell,+}z^\ell,  
\label{ALF_p} \\ 
G_{\ul p}(z) &= \sum_{\ell=1}^{p_-} g_{p_- -\ell,-}z^{-\ell}  + \sum_{\ell=0}^{p_+} 
g_{p_+ -\ell,+}z^\ell,  
 \label{ALG_p}  \\ 
H_{\ul p}(z) &= \sum_{\ell=0}^{p_- -1} h_{p_- -1-\ell,-}z^{-\ell}  + \sum_{\ell=1}^{p_+} h_{p_+ -\ell,+}z^\ell, 
 \label{ALH_p}  \\
K_{\ul p}(z) &= \sum_{\ell=0}^{p_-} g_{p_- -\ell,-}z^{-\ell}  +  \sum_{\ell=1}^{p_+} 
g_{p_+ -\ell,+}z^\ell 
= G_{\ul p}(z)+g_{p_-,-}-g_{p_+,+}.   \label{ALK_p}
\end{align}

The corresponding homogeneous quantities are defined by ($\ell\in\bbN_0$)
\begin{align} 
\begin{split}
\hatt F_{0,\mp} (z) & = 0, \quad 
\hatt F_{\ell,-}(z) = \sum_{k=1}^\ell \hat f_{\ell-k,-} z^{-k}, \quad 
\hatt F_{\ell,+}(z) =  \sum_{k=0}^{\ell-1}
 \hat f_{\ell-1-k,+}z^k,   \\ 
\hatt G_{0,-} (z) &  = 0,   \quad 
\hatt G_{\ell,-} (z) = \sum_{k=1}^\ell \hat g_{\ell-k,-}z^{-k},  \\ 
\hatt G_{0,+} (z) &  = \f{1}{2},   \quad 
\hatt G_{\ell,+} (z) = \sum_{k=0}^\ell  \hat g_{\ell-k,+}z^k, \\
\hatt H_{0,\mp} (z) &  = 0, \quad 
\hatt H_{\ell,-} (z) = \sum_{k=0}^{\ell-1} \hat h_{\ell-1-k,-} z^{-k}, \quad 
\hatt H_{\ell,+} (z) =  \sum_{k=1}^\ell \hat h_{\ell-k,+}z^k,      \label{ALhat_F_p} \\
\hatt K_{0,-} (z) & = \f{1}{2},  \quad 
\hatt K_{\ell,-} (z) = \sum_{k=0}^\ell \hat g_{\ell-k,-}z^{-k} 
= \hatt G_{\ell,-} (z)+\hat g_{\ell,-}, \\
\hatt K_{0,+} (z) & = 0,  \quad 
\hatt K_{\ell,+} (z) =  \sum_{k=1}^\ell \hat g_{\ell-k,+}z^k 
= \hatt G_{\ell,+} (z)-\hat g_{\ell,+}.  
\end{split}
\end{align}

The stationary zero-curvature relation \eqref{ALstatzc}, $0=U V_{\ul p} - V_{\ul p}^+ U$, is then equivalent to 
\begin{align}
 -\alpha(g_{p_+,+} + g_{p_-,-}^-) + f_{p_+ -1,+} - f_{p_- -1,-}^-&=0,  \lb{AL2.50}\\ 
 \beta(g_{p_+,+}^- + g_{p_-,-}) + h_{p_+ -1,+}^- - h_{p_- -1,-} &=0.  \lb{AL2.51}
\end{align}
Thus, varying $p_\pm \in \bbN_0$,  equations \eqref{AL2.50} and \eqref{AL2.51} give rise to the stationary Ablowitz--Ladik (AL) hierarchy which we introduce as follows
\begin{equation}\lb{ALstat}
\sAL_{\ul p}(\alpha, \beta) = \begin{pmatrix}
- \alpha(g_{p_+,+} + g_{p_-,-}^-) + f_{p_+ -1,+} - f_{p_- -1,-}^-\\  
\beta(g_{p_+,+}^- + g_{p_-,-}) + h_{p_+ -1,+}^- - h_{p_- -1,-}  \end{pmatrix}=0, \quad 
\ul p\in\bbN_0^2. 
\end{equation}
Explicitly (recalling $\gamma=1-\alpha\beta$ and taking $p_-=p_+$ for simplicity), 
\begin{align} \no
\sAL_{(0,0)} (\alpha, \beta) &=  \begin{pmatrix}  -c_{(0,0)} \alpha\\
c_{(0,0)}\beta\end{pmatrix} 
=0,\\ \no
\sAL_{(1,1)} (\alpha, \beta) &=  \begin{pmatrix} -\gamma (c_{0,-}\alpha^- + c_{0,+}\alpha^+) 
- c_{(1,1)} \alpha \\
 \gamma (c_{0,+}\beta^- + c_{0,-}\beta^+) +
c_{(1,1)} \beta\end{pmatrix}=0,\\ \no
\sAL_{(2,2)} (\alpha, \beta) &=  \begin{pmatrix}\begin{matrix}
-\gamma \big(c_{0,+}\alpha^{++} \gamma^+ + c_{0,-}\alpha^{--} \gamma^-
- \alpha (c_{0,+}\alpha^+\beta^- + c_{0,-}\alpha^-\beta^+)\\
- \beta (c_{0,-}(\alpha^-)^2 + c_{0,+}(\alpha^+)^2)\big)\end{matrix}\\[3mm] 
\begin{matrix}
 \gamma \big(c_{0,-}\beta^{++} \gamma^+ + c_{0,+}\beta^{--} \gamma^-
- \beta (c_{0,+}\alpha^+\beta^- + c_{0,-}\alpha^-\beta^+)\\
- \alpha (c_{0,+}(\beta^-)^2 + c_{0,-}(\beta^+)^2)\big)\end{matrix}\end{pmatrix}
 \\ & \quad+ \begin{pmatrix}
-\gamma (c_{1,-} \alpha^- + c_{1,+} \alpha^+) - c_{(2,2)} \alpha\\
 \gamma (c_{1,+} \beta^- + c_{1,-} \beta^+) + c_{(2,2)} \beta\end{pmatrix}
=0,  \, \text{ etc.,}
\end{align}
represent the first few equations of the stationary Ablowitz--Ladik hierarchy. 
Here we introduced  
\begin{equation}
c_{\ul p} = (c_{p_-,-} + c_{p_+,+})/2, \quad p_\pm \in\bbN_0.   \lb{ALdefcp}
\end{equation}
By definition, the set of solutions of \eqref{ALstat}, with $\ul p$ ranging in $\bbN_0^2$ and $c_{\ell,\pm}\in\bbC$, $\ell\in\bbN_0$, represents the class of algebro-geometric Ablowitz--Ladik solutions. 

Using \eqref{AL2.01}, one can show (cf.\ \cite{GesztesyHoldenMichorTeschl:2006}) that $g_{p_+,+} = g_{p_-,-}$ up to a lattice constant which can be set equal to zero without loss of generality. Thus, we will henceforth assume that
\begin{equation}
g_{p_+,+} = g_{p_-,-},   \lb{g+=g-}
\end{equation} 
which in turn implies that
\begin{equation}
K_{\ul p}= G_{\ul p}   \lb{ALK=G}
\end{equation}
and hence renders $V_{\ul p}$ in \eqref{AL_v} traceless in the stationary context. (We note that equations \eqref{g+=g-} and \eqref{ALK=G} cease to be valid in the time-dependent context, though.)

\bigskip
Next we turn to the time-dependent Ablowitz--Ladik hierarchy. For that purpose the coefficients 
$\alpha$ and $\beta$ are now considered as functions of both the lattice point and time. For each 
equation in the hierarchy, that is, for each $\ul p$, we introduce a deformation (time) parameter 
$t_{\ul p}\in\bbR$ in $\alpha, \beta$, replacing $\alpha(n), \beta(n)$ by $\alpha(n,t_{\ul p}), \beta(n,t_{\ul p})$. 
Moreover, the definitions \eqref{AL2.03},  \eqref{AL_v}, and \eqref{ALF_p}--\eqref{ALK_p}  of 
$U, V_{\ul p}$, and $F_{\ul p}, G_{\ul p}, H_{\ul p}, K_{\ul p}$, respectively, still apply. Imposing 
the zero-curvature relation
\begin{equation}
U_{t_{\ul p}} + U V_{\ul p} - V_{\ul p}^+ U =0, \quad \ul p\in\bbN_0^2,   \lb{ALzc p}
\end{equation}
then results in the equations
\begin{align}  \label{ALalphat}
\alpha_{t_{\ul p}} &= i \big(z F_{\ul p}^- + \alpha (G_{\ul p} + K_{\ul p}^-) - F_{\ul p}\big),    \\ \label{ALbetat}
\beta_{t_{\ul p}} &= - i \big(\beta (G_{\ul p}^- + K_{\ul p}) - H_{\ul p} 
+ z^{-1} H_{\ul p}^-\big), \\ \label{AL1,1r}
0 &= z (G_{\ul p}^- - G_{\ul p}) + z\beta F_{\ul p} + \alpha H_{\ul p}^-,   \\ \label{AL2,2r}
0 &= z \beta F_{\ul p}^- + \alpha H_{\ul p} + K_{\ul p}^- - K_{\ul p}.
\end{align}
Varying $\ul p \in \bbN_0^2$, the collection of evolution equations   
\begin{align}   \label{AL_p}
\begin{split}
& \AL_{\ul p} (\alpha, \beta) =
\begin{pmatrix}-i\alpha_{t_{\ul p}} 
- \alpha(g_{p_+,+} + g_{p_-,-}^-) + f_{p_+ -1,+} - f_{p_- -1,-}^-\\
  -i\beta_{t_{\ul p}}+ \beta(g_{p_+,+}^- + g_{p_-,-}) - h_{p_- -1,-} + h_{p_+ -1,+}^- \end{pmatrix}=0,  \\
& \hspace*{8.15cm} t_{\ul p}\in\bbR, \; \ul p\in\bbN_0^2,   
\end{split}
\end{align}
then defines the time-dependent Ablowitz--Ladik hierarchy. Explicitly, taking 
$p_-=p_+$ for simplicity, 
\begin{align} \no \lb{ALh2.58}
& \AL_{(0,0)} (\alpha, \beta) =  \begin{pmatrix} -i \alpha_{t_{(0,0)}}- c_{(0,0)}\alpha \\
-i\beta_{t_{(0,0)}}+c_{(0,0)}\beta \end{pmatrix} 
=0,\\ \no
& \AL_{(1,1)} (\alpha, \beta) =  \begin{pmatrix}  -i \alpha_{t_{(1,1)}}- \gamma (c_{0,-}\alpha^- + c_{0,+}\alpha^+) 
- c_{(1,1)} \alpha \\
-i\beta_{t_{(1,1)}}+ \gamma (c_{0,+}\beta^- + c_{0,-}\beta^+) +
c_{(1,1)} \beta\end{pmatrix}=0,\\ \no
& \AL_{(2,2)} (\alpha, \beta)  \\
&\quad =  \begin{pmatrix}\begin{matrix}-i \alpha_{t_{(2,2)}}-
\gamma \big(c_{0,+}\alpha^{++} \gamma^+ + c_{0,-}\alpha^{--} \gamma^-
- \alpha (c_{0,+}\alpha^+\beta^- + c_{0,-}\alpha^-\beta^+)\\
- \beta (c_{0,-}(\alpha^-)^2 + c_{0,+}(\alpha^+)^2)\big)\end{matrix}\\[3mm] 
\begin{matrix}-i\beta_{t_{(2,2)}}+
 \gamma \big(c_{0,-}\beta^{++} \gamma^+ + c_{0,+}\beta^{--} \gamma^-
- \beta (c_{0,+}\alpha^+\beta^- + c_{0,-}\alpha^-\beta^+)\\
- \alpha (c_{0,+}(\beta^-)^2 + c_{0,-}(\beta^+)^2)\big)\end{matrix}\end{pmatrix}
\no \\ 
 & \qquad+ \begin{pmatrix}
-\gamma (c_{1,-} \alpha^- + c_{1,+} \alpha^+) - c_{(2,2)} \alpha\\
 \gamma (c_{1,+} \beta^- + c_{1,-} \beta^+) + c_{(2,2)} \beta\end{pmatrix}
=0, \, \text{ etc.,}   
\end{align}
represent the first few equations of the time-dependent Ablowitz--Ladik hierarchy. 
Here we recall the definition of $c_{\ul p}$ in \eqref{ALdefcp}.

By \eqref{AL_p}, \eqref{ALg_l+}, and \eqref{ALg_l-},
the time derivative of $\gamma=1-\alpha \beta$ is given by
\begin{equation} \lb{AL2.14}
\gamma_{t_{\ul p}} = i \gamma \big((g_{p_+,+} - g_{p_+,+}^-) 
- (g_{p_-,-} - g_{p_-,-}^-) \big),
\end{equation}
or, alternatively,
\begin{equation} \lb{AL2.14a}
\gamma_{t_{\ul p}} = i \gamma \big(\alpha z^{-1} H_{\ul p}^{-}-\alpha H_{\ul p}
+\beta F_{\ul p}-z \beta F_{\ul p}^-\big),
\end{equation}
using \eqref{ALalphat}--\eqref{AL2,2r}.

\begin{remark} \lb{rAL2.14}
$(i)$ The special choices $\beta=\pm\ol\alpha$, $c_{0,\pm}=1$ lead to the discrete nonlinear Schr\"odinger hierarchy. In particular, choosing $c_{(1,1)}=-2$ yields the discrete nonlinear Schr\"odinger equation in its usual form (see, e.g., 
\cite[Ch.\ 3]{AblowitzPrinariTrubatch:2004} and the references cited therein), 
\begin{equation}
-i\alpha_t - (1 \mp |\alpha|^2)(\alpha^- + \alpha^+) + 2\alpha = 0,   
\end{equation}
as its first nonlinear element. The choice $\beta = \ol \alpha$ is called the {\it defocusing} case, $\beta = - \ol \alpha$ represents the {\it focusing} case of the discrete nonlinear Schr\"odinger hierarchy. \\
$(ii)$ The alternative choice $\beta = \ol \alpha$, $c_{0,\pm} = \mp i$, leads to the hierarchy of Schur flows. In particular, choosing $c_{(1,1)} = 0$ yields  
\begin{equation}
\alpha_t - (1 - |\alpha|^2)(\alpha^+ - \alpha^-) = 0   
\end{equation}
as the first nonlinear element of this hierarchy (cf.\  \cite{AmmarGragg:1994}, 
\cite{FaybusovichGekhtman:1999}, \cite{FaybusovichGekhtman:2000},  
\cite{Golinskii:2006}, \cite{MukaihiraNakamura:2002}, \cite{Simon:2007}).
\end{remark}

\section{Lax pairs for the AL hierarchy}  \lb{sALh3}

In this section we introduce Lax pairs for the AL hierarchy and prove the 
equivalence of the zero-curvature and Lax representation. This result is new. 

Throughout this section we suppose Hypothesis \ref{hAL2.1}. We start by relating the homogeneous coefficients $\hat f_{\ell,\pm}$, 
$\hat g_{\ell,\pm}$, and $\hat h_{\ell,\pm}$ to certain matrix elements of $L$, where $L$ will later be identified as the Lax difference expression associated with the 
Ablowitz--Ladik hierarchy. For this purpose it is useful to introduce the standard basis 
$\{\delta_m\}_{m\in\bbZ}$ in $\ell^2(\bbZ)$ by
\begin{equation}
\delta_m=\{\delta_{m,n}\}_{n\in\bbZ}, \; m\in\bbZ, \quad
\delta_{m,n}=\begin{cases} 1, &m=n, \\ 0, & m\neq n. \end{cases}
\lb{ALbasis}
\end{equation}
The scalar product in $\ell^2(\bbZ)$, denoted by $(\dott,\dott) $, 
is defined by  
\begin{equation}
(f,g) =\sum_{n\in \bbZ} \ol{f(n)}g(n), \quad f,g \in \ell^2(\bbZ).
\lb{ALsp}
\end{equation}

In the standard basis just defined, we introduce the difference expression $L$ by
\begin{align}
L &= \left(\begin{smallmatrix} \ddots &&\hspace*{-8mm}\ddots
&\hspace*{-10mm}\ddots &\hspace*{-12mm}\ddots
&\hspace*{-14mm}\ddots &&&
\raisebox{-3mm}[0mm][0mm]{\hspace*{-6mm}{\Huge $0$}}
\\
&0& -\alpha(0) \rho(-1) & -\beta(-1)\alpha(0) &
-\alpha(1)\rho(0) & \rho(0) \rho(1)
\\
&& \rho(-1) \rho(0) & \beta(-1) \rho(0) &
-\beta(0) \alpha(1) & \beta(0) \rho(1) & 0
\\
&&&0& -\alpha(2) \rho(1) & -\beta(1) \alpha(2) &
-\alpha(3) \rho(2) & \rho(2) \rho(3)
\\
&&\raisebox{-4mm}[0mm][0mm]{\hspace*{-6mm}{\Huge $0$}} &&
\rho(1) \rho(2) & \beta(1) \rho(2) & -\beta(2) \alpha(3)
& \beta(2) \rho(3) & 0
\\
&&&&&\hspace*{-14mm}\ddots &\hspace*{-14mm}\ddots
&\hspace*{-14mm}\ddots &\hspace*{-8mm}\ddots &\ddots
\end{smallmatrix}\right)    \lb{ALLop} \\ 
&= \Big( -\beta(n)\alpha(n+1)\delta_{m,n}  \no \\
& \qquad 
+ \big(\beta(n-1)\rho(n)\dodd(n)-\alpha(n+1)\rho(n)\deven(n) \big)\delta_{m,n-1} \no \\  
\lb{ALLdelta}
&\qquad+\big(\beta(n)\rho(n+1)\dodd(n)-\alpha(n+2)\rho(n+1)\deven(n) \big)\delta_{m,n+1} \\
&\qquad+\rho(n+1)\rho(n+2)\deven(n)\delta_{m,n+2}
+\rho(n-1)\rho(n)\dodd(n)\delta_{m,n-2} \Big)_{m,n\in\bbZ}  \no \\
&= \rho^- \rho \, \deven \, S^{--} + (\beta^-\rho \, \deven - \alpha^+\rho \, \dodd) S^- 
- \beta\alpha^+   \no \\ \lb{ALLrec}
& \quad + (\beta \rho^+ \, \deven - \alpha^{++} \rho^+ \, \dodd) S^+ 
+ \rho^+ \rho^{++} \, \dodd \, S^{++}, 
\end{align}
where $\deven$ and $\dodd$ denote the characteristic functions of the even and odd integers,
\begin{equation}
\deven = \chi_{_{2\bbZ}}, \quad \dodd = 1 - \deven = \chi_{_{2\bbZ +1}}.
\end{equation}
In particular, terms of the form $-\beta(n) \alpha(n+1)$ 
represent the diagonal $(n,n)$-entries, $n\in\bbZ$, in the infinite matrix
\eqref{ALLop}. In addition, we used the abbreviation
\begin{equation}
\rho = \gamma^{1/2} = (1-\alpha \beta)^{1/2}.  \lb{ALLPrho}
\end{equation}

Next, we introduce the unitary operator $U_{\tilde \varepsilon}$ in $\ell^2(\bbZ)$ by 
\begin{equation}
U_{\tilde \varepsilon} = \big( \tilde \varepsilon(n) \delta_{m,n} \big)_{(m,n)\in\bbZ^2}, 
\quad \tilde\varepsilon(n)\in\{1,-1\}, \; n\in\bbZ,     \lb{ALUeps}
\end{equation}
and the sequence $\varepsilon =\{\varepsilon(n)\}_{n\in\bbZ}\in \bbC^{\bbZ}$ by
\begin{equation}
\varepsilon(n) = \tilde\varepsilon(n-1)\tilde\varepsilon(n), \; n\in\bbZ.    
\end{equation}
Assuming $\alpha, \beta \in \ell^\infty(\bbZ)$, a straightforward computation then shows that 
\begin{equation}
\breve L_{\varepsilon} = U_{\tilde \varepsilon} \breve L U_{\tilde \varepsilon}^{-1},   
\lb{ALUnEq}
\end{equation}
where $L_{\varepsilon}$ is associated with the sequences 
$\alpha_\varepsilon=\alpha$, $\beta_\varepsilon=\beta$, and 
$\rho_\varepsilon= \varepsilon \rho$, and $\breve L$ and $\breve L_\varepsilon$ are the 
bounded operator realizations of $L$ and $L_\varepsilon$ in $\ell^2(\bbZ)$, respectively. Moreover, the recursion formalism in \eqref{AL0+}--\eqref{ALh_l-} yields coefficients which are polynomials in $\alpha$, $\beta$ and some of their shifts and hence depends only quadratically on $\rho$. As a result, the choice of square root of $\rho(n)$, $n\in\bbZ$, in \eqref{ALLPrho} is immaterial when introducing the AL hierarchy via the Lax equations \eqref{ALLaxtp}.  

The half-lattice (i.e., semi-infinite) version of
$L$ was recently rediscovered by Cantero, Moral, and Vel\'azquez
\cite{CanteroMoralVelazquez:2003} in the special case where 
$\beta=\ol{\alpha}$ (see also Simon \cite{Simon:2005}, \cite{Simon:2006} who coined the term CMV matrix in this context). The matrix representation of $L^{-1}$ is then obtained from that of $L$ in \eqref{ALLop} by taking the formal adjoint of $L$ and subsequently 
exchanging $\alpha$ and $\beta$
\begin{align} \lb{ALL-1delta}
L^{-1}&= \Big( -\alpha(n)\beta(n+1)\delta_{m,n}
+ \big(\alpha(n-1)\rho(n)\deven(n)  \\ \no
& \qquad -\beta(n+1)\rho(n)\dodd(n) \big)\delta_{m,n-1}\\ \no
&\qquad+\big(\alpha(n)\rho(n+1)\deven(n)-\beta(n+2)\rho(n+1)\dodd(n) \big)\delta_{m,n+1} \\ \no
&\qquad+\rho(n+1)\rho(n+2)\dodd(n)\delta_{m,n+2}
+\rho(n-1)\rho(n)\deven(n)\delta_{m,n-2} \Big)_{m,n\in\bbZ}\\
&= \rho^- \rho \, \dodd \, S^{--} + (\alpha^-\rho \, \dodd - \beta^+\rho \, \deven) S^- 
- \alpha \beta^+   \no \\ \lb{ALL-1rec}
& \quad + (\alpha \rho^+ \, \dodd - \beta^{++} \rho^+ \, \deven) S^+ 
+ \rho^+ \rho^{++} \, \deven \, S^{++}.
\end{align}
$L$ and $L^{-1}$ define bounded operators in 
$\ell^2(\bbZ)$ if $\alpha$ and $\beta$ are bounded sequences. However, this is 
of no importance in the context of Lemma \ref{lALh3.1} below as we only apply the five-diagonal matrices $L$ and $L^{-1}$ to basis vectors of the type $\delta_m$.

Next, we discuss a useful factorization of $L$. For this purpose we introduce the sequence of $2\times 2$ matrices $\theta(n)$, $n\in\bbZ$, by
\begin{equation} \label{ALtheta}
\theta(n) = \begin{pmatrix} -\alpha(n) & \rho(n) \\ \rho(n) &
\beta(n) \end{pmatrix}, \quad n \in \bbZ,
\end{equation}
and two difference expressions $D$ and $E$ by their
matrix representations in the standard basis \eqref{ALbasis} of $\ell^2(\bbZ)$ 
\begin{align} 
D = \left(\begin{smallmatrix} \ddots & & &
\raisebox{-1mm}[0mm][0mm]{\hspace*{-5mm}\Huge $0$}  \\ 
& \theta(2n-2) & & \\  & & \theta(2n) & &  \\ & 
\raisebox{1mm}[0mm][0mm]{\hspace*{-10mm}\Huge $0$} & &
\ddots
\end{smallmatrix}\right),   \quad 
E =  \left(\begin{smallmatrix} \ddots & & &
\raisebox{-1mm}[0mm][0mm]{\hspace*{-5mm}\Huge $0$}  \\ 
& \theta(2n-1) & & \\  & & \theta(2n+1) & &  \\ & 
\raisebox{1mm}[0mm][0mm]{\hspace*{-10mm}\Huge $0$} & &
\ddots
\end{smallmatrix}\right),   \label{ALDE}
\end{align}  
where
\begin{align}
\begin{split}
\begin{pmatrix}
D(2n-1,2n-1) & D(2n-1,2n) \\ D(2n,2n-1)   & D(2n,2n)
\end{pmatrix} & =  \theta(2n),  \\
\begin{pmatrix}
E(2n,2n) & E(2n,2n+1) \\ E(2n+1,2n)  & E(2n+1,2n+1)
\end{pmatrix} & =  \theta(2n+1),
\quad n\in\bbZ.
\end{split}
\end{align}
Then $L$ can be factorized into   
\begin{equation}  
L = DE.   \lb{ALfact}
\end{equation}
Explicitly, $D$ and $E$ are given by
\begin{align}
D&=\rho \, \deven \, S^- - \alpha^+ \, \dodd +\beta \, \deven + \rho^+ \, \dodd \, S^+,  
\lb{ALD} \\
E&=\rho \, \dodd \, S^- + \beta \, \dodd - \alpha^+ \, \deven + \rho^+ \, \deven \, S^+,  
\lb{ALE}
\end{align}
and their inverses are of the form
\begin{align}
D^{-1} &= \rho \, \deven \, S^- - \beta^+ \, \dodd +\alpha \, \deven + \rho^+ \, \dodd \, S^+,  
\lb{ALD-1} \\
E^{-1} &=\rho \, \dodd \, S^- + \alpha \, \dodd - \beta^+ \, \deven + \rho^+ \, \deven \, S^+.  
\lb{ALE-1}
\end{align}

The next result details the connections between $L$ and the recursion coefficients $f_{\ell,\pm}$, $g_{\ell,\pm}$, and $h_{\ell,\pm}$.  

\begin{lemma} \lb{lALh3.1}
Let $n\in\bbZ$. Then the homogeneous coefficients 
$\{\hat f_{\ell,\pm}\}_{\ell\in\bbN_0}$, $\{\hat g_{\ell,\pm}\}_{\ell\in\bbN_0}$, and 
$\{\hat h_{\ell,\pm}\}_{\ell\in\bbN_0}$ satisfy the following relations:
\begin{align} \lb{ALh3.5}
\hat f_{\ell,+}(n) &=  
(\delta_{n},E L^\ell \delta_{n}) \deven(n)+
(\delta_{n},L^\ell D\delta_{n}) \dodd(n),
\quad \ell\in\bbN_0,  \no \\
 \hat f_{\ell,-}(n) &=
(\delta_{n},D^{-1}L^{- \ell} \delta_{n}) \deven(n)+
(\delta_{n},L^{- \ell}E^{-1} \delta_{n}) \dodd(n), 
\quad \ell\in\bbN_0, \no \\
\hat g_{0,\pm} &= 1/2, \quad 
\hat g_{\ell,\pm}(n) = (\delta_{n},L^{\pm \ell} \delta_{n}) ,  \quad 
\ell\in\bbN,  \\
\hat h_{\ell,+}(n) &=
(\delta_{n},L^{ \ell}D \delta_{n}) \deven(n)+
(\delta_{n},EL^{ \ell} \delta_{n}) \dodd(n),  
\quad \ell\in\bbN_0, \no  \\
 \hat h_{\ell,-}(n) &=
(\delta_{n},L^{- \ell} E^{-1} \delta_{n}) \deven(n)+
(\delta_{n},D^{-1} L^{- \ell} \delta_{n}) \dodd(n)
\;\; \ell\in\bbN_0.  \no
\end{align}
\end{lemma}
\begin{proof}
Using \eqref{ALfact}--\eqref{ALE-1} we show that the sequences defined in \eqref{ALh3.5} satisfy the recursion 
relations of Lemma \ref{lALB.1} respectively relation \eqref{ALg_l+}. 
For $n$ even,
\begin{align}
\begin{split}
\hat g_{\ell,+}(n) - \hat g_{\ell,+}(n-1) &=(\delta_n, DEL^{\ell-1} \delta_{n}) - (\delta_{n-1}, DEL^{\ell-1} \delta_{n-1})\\
&=(D^*\delta_n, EL^{\ell-1} \delta_{n}) - (D^* \delta_{n-1}, EL^{\ell-1} \delta_{n-1})\\
&=\beta(n)(\delta_n, EL^{\ell-1} \delta_{n}) + \rho(n) (\delta_{n-1}, EL^{\ell-1} \delta_{n}) \\
&\quad + \alpha(n)(\delta_{n-1}, EL^{\ell-1} \delta_{n-1}) - \rho(n)(\delta_{n}, EL^{\ell-1} \delta_{n-1}) \\
&= \beta(n) \hat f_{\ell-1,+}(n) + \alpha(n) \hat h_{\ell-1,+}(n-1),
\end{split}
\end{align}
since $(E L^\ell)^\top=E L^\ell$ by \eqref{ALtheta}, \eqref{ALfact}. Moreover,
\begin{align} 
\begin{split}
\hat f_{\ell,+}(n) &= (\delta_n, EL^{\ell} \delta_{n}) 
= (E^*\delta_n, L^{\ell} \delta_{n})\\ 
&= - \alpha(n+1)(\delta_n, L^{\ell} \delta_{n}) 
+ \rho(n+1)(\delta_{n+1}, L^{\ell} \delta_{n})\\ 
&\quad + \alpha(n+1)(\delta_{n+1}, L^{\ell} \delta_{n+1})
- \alpha(n+1)(\delta_{n+1}, L^{\ell} \delta_{n+1})\\
&= \hat f_{\ell-1,+}(n+1)-\alpha(n+1)\big(\hat g_{\ell,+}(n+1)+\hat g_{\ell,+}(n)\big),\\ 
\hat h_{\ell,+}(n) &=(\delta_{n},L^{ \ell}D \delta_{n})=
\beta(n)(\delta_n, L^{\ell} \delta_{n}) 
+ \rho(n)(\delta_{n}, L^{\ell} \delta_{n-1})\\ 
&\quad + \beta(n)(\delta_{n-1}, L^{\ell} \delta_{n-1})
- \beta(n)(\delta_{n-1}, L^{\ell} \delta_{n-1})\\ 
&= \hat h_{\ell-1,+}(n-1)+\beta(n)\big(\hat g_{\ell,+}(n)+\hat g_{\ell,+}(n-1)\big),  
\end{split}
\end{align}
that is, the coefficients satisfy \eqref{ALB1.9}.   
The remaining cases follow analogously.
\end{proof}

Finally, we derive an explicit expression for the Lax pair for the Ablowitz--Ladik hierarchy,  but first we need some notation. Let $T$ be a bounded operator in $\ell^2(\bbZ)$. Given the standard basis \eqref{ALbasis} in $\ell^2(\bbZ)$, we represent $T$ by
\begin{equation}
T=\big(T(m,n)\big)_{(m,n)\in\bbZ^2}, \quad 
T(m,n)=(\delta_m,T \, \delta_n) , \quad (m, n) \in\bbZ^2. \lb{ALTop}
\end{equation}
Actually, for our purpose below, it is sufficient that $T$ is an $N$-diagonal matrix for some $N\in\bbN$. Moreover, we introduce the upper and lower triangular parts $T_\pm$ of $T$ by
\begin{equation}
T_\pm=\big(T_\pm (m,n)\big)_{(m,n)\in\bbZ^2}, \quad
T_\pm (m,n)=\begin{cases} T(m,n), &\pm(n-m)>0, \\ 0, & \text{otherwise.}
\end{cases}
\lb{ALTpm}
\end{equation}

Next, consider the finite difference expression $P_{\ul p}$ defined by 
\begin{align} \no
P_{\ul p}& = \f{i}{2} \sum_{\ell=1}^{p_+} c_{p_+ -\ell,+} \big( (L^\ell)_+ - (L^\ell)_- \big)
- \f{i}{2} \sum_{\ell=1}^{p_-} c_{p_- -\ell,-} \big( (L^{-\ell})_+ - (L^{-\ell})_- \big)  - \f{i}{2} c_{\ul p} \, Q_d,  \\
& \hspace{9cm} \ul p\in\bbN_0^2,    \lb{ALP_p} 
\end{align} 
with $L$ given by \eqref{ALLop} and $Q_d$ denoting the doubly infinite diagonal matrix
\begin{equation} 
Q_d=\big((-1)^k \delta_{k,\ell} \big)_{k,\ell \in\bbZ}.    \lb{ALQ_d}
\end{equation}

Before we prove that $(L,P_{\ul p})$ is indeed the Lax pair for the Ablowitz--Ladik hierarchy, we derive one more representation of $P_{\ul p}$ in terms of $L$. 

We denote by $\ell_0(\bbZ)$ the set of complex-valued sequences of compact support. If $R$ denotes a finite difference expression, then $\psi$ is called a weak solution of 
$R\psi = z\psi$, for some $z\in\bbC$, if the relation holds pointwise for each lattice point, that is, if $((R - z)\psi)(n) = 0$ for all $n\in \bbZ$.

\begin{lemma}  \lb{lALh3.2}
Let $\psi\in \ell^\infty_0(\bbZ)$. Then the difference expression  $P_{\ul p}$ defined in \eqref{ALP_p} acts on $\psi$ by  
\begin{align}
 (P_{\ul p} \psi)(n) 
& = i\bigg(-\sum_{\ell=1}^{p_-} f_{p_- -\ell,-}(n)(E L^{-\ell} \psi)(n) 
- \sum_{\ell=0}^{p_+ -1} f_{p_+ -1-\ell,+}(n)(E L^\ell \psi)(n) \no\\
&\qquad \; +\sum_{\ell=1}^{p_-} g_{p_- -\ell,-}(n)(L^{-\ell}\psi)(n) 
+ \sum_{\ell=1}^{p_+} g_{p_+ -\ell,+}(n)(L^{\ell}\psi)(n) \no\\
&\qquad \; + \f{1}{2}\big(g_{p_-,-}(n)+g_{p_+,+}(n)\big)\psi(n)\bigg)\dodd(n)  \lb{ALgPp} \\
&\quad +i\bigg(\sum_{\ell=0}^{p_- -1} h_{p_--1-\ell,-}(n)(D^{-1}L^{-\ell}\psi)(n) 
+ \sum_{\ell=1}^{p_+} h_{p_+-\ell,+}(n)(D^{-1}L^{\ell}\psi)(n) \no\\
&\qquad \quad - \sum_{\ell=1}^{p_-} g_{p_- -\ell,-}(n)(L^{-\ell}\psi)(n) 
- \sum_{\ell=1}^{p_+} g_{p_+-\ell,+}(n)(L^{\ell}\psi)(n) \no\\
&\qquad \quad - \f{1}{2}\big(g_{p_-,-}(n)+g_{p_+,+}(n)\big)\psi(n)\bigg)\deven(n), \quad 
n\in\bbZ.    \no 
\end{align}
In addition, if $u$ is a weak solution of $L u(z) = zu(z)$, then 
\begin{align}
& \big(P_{\ul p} u(z)\big)(n)  \no \\   
& \quad  =\bigg(- 
 iF_{\ul p}(z,n) (E u(z))(n) + \f{i}{2}\big(G_{\ul p}(z,n)
 + K_{\ul p}(z,n)\big)u(z,n)\bigg)\dodd(n) \no \\
& \qquad +\bigg( iH_{\ul p}(z,n) (D^{-1}u(z))(n) 
- \f{i}{2}\big(G_{\ul p}(z,n)+ K_{\ul p}(z,n)\big)u(z,n)\bigg)\deven(n), \no \\
& \hspace*{10cm} n\in\bbZ,   \lb{ALPp}
\end{align} 
in the weak sense. 
\end{lemma}
\begin{proof}
We consider the case where $n$ is even and use induction on
$\ul p=(p_-,p_+)$. The case $n$ odd is analogous.
For $\ul p=(0,0)$, the formulas \eqref{ALgPp} and \eqref{ALP_p} match. 
Denoting by $\hatt P_{\ul p}$ the corresponding homogeneous operator where all summation 
constants $c_{k,\pm}$, $k=1, \dots, p_\pm$, vanish, we have to show that
\begin{align} 
\begin{split}
i \hatt P_{\ul p} &= i \hatt P_{p_+ -1}^+ L - \hat h_{p_+-1,+}D^{-1}L + \f{1}{2} \big(\hat g_{p_+-1,+}L 
+ \hat g_{p_+,+} \big)   \\
&\quad+ i\hatt P_{p_--1}^- L^{-1} - \hat h_{p_--1,-}D^{-1} 
 + \f{1}{2} \big(\hat g_{p_--1,-}L^{-1} + \hat g_{p_-,-}\big),
 \end{split}  
\end{align}
where $\hatt P_j^\pm$ correspond to the powers of $L$ in \eqref{ALP_p}, 
$\hatt P_j^\pm =\f{i}{2} \big((L^{\pm j})_\pm - (L^{\pm j})_\mp\big)$. This can be done upon considering $(\delta_m, \hatt P_{\ul p} \delta_n) $ and making appropriate case distinctions $m=n$, $m>n$, and $m<n$.

Using \eqref{ALLdelta}, \eqref{ALL-1delta}, \eqref{ALfact}--\eqref{ALE-1}, \eqref{ALTpm}, and Lemma \ref{lALh3.1},
one verifies for instance in the case $m=n$,
\begin{align} \no
& (\delta_n, i\hatt P_{p_+}^+ \delta_n) \\ \no
&= (\delta_n, i\hatt P_{p_+-1}^+ L \delta_n) + \alpha(n+1)\hat h_{p_+-1,+}(n) \\ \no
&\quad + \f{1}{2}\big(\hat g_{p_+,+}(n) - \alpha(n+1)\beta(n)\hat g_{p_+-1,+}(n)\big)\\ \no
&= (\delta_n, \f{1}{2}\big((L^{p_+-1})_+ - (L^{p_+-1})_-\big)
\big(\alpha^{++}\rho^+\delta_{n-1} + \alpha^+\beta\delta_n   + \alpha^+\rho\delta_{n+1} 
- \rho^-\rho\delta_{n+2}\big))\\ \no
&\quad + \alpha(n+1)\hat h_{p_+-1,+}(n) 
 + \f{1}{2}\big(\hat g_{p_+,+}(n) - \alpha(n+1)\beta(n)\hat g_{p_+-1,+}(n)\big)\\ \no
&=-\f{1}{2}\alpha(n+1)\rho(n)(\delta_n,L^{p_+-1}\delta_{n-1})
+\f{1}{2}\alpha(n+2)\rho(n+1)(\delta_n,L^{p_+-1}\delta_{n+1})\\ \no
&\quad -\f{1}{2}\rho(n+1)\rho(n+2)(\delta_n,L^{p_+-1}\delta_{n+2})
+ \alpha(n+1)\hat h_{p_+-1,+}(n)\\ \no
&\quad + \f{1}{2}\big(\hat g_{p_+,+}(n) - \alpha(n+1)\beta(n)\hat g_{p_+-1,+}(n)\big)\\ \no
&=-\alpha(n+1)\rho(n)(\delta_n,L^{p_+-1}\delta_{n-1}) \\ \no  
&\quad - \alpha(n+1)\beta(n)\hat g_{p_+-1,+}(n) + \alpha(n+1)\hat h_{p_+-1,+}(n) \\
& =0,
\end{align}
since by Lemma \ref{lALh3.1}, 
\begin{align} \no
\begin{split}
\hat g_{p_+,+}(n) &= (\delta_{n},L^{p_+-1}L \delta_{n}),     \\  
\hat h_{p_+-1,+}(n) &= (\delta_n, L^{p_+-1}D\delta_n)=\beta(n)(\delta_n, L^{p_+-1}\delta_n)
+\rho(n)(\delta_n, L^{p_+-1}\delta_{n-1}). 
\end{split}
\end{align}
Similarly,
\begin{align} \no
&(\delta_n, i\hatt P_{p_-}^- \delta_n)\\ \no
&= (\delta_n, i\hatt P_{p_--1}^- L^{-1} \delta_n) - \alpha(n)\hat h_{p_--1,-}(n) 
+ \f{1}{2}\big(\hat g_{p_-,-}(n) - \alpha(n)\beta(n+1)\hat g_{p_--1,-}(n)\big)\\ \no
&= (\delta_n, \f{1}{2}\big( (L^{1-p_-})_+ - (L^{1-p_-})_-\big)
\big( \rho^+\rho^{++}\delta_{n-2}+\alpha\rho^+\delta_{n-1} -\alpha\beta^+\delta_n  
+ \alpha^-\rho\delta_{n+1} \big))\\ \no
&\quad - \alpha(n)\hat h_{p_--1,-}(n) 
+ \f{1}{2}\big(\hat g_{p_-,-}(n) - \alpha(n)\beta(n+1)\hat g_{p_--1,-}(n)\big)\\ \no
&= -\f{1}{2}\rho(n-1)\rho(n)(\delta_n,L^{1-p_-}\delta_{n-2})
-\f{1}{2}\alpha(n-1)\rho(n)(\delta_n,L^{1-p_-}\delta_{n-1})\\ \no
&\quad+\f{1}{2}\alpha(n)\rho(n+1)(\delta_n,L^{1-p_-}\delta_{n+1})
- \alpha(n)\hat h_{p_--1,-}(n)\\ \no
&\quad + \f{1}{2}\big(\hat g_{p_-,-}(n) - \alpha(n)\beta(n+1)\hat g_{p_--1,-}(n)\big)\\ \no
&=\alpha(n)\rho(n+1)(\delta_n,L^{1-p_-}\delta_{n+1})
- \alpha(n)\beta(n+1)\hat g_{p_--1,-}(n) - \alpha(n)\hat h_{p_--1,-}(n)\\  
&=0,
\end{align}
where we used Lemma \ref{lALh3.1} and \eqref{ALh_l-} at $\ell=p_--2$
for the last equality. This proves the case $m=n$. The remaining
cases $m>n$ and $m<n$ are settled in a similar fashion.

Equality \eqref{ALPp} then follows from $L u(z) = z u(z)$ and 
\eqref{ALF_p}--\eqref{ALK_p}.
\end{proof}

Next, we introduce the difference expression $P^\top_{\ul p}$ by
\begin{align} 
\begin{split}
P^\top_{\ul p}& = -\f{i}{2} \sum_{\ell=1}^{p_+} c_{p_+ -\ell,+} \big( ((L^\top)^\ell)_+  
 - ((L^\top)^\ell)_- \big)   \\
& \quad + \f{i}{2} \sum_{\ell=1}^{p_-} c_{p_- -\ell,-} \big( ((L^\top)^{-\ell})_+ 
- ((L^\top)^{-\ell})_- \big)  - \f{i}{2} c_{\ul p} \, Q_d,  \quad  
\ul p\in\bbN_0^2,       \lb{ALP_ptop} 
\end{split}
\end{align} 
with $L^\top=ED$ the difference expression associated with the transpose of the infinite matrix \eqref{ALLop} in the standard basis of $\ell^2(\bbZ)$ and $Q_d$ denoting the doubly infinite diagonal matrix in \eqref{ALQ_d}. Here we used 
\begin{equation}
(M_+)^\top = (M^\top)_-, \quad (M_-)^\top = (M^\top)_+
\end{equation}
for a finite difference expression $M$ in the standard basis of $\ell^2(\bbZ)$. 

For later purpose in Section \ref{sALh5} we now mention the analog of Lemma 
\ref{lALh3.2} for the difference expression $P_{\ul p}^\top$ without proof:

\begin{lemma}  \lb{lALh3.2top}
Let $\chi\in \ell^\infty_0(\bbZ)$. Then the difference expression  $P_{\ul p}^\top$ defined in \eqref{ALP_ptop} acts on $\chi$ by  
\begin{align}
(P_{\ul p}^\top \chi)(n) 
& = i\bigg(-\sum_{\ell=0}^{p_- -1} h_{p_--1-\ell,-}(n)(E^{-1}(L^\top)^{-\ell}\chi)(n) \no \\
&\qquad \;\, - \sum_{\ell=1}^{p_+} h_{p_+-\ell,+}(n)(E^{-1}(L^\top)^{\ell}\chi)(n) \no\\
&\qquad \;\, + \sum_{\ell=1}^{p_-} g_{p_- -\ell,-}(n)((L^\top)^{-\ell}\chi)(n) 
+ \sum_{\ell=1}^{p_+} g_{p_+-\ell,+}(n)((L^\top)^{\ell}\chi)(n) \no\\
&\qquad \;\, + \f{1}{2}\big(g_{p_-,-}(n)+g_{p_+,+}(n)\big)\chi(n)\bigg)\dodd(n)  \lb{ALgPpTop}  \\   
&\quad +i\bigg(\sum_{\ell=1}^{p_-} f_{p_- -\ell,-}(n)(D (L^\top)^{-\ell} \chi)(n) \no \\
& \qquad \quad + \sum_{\ell=0}^{p_+ -1} f_{p_+ -1-\ell,+}(n)(D (L^\top)^\ell \chi)(n) \no\\
&\qquad \quad -\sum_{\ell=1}^{p_-} g_{p_- -\ell,-}(n)((L^\top)^{-\ell}\chi)(n) 
- \sum_{\ell=1}^{p_+} g_{p_+ -\ell,+}(n)((L^\top)^{\ell}\chi)(n) \no\\
&\qquad \quad - \f{1}{2}\big(g_{p_-,-}(n)+g_{p_+,+}(n)\big)\chi(n)\bigg)\deven(n), \quad 
n\in\bbZ. \no  
\end{align}
In addition, if $\chi$ is a weak solution of $L^\top v(z) = z v(z)$, then 
\begin{align}
& \big(P_{\ul p}^\top v(z)\big)(n)  \no \\   
& \quad  =-i\bigg(
 H_{\ul p}(z,n) (E^{-1}v(z))(n) - \f{1}{2}\big(G_{\ul p}(z,n) 
 + K_{\ul p}(z,n)\big)v(z,n)\bigg)\dodd(n) \no \\
& \qquad +i\bigg( F_{\ul p}(z,n) (D v(z))(n) - \f{1}{2}\big(G_{\ul p}(z,n) 
+ K_{\ul p}(z,n)\big)v(z,n)\bigg)\deven(n), \no \\
& \hspace*{10cm} n\in\bbZ,   \lb{ALPpTop}
\end{align} 
in the weak sense. 
\end{lemma}

Given these preliminaries, one can now prove the following result, the proof of which is  based on fairly tedious computations. We present them here in some detail as these results have not appeared in print before.

\begin{theorem} \lb{tALh3.3}
Assume Hypothesis \ref{hAL2.1}. Then, for each $\ul p\in\bbN_0^2$, the $\ul p$th stationary Ablowitz--Ladik equation $\sAL_{\ul p}(\alpha,\beta)=0$ in \eqref{ALstat} is equivalent to
the vanishing of the commutator of $P_{\ul p}$ and $L$, 
\begin{equation}
[P_{\ul p},L]=0.
\end{equation}
In addition, the $\ul p$th time-dependent Ablowitz--Ladik equation $\AL_{\ul p} (\alpha, \beta) = 0$ in \eqref{AL_p} is equivalent to the Lax commutator equations  
\begin{equation}\label{ALLaxtp}
L_{t_{\ul p}}(t_{\ul p}) - [P_{\ul p}(t_{\ul p}), L(t_{\ul p})] =0, \quad t_{\ul p}\in\bbR.   
\end{equation}
In particular, the pair of difference expressions $(L,P_{\ul p})$ represents the Lax pair for the Ablowitz--Ladik hierarchy of nonlinear differential-difference evolution equations.
\end{theorem}
\begin{proof} 
Let $f\in\ell_0(\bbZ)$. To curb the length of this proof we will only consider the case $n$ even. We apply formulas \eqref{ALgPp} to compute the commutator $([P_{\ul p},L]f)(n)$ by rewriting $D^{-1}L^\ell=EL^{\ell-1}$ and using \eqref{ALLrec}, \eqref{ALD}, and 
\eqref{ALE}. This yields 
\begin{align}  
&i([P_{\ul p},L]f)(n) \no \\ \no
&= \bigg(\sum_{\ell=1}^{p_+}\rho^-\rho (g_{p_+-\ell,+}^- - g_{p_+-\ell,+}^{--}) \\ \no
&\qquad- \sum_{\ell=0}^{p_+-1}\rho^-\rho (\alpha^- h_{p_+-1-\ell,+}^{--} +\beta^-f_{p_+-1-\ell,+}^-)\bigg)(L^\ell f)(n-2)
\\ \no
&\quad + \bigg(\sum_{\ell=1}^{p_+}\rho \big(2\beta^-g_{p_+-\ell,+}^-
- h_{p_+-\ell,+}^-\big) \\ \no
&\qquad \quad + \sum_{\ell=0}^{p_+-1}\rho\big((\rho^-)^2 h_{p_+-1-\ell,+}^{--} 
- (\beta^-)^2 f_{p_+-1-\ell,+}^-\big)\bigg) (L^\ell f)(n-1)
\\ \no  
&\quad + \bigg(\sum_{\ell=1}^{p_+} \rho^+\big(\beta(g_{p_+-\ell,+}^+ +g_{p_+-\ell,+}^-) 
- h_{p_+-\ell, +}\big) \\ \no
&\qquad \quad + \sum_{\ell=0}^{p_+-1} \rho^+ \big(h_{p_+-1-\ell,+}^- 
- \beta\beta^+ f_{p_+-1-\ell,+}^+ - \beta\alpha^+ h_{p_+-1-\ell,+}\big) \bigg)(L^\ell f)(n+1)
\\ \no
&\quad + \bigg(\sum_{\ell=2}^{p_++1}\big(g_{p_++1-\ell,+}-g_{p_++1-\ell,+}^-\big)\\ \no
&\qquad \quad + \sum_{\ell=1}^{p_+}
\Big(\alpha^+\big(\beta(g_{p_+-\ell,+}-g_{p_+-\ell,+}^-)
+h_{p_+-\ell,+}\big)-\alpha h_{p_+-\ell,+}^-\Big)\\ \no
&\qquad \quad + \sum_{\ell=0}^{p_+-1}\Big(\beta (\alpha^+)^2 h_{p_+-1-\ell,+} 
- \beta (\rho^+)^2 f_{p_+-1-\ell,+}^+-\alpha^+ h_{p_+-1-\ell,+}^-\Big)\bigg) (L^\ell f)(n)
\\ \no
&\quad+\sum_{\ell=1}^{p_-}\rho^-\rho\big(g_{p_--\ell,-}^- -g_{p_--\ell,-}^{--}-
\beta^- f_{p_--\ell,-}^- -\alpha^-h_{p_--\ell,-}^{--}\big)(L^{-\ell}f)(n-2)
\\ \no
&\quad + \bigg(\sum_{\ell=1}^{p_-}\rho\Big(\beta^-\big(2g_{p_--\ell,-}^- - \beta^-f_{p_--\ell,-}^-
\big) + \rho(\rho^-)^2 h_{p_--\ell,-}^{--}\Big)\\ \no
&\qquad \quad - \sum_{\ell=0}^{p_--1}\rho h_{p_--1-\ell,-}^- \bigg)(L^{-\ell}f)(n-1)
\\ \no
&\quad + \bigg( \sum_{\ell=1}^{p_-}\Big(\beta\rho^+\big(g_{p_--\ell,-}^+ +g_{p_--\ell,-}^-
-\alpha^+ h_{p_--\ell,-}-\beta^+ f_{p_--\ell,-}^+\big)+\rho^+ h_{p_--\ell,-}^-\Big)\\ \no
&\qquad \quad -\sum_{\ell=0}^{p_--1}\rho^+h_{p_--1-\ell,-}\bigg)(L^{-\ell}f)(n+1)
\\ \no
&\quad + \bigg(\sum_{\ell=1}^{p_-}\Big(\beta\alpha^+(g_{p_--\ell,-}-g_{p_--\ell,-}^- +
\alpha^+ h_{p_--\ell,-}+\beta^+ f_{p_--\ell,-}^+)- \beta f_{p_--\ell,-}^+ \no \\
& \qquad \qquad \quad \; -\alpha^+ h_{p_--\ell,-}^-\Big)\no \\ \no
&\qquad \quad + \sum_{\ell=0}^{p_--1}\big(g_{p_--1-\ell,-}-g_{p_--1-\ell,-}^- 
-\alpha h_{p_--1-\ell,-}^- + \alpha^+ h_{p_--1-\ell,-}\big)\bigg)(L^{-\ell}f)(n)
\\ \no
&\quad + \f{1}{2}\Big(
\beta\rho^+ \big(g_{p_-,-}^+ +g_{p_+,+}^+ +g_{p_-,-}+g_{p_+,+}\big)f(n+1)\\ \no
&\qquad \quad \; + \beta^-\rho \big(g_{p_-,-}^- +g_{p_+,+}^- +g_{p_-,-}+g_{p_+,+}\big)f(n-1)\\ 
&\qquad \quad \; - \rho^-\rho \big(g_{p_-,-}^{--} +g_{p_+,+}^{--}-g_{p_-,-}-g_{p_+,+}\big)f(n-2)\Big),
\end{align}
where we added the terms
\begin{align}  \label{ALh3.33}
0&=-\sum_{\ell=1}^{p_+}g_{p_+-\ell,+}^-(L^{\ell+1}f)(n)+\sum_{\ell=1}^{p_+}g_{p_+-\ell,+}^-
(L^{\ell+1}f)(n)  \no \\ 
&=-\sum_{\ell=2}^{p_++1}g_{p_++1-\ell,+}^-(L^{\ell}f)(n)+\sum_{\ell=1}^{p_+}g_{p_+-\ell,+}^-
L(L^{\ell}f)(n),  \no  \\[2mm]
0&=-\sum_{\ell=1}^{p_+}h_{p_+-\ell,+}^-(D^{-1}L^{\ell}f)(n)+\sum_{\ell=1}^{p_+}h_{p_+-\ell,+}^-
(EL^{\ell-1}f)(n)  \no \\ 
&=-\sum_{\ell=1}^{p_+}h_{p_+-\ell,+}^-\Big(\alpha(L^{\ell}f)(n)+\rho(L^{\ell}f)(n-1)\Big)  \no \\ 
&\quad+\sum_{\ell=0}^{p_+-1}h_{p_+-1-\ell,+}^-\Big(-\alpha^+(L^{\ell}f)(n)+\rho^+
(L^{\ell}f)(n+1)\Big),   \\[2mm] 
0&=-\sum_{\ell=1}^{p_-}g_{p_--\ell,-}^-(L^{-\ell+1}f)(n)+\sum_{\ell=1}^{p_-}g_{p_--\ell,-}^-
(L^{-\ell+1}f)(n)  \no \\ 
&=-\sum_{\ell=0}^{p_--1}g_{p_--1-\ell,-}^-(L^{-\ell}f)(n)+\sum_{\ell=1}^{p_-}g_{p_--\ell,-}^-
L(L^{-\ell}f)(n),  \no  \\[2mm] 
0&=-\sum_{\ell=1}^{p_-}h_{p_--\ell,-}^-(D^{-1}L^{-\ell+1}f)(n)+\sum_{\ell=1}^{p_-}h_{p_--\ell,-}^-
(EL^{-\ell}f)(n)  \no \\ 
&=-\sum_{\ell=0}^{p_--1}h_{p_--1-\ell,-}^-\Big(\alpha(L^{-\ell}f)(n)+\rho(L^{-\ell}f)(n-1)\Big) \no \\ 
&\quad+\sum_{\ell=1}^{p_-}h_{p_--\ell,-}^-\Big(-\alpha^+(L^{-\ell}f)(n)+\rho^+
(L^{-\ell}f)(n+1)\Big).  \no
\end{align}
Next we apply the recursion relations \eqref{AL0+}--\eqref{ALh_l-}. In addition, we
also use
\begin{align}\no
\alpha^+ h_{p_--\ell,-} + \beta^+ f_{p_--\ell,-}^+&=
\alpha^+\big(h_{p_--1-\ell,-}^+ - \beta^+(g_{p_--\ell,-}^+ + g_{p_--\ell,-})\big)\\ \no
&\quad + \beta^+\big(f_{p_--1-\ell,-} + \alpha^+(g_{p_--\ell,-}^+ + g_{p_--\ell,-})\big)\\ 
&=g_{p_--\ell,-}^+ - g_{p_--\ell,-}. 
\end{align}
This implies, 
\begin{align}  
&i([P_{\ul p},L]f)(n) \no \\ \no
&= 
\sum_{\ell=1}^{p_+-1}\rho^-\rho\Big(g_{p_+-\ell,+}^- - g_{p_+-\ell,+}^{--} -\alpha^- h_{p_+-1-\ell,+}^{--} 
-\beta^-f_{p_+-1-\ell,+}^-\Big)(L^\ell f)(n-2)\\ \no
&\quad + \sum_{\ell=1}^{p_+-1}\Big(\beta^-\rho\big(g_{p_+-\ell,+}^- - g_{p_+-\ell,+}^{--} 
 -\alpha^- h_{p_+-1-\ell,+}^{--}-\beta^-f_{p_+-1-\ell,+}^-\big)\\ \no 
&\qquad \qquad \;\;\, + \rho \big(\beta^-(g_{p_+-\ell,+}^- +g_{p_+-\ell,+}^{--}) +  h_{p_+-1-\ell,+}^{--} -  h_{p_+-\ell,+}^-\big)\Big)(L^\ell f)(n-1)\\ \no  
&\quad + \sum_{\ell=1}^{p_+-1}\Big(\beta \rho^+\big(g_{p_+-\ell,+}^+ -g_{p_+-\ell,+} -
\alpha^+ h_{p_+-1-\ell,+}- \beta^+ f_{p_+-1-\ell,+}^+ \big)  \\ \no
&\qquad \qquad \;\;\, + \rho^+\big(\beta (g_{p_+-\ell,+} +g_{p_+-\ell,+}^-)  + h_{p_+-1-\ell,+}^-  - h_{p_+-\ell,+}\big)\Big)
(L^\ell f)(n+1)\\ \no
&\quad + \bigg(\sum_{\ell=1}^{p_+-1}\big(g_{p_++1-\ell,+}-g_{p_++1-\ell,+}^- -\alpha h_{p_+-\ell,+}^- 
+ \beta\alpha^+ (g_{p_+-\ell,+}^+ + g_{p_+-\ell,+}) \no \\
& \hspace*{1.95cm} - \beta f_{p_+-1-\ell,+}^+\big) \\ \no
&\qquad \quad +  \sum_{\ell=1}^{p_+-1}\alpha^+\big(\beta(-g_{p_+-\ell,+}-g_{p_+-\ell,+}^-)
+h_{p_+-\ell,+}-h_{p_+-1-\ell,+}^-\big)\\ \no
&\qquad \quad + \sum_{\ell=0}^{p_+-1}\beta \alpha^+ \big(g_{p_+-\ell,+}-g_{p_+-\ell,+}^+ + \alpha^+ h_{p_+-1-\ell,+} 
+ \beta^+ f_{p_+-1-\ell,+}^+  \big)\bigg) (L^\ell f)(n)\\ \no
&\quad+\sum_{\ell=1}^{p_-}\rho^-\rho \big(g_{p_--\ell,-}^- -g_{p_--\ell,-}^{--}-
\beta^-f_{p_--\ell,-}^- -\alpha^-h_{p_--\ell,-}^{--}\big)(L^{-\ell}f)(n-2)\\ \no
&\quad + \sum_{\ell=1}^{p_--1}\Big(\beta^-\rho\big(g_{p_--\ell,-}^- - g_{p_--\ell,-}^{--}  - \beta^- f_{p_--\ell,-}^- -
\alpha^- h_{p_--\ell,-}^{--}\big)\\ \no
&\qquad \qquad \;\;\,+ \rho \big( \beta^-(g_{p_--\ell,-}^- + g_{p_--\ell,-}^{--}) + h_{p_--\ell,-}^{--}-h_{p_--1-\ell,-}^- \big)\Big)(L^{-\ell}f)(n-1)\\ \no
&\quad + \sum_{\ell=1}^{p_--1}\Big(\beta \rho^+ \big(g_{p_--\ell,-}^+ - g_{p_--\ell,-} -\alpha^
+h_{p_--\ell,-}-\beta^+f_{p_--\ell,-}^+\big)\\\no
&\qquad \qquad \;\;\, +\rho^+\big( \beta(g_{p_--\ell,-} +g_{p_--\ell,-}^-) + h_{p_--\ell,-}^- -h_{p_--1-\ell,-}\big)\Big)(L^{-\ell}f)(n+1)\\ \no
&\quad + \bigg(\sum_{\ell=1}^{p_--1}\Big(g_{p_--1-\ell,-}-g_{p_--1-\ell,-}^- -\alpha h_{p_--1-\ell,-}^-\big)\\ \no
&\qquad \quad + \sum_{\ell=1}^{p_-} \beta \big( \alpha^+(g_{p_--\ell,-}^+ +g_{p_--\ell,-}) - f_{p_--\ell,-}^+ \big)\\ \no
&\qquad \quad + \sum_{\ell=1}^{p_-} \beta \alpha^+\big(g_{p_--\ell,-}-g_{p_--\ell,-}^+ +
\alpha^+h_{p_--\ell,-}+\beta^+ f_{p_--\ell,-}^+\big)\\ \no
&\qquad \quad + \sum_{\ell=1}^{p_--1} \alpha^+ \big( \beta( -g_{p_--\ell,-} -g_{p_--\ell,-}^-) 
-h_{p_--\ell,-}^- +h_{p_--1-\ell,-} \big)\bigg)(L^{-\ell}f)(n)\\ \no
&\quad + \rho^-\rho \big(g_{0,+}^- - g_{0,+}^{--} \big)(L^{p_+}f)(n-2) 
- \rho^-\rho \big(\alpha^-h_{p_+-1,+}^{--}+\beta^-f_{p_+-1,+}^-\big)f(n-2)\\ \no
&\quad + \rho\big(2\beta^-g_{0,+}^- -h_{0,+}^-\big)(L^{p_+}f)(n-1) 
+ \rho\big((\rho^-)^2h_{p_+-1,+}^{--}-(\beta^-)^2 f_{p_+-1,+}^-\big)f(n-1)\\ \no
&\quad+ \rho^+\big(\beta(g_{0,+}^+ +g_{0,+}^-) - h_{0,+}\big)(L^{p_+}f)(n+1)\\ \no
&\quad  -\rho^+\big(\beta(\alpha^+ h_{p_+-1,+} + \beta^+f_{p_+-1,+}^+)-h_{p_+-1,+}^-\big)f(n+1)\\ \no
&\quad + (g_{0,+}-g_{0,+}^-)(L^{p_++1}f)(n)\\ \no
&\quad +\big( g_{1,+}-g_{1,+}^- -\alpha h_{0,+}^- + \beta \alpha^+(g_{0,+}-g_{0,+}^-)
+\alpha^+h_{0,+}\big)(L^{p_+}f)(n)\\ \no
&\quad + \big(g_{p_--1,-} - g_{p_--1,-}^- - \alpha h_{p_--1,-}^- - \beta f_{p_--1,-}\big)f(n)\\ \no
&\quad + \beta\big(\alpha^+( g_{p_+,+}^+ -  g_{p_+,+})+  f_{p_--1,-} -  f_{p_+-1,+}^+\big)f(n)\\ \no
&\quad +\alpha^+\big(\beta(\beta  f_{p_+-1,+}+\alpha  h_{p_+-1,+}^-) + h_{p_--1,-} - h_{p_+-1,+}^-\big)f(n) \\ \no
&\quad - \big(\beta\alpha^+(g_{0,-}+g_{0,-}^-)-\beta f_{-1,-}+\alpha^+h_{0,-}^-\big)(L^{-p_-}f)(n)\\ \no
&\quad - (\alpha h_{p_+-1,+}^- + \beta f_{p_+-1,+})\big(\beta \rho^+ f(n+1) + \beta^-\rho f(n-1)
+ \rho^-\rho f(n-2)\big)\\ \no
&\quad + \Big(\beta^-\rho\big(g_{0,-}^- +g_{0,-}^- - \beta^- f_{0,-}^-\big) 
+(\rho^-)^2\rho h_{0,-}^{--} \Big)(L^{-p_-}f)(n-1) \\ \no
&\quad + \Big(\beta\rho^+\big(g_{0,-}^+ + g_{0,-}^- -\beta^+f_{0,-}^+ 
- \alpha^+h_{0,-}\big) + \rho^+h_{0,-}^- \Big)(L^{-p_-}f)(n+1)\\ \no
&\quad - \rho h_{p_--1,-}^{-}f(n-1) -\rho^+ h_{p_--1,-}f(n+1)\\ \no
&\quad + \f{1}{2}\Big(
\beta\rho^+ \big(g_{p_-,-}^+ +g_{p_+,+}^+ +g_{p_-,-}+g_{p_+,+}\big)f(n+1)\\ \no
&\qquad \quad \;+ \beta^-\rho \big(g_{p_-,-}^- +g_{p_+,+}^- +g_{p_-,-}+g_{p_+,+}\big)f(n-1)\\ \no
&\qquad \quad \; - \rho^-\rho \big(g_{p_-,-}^{--} +g_{p_+,+}^{--}-g_{p_-,-}-g_{p_+,+}\big)f(n-2)\Big)\\ \no
&=
 \f{\rho^-\rho }{2}\big(g_{p_+,+}^{--} -g_{p_-,-}^{--}+g_{p_-,-} -g_{p_+,+}\big)f(n-2)\\ \no
&\quad +\Big(\rho\big(\beta^-(g_{p_+,+}^{--}+g_{p_-,-}^-)- h_{p_--1,-}^-+ h_{p_+-1,+}^{--}\big)\\ \no
&\qquad \;\;\, +\f{\beta^-\rho}{2}\big(g_{p_+,+}^{-} +g_{p_-,-}-g_{p_+,+} -g_{p_-,-}^-\big)\Big)f(n-1)\\ \no
&\quad + \Big(\beta\big(\alpha^+( g_{p_+,+}^+ +  g_{p_-,-})+  f_{p_--1,-} -  f_{p_+-1,+}^+\big)\\ \no
&\qquad \;\;\, -\alpha^+\big(\beta( g_{p_+,+}^- +  g_{p_-,-}) - h_{p_--1,-} + h_{p_+-1,+}^-\big)\Big)f(n) \\ \no
&\quad +\Big(\rho^+\big(\beta(g_{p_+,+}^{-}+g_{p_-,-})- h_{p_--1,-}+ h_{p_+-1,+}^{-}\big)\\ 
&\qquad \;\;\, +\f{\beta \rho^+}{2}\big(g_{p_+,+} +g_{p_-,-}^+-g_{p_+,+}^+ -g_{p_-,-}\big)\Big)f(n+1),
\end{align}
where we also used \eqref{ALadd}.

Comparing coefficients
finally shows that \eqref{ALLaxtp} is equivalent to
\begin{align} \label{ALrmrtp}
(\rho^-\rho)_{t_{\ul p}} &= \rho^- \rho (C^- + C), \\ \label{ALarmtp}
(\alpha \rho^-)_{t_{\ul p}} &=\rho^- A + \alpha \rho^- C^-,    \\ \label{ALbrptp}
(\beta \rho^+)_{t_{\ul p}} &=\rho^+ B + \beta \rho^+ C^+,  \\
(\alpha^+\beta)_{t_{\ul p}} & = \beta A^+ + \alpha^+ B,
\end{align}
where
\begin{align}
A &= i \big( \alpha(g_{p_+,+} + g_{p_-,-}^-) - f_{p_+-1,+} + f_{p_--1,-}^- \big),  \\
B &= i \big( -\beta(g_{p_+,+}^- + g_{p_-,-}) + h_{p_--1,-} - h_{p_+-1,+}^- \big),    \\
C &= \frac{i}{2}\big(g_{p_+,+}+g_{p_-,-}^- -g_{p_+,+}^- -g_{p_-,-}\big).   
\end{align}
In particular, \eqref{AL_p} implies \eqref{ALLaxtp} since, by \eqref{AL2.14},
\begin{equation}
\rho_{t_{\ul p}}=\frac{i}{2}\rho\big(g_{p_+,+}+g_{p_-,-}^- -g_{p_+,+}^- -g_{p_-,-}\big).
\end{equation}
To prove the converse assertion (i.e., that \eqref{ALLaxtp} implies \eqref{AL_p}), we 
argue as follows: Rewriting \eqref{ALarmtp} and 
\eqref{ALbrptp} using $\rho=\gam^{1/2}=(1-\alpha\beta)^{1/2}$ and \eqref{ALrmrtp} yields
\begin{align}
\begin{split}
(1 + \frac{\alpha\beta}{2\gamma}) \alpha_{t_{\ul p}} + \frac{\alpha^2}{2\gamma} \beta_{t_{\ul p}}
&= A - \alpha C,\\
\frac{\beta^2}{2\gamma} \alpha_{t_{\ul p}} + (1 + \frac{\alpha\beta}{2\gamma}) \beta_{t_{\ul p}}
&= B - \beta C.
\end{split}
\end{align}
This linear system is uniquely solvable since its determinant equals $\gamma^{-1}$ and
the solution reads
\begin{align}
\begin{split}
\alpha_{t_{\ul p}} &= A - \frac{\alpha}{2} \big( \beta A + \alpha B + 2 \gamma C \big),\\
\beta_{t_{\ul p}} &= B - \frac{\beta}{2} \big( \beta A + \alpha B + 2 \gamma C \big).
\end{split}
\end{align}
Using \eqref{ALg_l+} and \eqref{ALg_l-} it is straightforward to check that
$\beta A + \alpha B + 2 \gamma C =0$ which shows that the converse assertion also holds.
\end{proof}

The Ablowitz--Ladik Lax pair in the special defocusing case, where $\beta=\ol{\alpha}$, in the finite-dimensional context, was recently discussed by Nenciu \cite{Nenciu:2006}.

\section{Green's functions and high- and low-energy expansions} \lb{sALh4}

In this section we discuss the Green's function of an $\ell^2(\bbZ)$-realization of the difference expression $L$ and systematically derive 
high- and low-energy expansions of solutions of an associated Riccati-type equation. 

Throughout this section we make the following strengthened assumptions on the 
coefficients $\alpha$ and $\beta$.

\begin{hypothesis} \lb{hALh4.1} 
Suppose that $\alpha, \beta$ satisfy
\begin{equation}
\alpha, \beta\in \ell^{\infty}(\bbZ), \quad  \alpha(n)\beta(n)\notin \{0,1\}, \; n\in\bbZ. 
\lb{ALh4.1}
\end{equation} 
\end{hypothesis}

Given Hypothesis \ref{hALh4.1} we introduce the $\ell^2(\bbZ)$-realization $\breve L$ of the difference expression $L$ in \eqref{ALLrec} by 
\begin{equation}
\breve Lf = Lf, \quad f\in\dom\big(\breve L\big) = \ell^2(\bbZ),    \lb{ALh4.2}
\end{equation} 
and similarly introduce the $\ell^2(\bbZ)$-realizations of the difference expression $D$, $E$, $D^{-1}$, and $E^{-1}$ in \eqref{ALD}--\eqref{ALE-1} by
\begin{align}
\breve D f &= Df, \quad f\in\dom\big(\breve D\big) = \ell^2(\bbZ),   \lb{ALh4.3} \\ 
\breve E f &= Ef, \quad f\in\dom\big(\breve E\big) = \ell^2(\bbZ),   \lb{ALh4.4} \\
{\breve D}^{-1} f &= D^{-1}f, \quad f\in\dom\big({\breve D}^{-1}\big) = \ell^2(\bbZ),  
\lb{ALh4.5}  \\
{\breve E}^{-1} f &= E^{-1}f, \quad f\in\dom\big({\breve E}^{-1}\big) = \ell^2(\bbZ). \lb{ALh4.6} 
\end{align}

The following elementary result shows that these $\ell^2(\bbZ)$-realizations are  meaningful; it will be used in the proof of Lemma \ref{lALh4.3} below.

\begin{lemma} \lb{lALh4.2} 
Assume Hypothesis \ref{hALh4.1}. Then the operators $\breve D, {\breve D}^{-1}, \breve E, {\breve E}^{-1}, \breve L$, and ${\breve L}^{-1}$ are bounded on $\ell^2(\bbZ)$. 
In addition, $\big(\breve L-z\big)^{-1}$ is norm analytic with respect to $z$ in an open neighborhood of $z=0$, and $\big(\breve L-z\big)^{-1} 
= -z^{-1}\big(I - z^{-1} \breve L\big)^{-1}$ is analytic with respect to $1/z$ in an open neighborhood of $1/z=0$.
\end{lemma}
\begin{proof}
By Hypothesis \ref{hALh4.1}, $\rho^2=1-\alpha\beta$, and \eqref{ALD}--\eqref{ALE-1},  one infers that $\breve D$, $\breve E$, ${\breve D}^{-1}$, ${\breve E}^{-1}$ are bounded operators 
on $\ell^2(\bbZ)$ whose norms are bounded by 
\begin{align}
\big\|\breve D\big\|, \big\|\breve E\big\|, \big\|{\breve D}^{-1}\big\|, \big\|{\breve E}^{-1}\big\| 
& \leq 2\|\rho\|_{\infty} + \|\alpha\|_{\infty} + \|\beta\|_{\infty}   \no \\
& \leq 2(1+ \|\alpha\|_{\infty} + \|\beta\|_{\infty}).   \lb{ALh4.7}
\end{align}
Since by \eqref{ALfact},
\begin{equation}
\breve L = \breve D \breve E, \quad {\breve L}^{-1} = {\breve E}^{-1} {\breve D}^{-1},
\end{equation}
the assertions of Lemma \ref{lALh4.2} are evident (alternatively, one can of course invoke 
\eqref{ALLrec} and \eqref{ALL-1rec}).
\end{proof}

To introduce the Green's function of $\breve L$, we need to digress a bit. Introducing the transfer matrix $T(z,\dott)$ associated with $L$ by
\begin{equation}
T(z,n) = \begin{cases} \rho(n)^{-1} \begin{pmatrix} \alpha(n) & z \\ z^{-1} & \beta(n)
 \end{pmatrix}, & \text{$n$ odd,} \\[4mm]
 \rho(n)^{-1} \begin{pmatrix} \beta(n) & 1 \\ 1 & \alpha(n)
 \end{pmatrix}, & \text{$n$ even,} \end{cases}  \quad z\in\bbC\setminus\{0\}, \;
 n\in\bbZ,  \lb{ALhTM}
\end{equation}
recalling that $\rho = \gamma^{1/2}=(1-\alpha\beta)^{1/2}$, one then verifies that 
(cf.\ \eqref{AL2.03})
\begin{equation}
T(z,n) = A(z,n) z^{-1/2} \rho(n)^{-1} U(z,n) A(z,n-1)^{-1}, \quad z\in\bbC\setminus\{0\}, \; 
n\in\bbZ.   \lb{ALhTU}
\end{equation} 
Here we introduced 
\begin{equation}
A(z,n) = \begin{cases} \begin{pmatrix} z^{1/2} & 0 \\ 0 & z^{-1/2}
 \end{pmatrix}, & \text{$n$ odd,} \\[4mm] 
 \begin{pmatrix} 0 & 1 \\ 1 & 0
 \end{pmatrix}, & \text{$n$ even,} \end{cases}  \quad 
 z\in\bbC\setminus\{0\}, \; n \in \bbZ.  \lb{ALhAm}
\end{equation}

Next, we consider a fundamental system of solutions 
\begin{equation} 
\Psi_{\pm}(z,\dott)=\begin{pmatrix}\psi_{1,\pm}(z,\dott) \\ \psi_{2,\pm}(z,\dott) 
\end{pmatrix} 
\end{equation}
of 
\begin{equation}
U(z)\Psi^-_{\pm}(z) = \Psi_{\pm}(z), \quad z\in\bbC\setminus \big(\spec\big(\breve L\big) 
\cup\{0\}\big), 
\lb{ALh4.13a}
\end{equation}
with $\spec\big(\breve L\big)$ denoting the spectrum of $\breve L$ and $U$ given by 
\eqref{AL2.03}, such that
\begin{equation}
\det(\Psi_-(z),\Psi_+(z)) \neq 0.     \lb{ALhpsipm}
\end{equation}
The precise form of $\Psi_\pm$ will be chosen as a consequence of \eqref{ALh4.20} below. Introducing in addition,
\begin{align}
\begin{split}
& \begin{pmatrix} u_{\pm}(z,n) \\ v_{\pm}(z,n) \end{pmatrix} 
=  C_\pm z^{-n/2}  \bigg(\prod_{n'=1}^n \rho(n')^{-1}\bigg)
A(z,n) \begin{pmatrix} \psi_{1,\pm}(z,n) \\ \psi_{2,\pm}(z,n) \end{pmatrix},  \\  
& \hspace*{4.5cm} z\in\bbC\setminus \big(\spec\big(\breve L\big)\cup\{0\}\big), 
\; n\in\bbZ,   \lb{ALhuv}
\end{split} 
\end{align}
for some constants $C_\pm \in\bbC\setminus\{0\}$, \eqref{ALhTU} and \eqref{ALhuv} yield 
\begin{equation}
T(z) \begin{pmatrix} u^-_{\pm}(z) \\ v^-_{\pm}(z) \end{pmatrix}  
= \begin{pmatrix} u_{\pm}(z) \\ v_{\pm}(z) \end{pmatrix}.    \lb{ALhTuv} 
\end{equation}
Moreover, one can show (cf.\ \cite{GesztesyZinchenko:2006}) that 
\begin{align}
L u_{\pm}(z) &= z u_{\pm}(z), \quad L^\top v_{\pm}(z) = z v_{\pm}(z), \lb{ALh4.19a} \\
D v_{\pm}(z) &= u_{\pm}(z), \quad E u_{\pm}(z) = z v_{\pm}(z), \lb{ALh4.19b}
\end{align}
where 
\begin{equation}\lb{ALh4.19c}
L = D E, \quad L^\top=E D,  
\end{equation}
and hence $L^\top$ represents the difference expression associated with the transpose of the infinite matrix $L$ (cf.\ \eqref{ALLop}) in the standard basis of $\ell^2(\bbZ)$. Next, we choose $\Psi_\pm(z)$ in such a manner (cf.\ again \cite{GesztesyZinchenko:2006}), such that for all $n_0\in\bbZ$, 
\begin{equation}
\begin{pmatrix} u_{\pm}(z,\dott) \\ v_{\pm}(z,\dott) \end{pmatrix} \in 
\ell^2([n_0,\pm\infty)\cap\bbZ)^2, \quad z\in\bbC\setminus \big(\spec\big(\breve L\big)
\cup\{0\}\big). 
\lb{ALh4.20}
\end{equation}
Since by hypothesis $z\in\bbC\setminus \spec\big(\breve L\big)$,
$(u_+ (z,\dott), v_+ (z,\dott))^\top$ and $(u_- (z,\dott), v_-
(z,\dott))^\top$ are linearly independent since otherwise $z$ would
be an eigenvalue of $\breve L$. This is of course consistent with
\eqref{ALhpsipm} and \eqref{ALhuv}.

The Green's function of $\breve L$, the $\ell^2(\bbZ)$-realization of the Lax difference expression $L$, is then of the form 
\begin{align}
& G(z,n,n') = \big(\delta_n,\big(\breve L-z\big)^{-1}\delta_{n'}\big)   \lb{ALh4.21} \\
& \quad =\f{-1}{z \det \begin{pmatrix} u_+(z,0) & u_-(z,0) \no \\ 
v_+(z,0) & v_-(z,0) \end{pmatrix}}  \no \\
& \qquad \times  \begin{cases} v_-(z,n') u_+(z,n), & \text{$n' < n$ or 
$n=n'$ even}, \\
 v_+(z,n') u_-(z,n), & \text{$n' > n$ or $n=n'$ odd}, \end{cases} 
 \quad n, n' \in \bbZ,    \no \\ 
 & \quad = -\f{1}{4z} \f{(1-\phi_+(z,0))(1-\phi_-(z,0))}{\phi_+(z,0) - \phi_-(z,0)}  \no \\
 & \qquad \times  \begin{cases} v_-(z,n') u_+(z,n), & \text{$n' < n$ or 
$n=n'$ even}, \\
 v_+(z,n') u_-(z,n), & \text{$n' > n$ or $n=n'$ odd}, \end{cases} 
 \quad n, n' \in \bbZ,     \lb{ALhGreen's}  \\
& \hspace*{6.05cm}  z\in\bbC\setminus \big(\spec\big(\breve L\big)\cup\{0\}\big).  \no
 \end{align}

Introducing
\begin{equation}
\phi_{\pm} (z,n) = \f{\psi_{2,\pm}(z,n)}{\psi_{1,\pm}(z,n)}, \quad 
z\in\bbC\setminus \big(\spec\big(\breve L\big)\cup\{0\}\big), \; n\in\bbN, 
\lb{ALhphipm}
\end{equation}
then \eqref{ALh4.13a} implies that $\phi_{\pm}$ satisfy the Riccati-type equation
\begin{equation}
\alpha \phi_{\pm}\phi^-_{\pm} - \phi^-_{\pm} +z\phi_{\pm} = z \beta,  \lb{ALRicc}
\end{equation}
and one introduces in addition,
\begin{align}
\frf&= \f{2}{\phi_+ - \phi_-},   \lb{ALhgothf}  \\
\frg&= \f{\phi_+ + \phi_-}{\phi_+ - \phi_-},    \lb{ALhgothg}  \\
\frh&= \f{2\phi_+  \phi_-}{\phi_+ - \phi_-}.    \lb{ALhgothh}
\end{align}
Using the Riccati-type equation \eqref{ALRicc} and its consequences,
\begin{align}
\alpha(\phi_+\phi^-_+ - \phi_- \phi^-_-) - (\phi^-_+ - \phi_-^-) + 
z (\phi_+ - \phi_-) &= 0,   \lb{ALhRicc1} \\
\alpha(\phi_+\phi^-_+ + \phi_- \phi^-_-) - (\phi^-_+ + \phi_-^-) + 
z (\phi_+ + \phi_-) &=2z \beta,   \lb{ALhRicc2}
\end{align}
one then derives the identities
\begin{align} 
 z(\frg^- - \frg) + z \beta \frf +\alpha \frh^- & =0,    \lb{ALhlin1} \\
 z \beta \frf^- +\alpha \frh - \frg + \frg^- & =0,    \lb{ALhlin2} \\
- \frf + z \frf^- + \alpha (\frg + \frg^-) & =0,     \lb{ALhlin3} \\
z \beta (\frg^- + \frg) -z \frh + \frh^- & =0,      \lb{ALhlin4} \\
\frg^2 -\frf \frh & = 1.  \lb{ALhquadr}
\end{align}

For the connection between $\frf$, $\frg$, and $\frh$ and the Green's function of $L$ one finally obtains
\begin{align}
\frf (z,n) &= -2 \alpha(n)(zG(z,n,n)+1) - 2 \rho(n) z
\begin{cases} G(z,n-1,n), & \text{$n$ even,} \\ G(z,n,n-1), & \text{$n$ odd,} \end{cases}   
\no \\
\frg (z,n) &= - 2z G(z,n,n) -1,   \lb{ALhGreen} \\
\frh (z,n) &= -2 \beta(n) zG(z,n,n) - 2 \rho(n) z
\begin{cases} G(z,n,n-1), & \text{$n$ even,} \\ G(z,n-1,n), & \text{$n$ odd,} \end{cases} 
\no
\end{align}
illustrating the spectral theoretic content of $\frf$, $\frg$, and $\frh$.

We are particularly interested in the asymptotic expansion of $\phi_\pm$ in a neighborhood of the points $z=0$ and $1/z=0$ and turn to this topic next. 

\begin{lemma} \lb{lALh4.3}
Assume that $\alpha,\beta$ satisfy Hypothesis \ref{hALh4.1}. Then $\phi_{\pm}$ have the following 
convergent expansions with respect to $1/z$ around $1/z=0$ and with respect to $z$ around $z=0$,
\begin{align}
\phi_{\pm}(z) & \underset{z\to\infty}{=} 
\begin{cases}
\sum_{j=0}^\infty \phi_{j,+}^\infty z^{-j},\\[1mm]
\sum_{j=-1}^\infty \phi_{j,-}^\infty z^{-j},
\end{cases}\lb{ALh2.8} \\
\phi_{\pm}(z) & \underset{z\to0}{=} 
\begin{cases}
\sum_{j=0}^\infty \phi_{j,+}^0 z^{j},\\[1mm]
\sum_{j=1}^\infty \phi_{j,-}^0 z^{j},
\end{cases}\lb{ALh2.8Z}
\end{align}
where
\begin{align}
\phi_{0,+}^\infty &=\beta, \quad \phi_{1,+}^\infty =\beta^-\gamma,  \no \\
\phi_{j+1,+}^\infty &=  (\phi_{j,+}^\infty)^- -\alpha\sum_{\ell=0}^{j} (\phi_{j-\ell,+}^\infty)^- \phi_{\ell,+}^\infty, 
\quad j\in\bbN,\lb{ALhphiinfp}\\
\phi_{-1,-}^\infty&=-\frac1{\alpha^+}, \quad \phi_{0,-}^\infty=\frac{\alpha^{++}}{(\alpha^+)^2}\gamma^+, \no \\
\phi_{j+1,-}^\infty
&=- \frac{\alpha^{++}}{\alpha^+}\phi_{j,-}^\infty + \alpha^{++} \sum_{\ell=0}^{j} 
\phi_{j-\ell,-}^\infty (\phi_{\ell,-}^\infty)^+, \quad j\in\bbN_0, \lb{ALhphiinfm} \\
\phi_{0,+}^0 &=\frac1\alpha, \quad \phi_{1,+}^0 =-\frac{\alpha^-}{\alpha^2}\gamma,  \no \\ 
\phi_{j+1,+}^0 &=  (\phi_{j,+}^0)^+ +\alpha^+\sum_{\ell=0}^{j+1} (\phi_{j+1-\ell,+}^0)^+ \phi_{\ell,+}^0, 
\quad j\in\bbN, \lb{ALhphizerop}\\
\phi_{1,-}^0 &=-\beta^+,  \no \\  
\phi_{j+1,-}^0 &=(\phi_{j,-}^0)^+ +\alpha^+\sum_{\ell=1}^{j} \phi_{j+1-\ell,-}^0 (\phi_{\ell,-}^0)^+, 
\quad j\in \bbN. \lb{ALhphizerom}
\end{align}
\end{lemma}
\begin{proof}
Since
\begin{equation}
\phi_\pm = \f{\frg \pm 1}{\frf},
\end{equation}
combining Lemma \ref{lALh4.2}, \eqref{ALh4.21} and \eqref{ALhGreen} proves that 
$\phi_\pm$ has a convergent expansion with respect to $z$ and $1/z$ in a 
neighborhood of $z=0$ and $1/z=0$, respectively.  

The explicit expansion coefficients  $\phi_{j,\pm}^\infty$ are then readily derived by making the ansatz
\begin{equation}
\phi_{\pm}\underset{z\to\infty}{=}\sum_{j=-1}^\infty \phi_{j,\pm}^\infty z^{-j}.    \lb{ALh4.42}
\end{equation}
Inserting \eqref{ALh4.42} into the Riccati-type equation \eqref{ALRicc} one finds
\begin{equation}
0=\alpha \phi_{\pm}\phi_{\pm}^- -\phi_{\pm}^- +z(\phi_{\pm}-\beta)
= (\alpha\phi_{-1,\pm}\phi_{-1,\pm}^- +\phi_{-1,\pm})z^2+ \Oh(z^2),
\end{equation}
which yields the case distinction above and the formulas for $\phi_{j,\pm}^\infty$. The corresponding expansion coefficients $ \phi_{j,\pm}^0$ are obtained analogously by 
making the ansatz $\phi_{\pm}\underset{z\to 0}{=}
\sum_{j=0}^\infty \phi_{j,\pm}^0 z^{j}$.
\end{proof}

For the record we list a few explicit expressions:
\begin{align}
\phi_{0,+}^\infty&=\beta, \no \\
\phi_{1,+}^\infty&=\beta^-\gamma, \no \\
\phi_{2,+}^\infty&=\gamma\big(-\alpha(\beta^{-})^2+\beta^{--}\gamma^{-} \big), \no \\
\phi_{-1,-}^\infty&=-\frac1{\alpha^+},\no \\
\phi_{0,-}^\infty&=\frac{\alpha^{++}}{(\alpha^+)^2}\gamma^+, \no \\
\phi_{1,-}^\infty&= \frac{\gamma^+}{(\alpha^+)^3}\big(\alpha^+\alpha^{+++}
\gamma^{++} - (\alpha^{++})^2\big), \no \\
\phi_{0,+}^0 &=\frac1\alpha, \no \\
\phi_{1,+}^0 &=-\frac{\alpha^-}{\alpha^2}\gamma, \no \\
\phi_{2,+}^0 &=\frac{\gamma}{\alpha^3}\big((\alpha^{-})^2- \alpha^{--}\alpha \gamma^-\big), \no \\
\phi_{1,-}^0 &=-\beta^+,\no \\
\phi_{2,-}^0 &= -\gamma^{+}\beta^{++} , \no \\
\phi_{3,-}^0 &= \gamma^{+}\big(\alpha^{+}(\beta^{++})^2-\gamma^{++}\beta^{+++} \big), \no \\ 
\phi_{4,-}^0 &= \gamma^{+}\Big(-(\alpha^{+})^2(\beta^{++})^3 \no\\
&\quad +\gamma^{++}\big(2\alpha^+\beta^{++}\beta^{+++}+\alpha^{++}(\beta^{+++})^2
-\gamma^{+++}\beta^{++++} \big)\Big),
\text{ etc.}  \no 
\end{align}

Later on we will also need the convergent expansions of 
$\ln(z+\alpha^+ \phi_\pm (z))$ with 
respect to $z$ and $1/z$. We will separately provide all four expansions of 
$\ln(z+\alpha^+ \phi_\pm (z))$ around $1/z=0$ and $z=0$ and repeatedly use the general formula
\begin{equation}
\ln\bigg(1+\sum_{j=1}^\infty \omega_j z^{\pm j}\bigg) 
=  \sum_{j=1}^\infty \sigma_j z^{\pm j},    \lb{ALh4.12A}
\end{equation}
where
\begin{equation}
\sigma_1=\omega_1, \qquad 
\sigma_j=\omega_j-\sum_{\ell=1}^{j-1}\frac{\ell}{j}\omega_{j-\ell}\sigma_\ell, \quad j\ge 2, 
\end{equation}
and $|z|$ as $|z|\to 0$, respectively, $1/|z|$ as $|z|\to\infty$, are assumed to be sufficiently small in \eqref{ALh4.12A}. 
We start by expanding $\phi_+$ around $1/z=0$
\begin{align}
\ln(z+\alpha^+ \phi_+ (z))
&=\ln\bigg(z+\alpha^+\sum_{j=0}^\infty \phi_{j,+}^\infty z^{-j} \bigg)\no\\
&=\ln\bigg(1+\alpha^+\sum_{j=0}^\infty \phi_{j,+}^\infty z^{-j-1}\bigg)\no\\
&=\ln\bigg(1+\alpha^+\sum_{j=1}^\infty \phi_{j-1,+}^\infty z^{-j}\bigg)\no\\
&=\sum_{j=1}^\infty\rho_{j,+}^\infty z^{-j}, \lb{ALh4.13}
\end{align}
where
\begin{equation} \lb{ALhrhoinfp}
\rho_{1,+}^\infty =\alpha^+\phi_{0,+}^\infty, \quad 
\rho_{j,+}^\infty =\alpha^+\bigg(\phi_{j-1,+}^\infty -\sum_{\ell=1}^{j-1}\frac{\ell}{j}\phi_{j-1-\ell,+}^\infty 
\rho_{\ell,+}^\infty \bigg),  \quad j\geq 2.
\end{equation}
An expansion of $\phi_-$ around $1/z=0$ yields
\begin{align}
\ln(z+\alpha^+ \phi_- (z))
&=\ln\bigg(z+\alpha^+\sum_{j=-1}^\infty \phi_{j,-}^\infty z^{-j} \bigg)\no\\
&= \ln\bigg( \frac{\alpha^{++}\gamma^+}{\alpha^+}\bigg)
+\ln\bigg(1+\frac{(\alpha^+)^2}{\alpha^{++}\gamma^+}\sum_{j=1}^\infty \phi_{j,-}^\infty z^{-j}\bigg) \no\\
&=\ln\bigg( \frac{\alpha^{++}}{\alpha^+}\bigg)+ \ln\gamma^+
+\ln\bigg(1+\frac{(\alpha^+)^2}{\alpha^{++}\gamma^+}\sum_{j=1}^\infty \phi_{j,-}^\infty z^{-j}\bigg)\no\\
&=\ln\bigg( \frac{\alpha^{++}}{\alpha^+}\bigg)+\ln\gamma^+
+ \sum_{j=1}^\infty\rho_{j,-}^\infty z^{-j},  \lb{ALh4.15}
\end{align}
where
\begin{equation} \lb{ALhrhoinfm}
\rho_{1,-}^\infty =\frac{(\alpha^+)^2}{\alpha^{++}\gamma^+}\phi_{1,-}^\infty, \quad 
\rho_{j,-}^\infty =\frac{(\alpha^+)^2}{\alpha^{++}\gamma^+}\bigg(\phi_{j,-}^\infty -\sum_{\ell=1}^{j-1}
\frac{\ell}{j}\phi_{j-\ell,-}^\infty \rho_{\ell,-}^\infty \bigg),  \quad j\geq 2.
\end{equation}
For the expansion of $\phi_+$ around $z=0$ one gets 
\begin{align}
 \ln(z + \alpha^+\phi_+ (z))&=
 \ln\bigg(z+\alpha^+\sum_{j=0}^\infty \phi_{j,+}^0 z^j\bigg) \no\\
&= \ln\bigg(\frac{\alpha^+}{\alpha}\bigg) +
 \ln\bigg(1 + \frac{\alpha}{\alpha^+}\big(1+\alpha^+\phi_{1,+}^0\big)z 
+\alpha \sum_{j=2}^\infty\phi_{j,+}^0z^j \bigg)\no\\
&= \ln\bigg(\frac{\alpha^+}{\alpha}\bigg) +
 \sum_{j=1}^\infty\rho_{j,+}^0z^j,  \lb{ALh4.17}
\end{align}
where
\begin{align} \lb{ALhrhozerop}
\begin{split}
\rho_{1,+}^0&=\frac{\alpha}{\alpha^+}\big(1+\alpha^+\phi_{1,+}^0\big), 
\quad \rho_{2,+}^0=\alpha\phi_{2,+}^0 - \frac12 (\rho_{1,+}^0)^2,  \\
\rho_{j,+}^0&=\alpha\bigg(\phi_{j,+}^0-\sum_{j=1}^{j-2}\frac\ell j \phi_{j-\ell,+}^0\rho_{\ell,+}^0\bigg)
-\frac{j-1}{j} \rho_{1,+}^0 \rho_{j-1,+}^0, \quad j\geq3.
\end{split}
\end{align}
Finally, the expansion of $\phi_-$ around $z=0$ is given by
\begin{align}
\ln(z + \alpha^+\phi_- (z))&= \ln\bigg(z+\alpha^+\sum_{j=1}^\infty \phi_{j,-}^0z^j\bigg) \no\\
&=\ln\bigg(\gamma^+z +\alpha^+ \sum_{j=2}^\infty\phi_{j,-}^0z^j \bigg) \no\\
&=\ln z + \ln \gamma^+
+ \ln \bigg(1+\frac{\alpha^+}{\gamma^+}\sum_{j=1}^\infty \phi_{j+1,-}^0z^j\bigg) \no\\
&=\ln z + \ln \gamma^++ \sum_{j=1}^\infty\rho_{j,-}^0z^j,  \lb{ALh4.19}
\end{align}
where
\begin{equation}  \lb{ALhrhozerom}
\rho_{1,-}^0=\frac{\alpha^+}{\gamma^+}\phi_{2,-}^0, \quad \rho_{j,-}^0=\frac{\alpha^+}{\gamma^+}
\bigg(\phi_{j+1,-}^0-\sum_{\ell=1}^{j-1}\frac{\ell}{j}\phi_{j+1-\ell,-}^0\rho_{\ell,-}^0\bigg),  \quad j\geq2.
\end{equation}
Explicitly, the first expansion coefficients are given by
\begin{align} \lb{ALh4.55}
\begin{split}
\rho_{1,+}^\infty&=\alpha^+\beta, \\ 
\rho_{2,+}^\infty&=- \tfrac{1}{2}(\alpha^+\beta)^2+\gamma\alpha^+\beta^-, \\ 
\rho_{3,+}^\infty&= \tfrac{1}{3}(\alpha^+\beta)^3-
\gamma \big(\gamma^- \alpha^+\beta^{--}-(\alpha^+)^2\beta^-\beta -\alpha\alpha^+(\beta^-)^2\big), \\ 
\rho_{1,-}^\infty&=-\alpha^{+++}\beta^{++} + (S^+-I) \frac{\alpha^{++}}{\alpha^+}, \\
\rho_{1,+}^0&=\alpha^-\beta + (S^+-I) \frac{\alpha^-}{\alpha}, \\
\rho_{1,-}^0&=-\alpha^+\beta^{++}, \\
\rho_{2,-}^0&= \tfrac{1}{2}(\alpha^+\beta^{++})^2-\gamma^{++}\alpha^+\beta^{+++}, \\
\rho_{3,-}^0&= - \tfrac{1}{3}(\alpha^+\beta^{++})^3\\
&\quad +
\gamma^{++} \big(-\gamma^{+++} \alpha^+\beta^{++++}+(\alpha^+)^2\beta^{++}
\beta^{+++} +\alpha^+\alpha^{++}(\beta^{+++})^2\big), \, \text{ etc.}
\end{split}
\end{align}

The next result shows that $\hat g_{j,+}$ and $\pm j \rho_{j,\pm}^\infty$, respectively 
$\hat g_{j,-}$ and $\pm j \rho_{j,\pm}^0$, are equal up to terms 
that are total differences, that is, are of the form $(S^+ -I)d_{j,\pm}$ for some sequence 
$d_{j,\pm}$. The exact form of $d_{j,\pm}$ will not be needed later.
In the proof we will heavily use the equations \eqref{ALhlin1}--\eqref{ALhquadr}. 

\begin{lemma} \lb{lALh4.4}
Suppose Hypothesis~\ref{hALh4.1} holds. Then
\begin{align}
\hat g_{j,+}&=- j\rho_{j,+}^\infty +(S^+ - I) d_{j,+}= j\rho_{j,-}^\infty +(S^+ - I) e_{j,+}, \quad j\in\bbN, 
\lb{ALh4.23} \\
\hat g_{j,-}&=- j\rho_{j,+}^0 +(S^+ - I) d_{j,-}= j\rho_{j,-}^0 +(S^+ - I) e_{j,-}, \quad j\in\bbN,\lb{ALh4.24}
\end{align}
for some polynomials $d_{j,\pm}$,  $e_{j,\pm}$, $j\in\bbN$, in $\alpha$ and $\beta$ and certain shifts thereof.
\end{lemma}
\begin{proof} 
We consider the case for $\hat g_{j,+}$ first. Our aim is to show that 
\begin{equation} \lb{ALh4.25}
\frac{d}{dz}\ln(z+\alpha^+ \phi_+)= - \frac{1}{2z}\frg + \frac{1}{2z} + (S^+ -I)K+(S^+ -I)M,
\end{equation}
where
\begin{equation}
K= \f12 \bigg(\frg \frac{\dot \frf}{\frf} - \dot \frg\bigg), \quad
M= \f12 \frac{\dot \frf}{\frf},
\end{equation}
which implies \eqref{ALh4.23} by \eqref{ALh4.13}. Here $\dott$ denotes $d/dz$. 

Since $\phi_+= (\frg+1)/\frf$,
\begin{align}\lb{ALh4.33}
&\frac{d}{dz}\ln(z+\alpha^+ \phi_+)=\frac{1+\alpha^+ \dot \phi_+}{z+\alpha^+ \phi_+}
=\frac{\frf^2 + \alpha^+ (\frf \dot \frg - \dot \frf \frg  - \dot \frf)}{\frf(z  \frf + \alpha^+ \frg + \alpha^+)}
\frac{z  \frf + \alpha^+ \frg - \alpha^+}{z  \frf + \alpha^+ \frg - \alpha^+}.
\end{align}
Next we treat the denominator of \eqref{ALh4.33} using \eqref{ALhlin1}, \eqref{ALhlin3},
\begin{align} \lb{ALh4.34}
\frf\big((z  \frf + \alpha^+ \frg)^2 - (\alpha^+)^2\big)&=
\frf\big((z  \frf + \alpha^+ \frg)^2 - (\alpha^+)^2(\frg^2 - \frf\frh)\big) \no \\
&=z \frf^2\Big(z  \frf + \alpha^+\frg  + \alpha^+\frg +(\alpha^+)^2\f1z\frh \Big)\no \\
&=z \frf^2\big(\frf^+ - \alpha^+\frg^+  + \alpha^+\frg^+ -\alpha^+\beta^+\frf^+ \big)\no \\
&=z \gamma^+\frf^2\frf^+.
\end{align}
Expanding the numerator in \eqref{ALh4.33} and applying 
\eqref{ALhlin1}, \eqref{ALhlin3}, and their derivatives with respect to $z$ 
as well as $2 \frg \dot \frg =  \dot \frf \frh + \frf \dot \frh$ yields
\begin{align} \lb{ALh4.35}
&\big(z  \frf + \alpha^+ \frg - \alpha^+\big)
\big(\frf^2 + \alpha^+ (\frf \dot \frg - \dot \frf \frg  - \dot \frf)\big)
\no \\
&=\frf\big(z\frf^2+z\alpha^+(\frf \dot \frg-\dot \frf \frg) + \alpha^+\frf\frg 
+\tfrac12(\alpha^+)^2(\frf\dot\frh - \dot\frf\frh)-\alpha^+(\frf +z\dot\frf+\alpha^+\dot\frg)\big)
\no\\
&=\frac\frf2\Big(2z\frf^2 + z\alpha^+\frf \dot \frg +z\frf(-\alpha^+\dot\frg^+-z\dot\frf-\frf+\dot\frf^+)
-z\alpha^+\dot \frf \frg +z\dot\frf(\alpha^+\frg^++z\frf-\frf^+)\no\\
&\quad + \alpha^+\frf\frg + \frf(-\alpha^+\frg^+-z\frf+\frf^+) 
+\alpha^+\frf(-\beta^+\frf^+ - z\beta^+\dot\frf^+-\frg+\frg^+-z\dot\frg +z\dot\frg^+)\no\\
&\quad +\alpha^+\dot\frf(z\beta^+\frf^++z\frg-z\frg^+) -2\alpha^+(\frf +z\dot\frf+\alpha^+\dot\frg)\Big)
\no\\
&= \frac\frf2\Big(\gamma^+\frf\frf^++z\gamma^+\frf\dot\frf^+-z\gamma^+\dot\frf\frf^+-2\alpha^+(\frf +z\dot\frf+\alpha^+\dot\frg)\Big).
\end{align}
In summary,
\begin{equation} \lb{ALh4.36}
\frac{d}{dz}\ln(z+\alpha^+ \phi_+)= \frac{1}{2z} +(S^+ -I)M - 
\frac{\alpha^+}{z\gamma^+\frf\frf^+}\big(\frf +z\dot\frf+\alpha^+\dot\frg\big).
\end{equation}
We multiply the numerator on the right-hand side by $-2=-2(\frg^2-\frf\frh)$ and use again 
\eqref{ALhlin1}, \eqref{ALhlin3}, and their derivatives:
\begin{align}
&2\alpha^+\big(\frf\frh- \frg^2\big)\big(\frf +z\dot\frf+\alpha^+\dot\frg\big)
\no\\
&=2\alpha^+\frf\frh(\frf +z\dot\frf+\alpha^+\dot\frg) 
-2\alpha^+\frf\frg^2 - 2z\alpha^+\dot\frf \frg^2 - (\alpha^+)^2\frg(\dot\frf\frh+\frf\dot\frh)
\no\\
&=\alpha^+\frf\frh(\frf +z\dot\frf)+\alpha^+\frf\dot\frg(-z\beta^+\frf^+ - z\frg +z\frg^+)\no\\
&\quad +\frf(-z\beta^+\frf^+ - z\frg +z\frg^+)(\dot\frf^+-\alpha^+\dot\frg^+) \no\\
&\quad -\alpha^+\frf\frg^2 +\frf\frg(\alpha^+\frg^++z\frf-\frf^+) - z \alpha^+\dot\frf\frg^2
+z\dot\frf\frg(\alpha^+\frg^++z\frf-\frf^+) \no\\
&\quad + \alpha^+\dot\frf\frg(z\beta^+\frf^+ + z\frg -z\frg^+) + 
\alpha^+\frf\frg(\beta^+\frf^+ +z\beta^+\dot\frf^++\frg-\frg^++z\dot\frg-z\dot\frg^+) \no\\
&=\alpha^+\frf\frh(\frf +z\dot\frf)
+\alpha^+\frf\dot\frg(-z\beta^+\frf^+ - z\frg +z\frg^+)
+\frf\dot\frf^+(-z\beta^+\frf^+ - z\frg +z\frg^+) \no\\
&\quad +\alpha^+\frf\dot\frg^+(z\beta^+\frf^+ + z\frg -z\frg^+) 
+z\frf^2\frg -\gamma^+\frf\frf^+\frg +z^2\frf\dot\frf\frg\no\\
&\quad -z\gamma^+\dot\frf\frf^+\frg 
 + z\beta^+\frf\dot\frf^+(-\alpha^+\frg^+-z\frf+\frf^+)
 +z\alpha^+\frf\frg(\dot\frg-\dot\frg^+) \no\\
&=\alpha^+\frf\frh(\frf +z\dot\frf) -z\alpha^+\beta^+\frf\frf^+(\dot\frg-\dot\frg^+)
- z\frf (\alpha^+\frg + z\frf - \frf^+)(\dot\frg - \dot\frg^+)\no\\
&\quad-z\frf\dot\frf^+\frg 
+z\frf^2\frg  -\gamma^+\frf\frf^+\frg + z^2\frf\dot\frf\frg 
-z\gamma^+\dot\frf\frf^+\frg + z\gamma^+\frf\dot\frf^+\frg^+ -z^2\beta^+\frf^2\dot\frf^+\no\\
&=z\alpha^+\frf\dot\frf\frh -z\frf^2(z\beta^+\dot\frf^+ + z\dot\frg -z\dot\frg^+ -\tfrac1z\alpha^+\frh)
+z\frf\frg(\frf+z\dot\frf - \alpha^+\dot\frg - \dot\frf^++\alpha^+\dot\frg^+) \no\\
&\quad +z\gamma^+\frf\frf^+(\dot\frg-\dot\frg^+)
-\gamma^+\frf\frf^+\frg -z\gamma^+\dot\frf\frf^+\frg + z\gamma^+\frf\dot\frf^+\frg^+\no\\
&=z\alpha^+\frf(\dot\frf\frh +\frf\dot\frh -2\frg\dot\frg)
+z\gamma^+\frf\frf^+(\dot\frg-\dot\frg^+) -\gamma^+\frf\frf^+\frg -z\gamma^+\dot\frf\frf^+\frg + z\gamma^+\frf\dot\frf^+\frg^+\no\\
&=z\gamma^+\frf\frf^+(\dot\frg-\dot\frg^+) -\gamma^+\frf\frf^+\frg -z\gamma^+\dot\frf\frf^+\frg + z\gamma^+\frf\dot\frf^+\frg^+.
\end{align}
Inserting this in \eqref{ALh4.36} finally yields \eqref{ALh4.25}. 
The result for $\hat g_{j,-}$ is derived similarly starting from $\phi_-=(\frg-1)/\frf$.
\end{proof}

\section{Local conservation laws} \lb{sALh5}

Throughout this section (and with the only exception of Theorem \ref{tALh5.5}) we make the following assumption:
\begin{hypothesis} \lb{hALh5.1}
Suppose that $\alpha, \beta \colon \bbZ\times\bbR \to \bbC$ satisfy
\begin{align}
\begin{split}
& \sup_{(n,t_{\ul p})\in\bbZ\times\bbR}\big(|\alpha(n,t_{\ul p})|
+|\beta(n,t_{\ul p})|\big) < \infty,  \lb{ALh5.1} \\
& a(n,\dott), \, b(n,\dott) \in C^1(\bbR), \; n\in\bbZ,  \quad 
 \alpha(n,t_{\ul p})\beta(n,t_{\ul p})\notin\{0,1\}, \; (n,t_{\ul p})\in\bbZ\times\bbR.  
 \end{split}
\end{align}
\end{hypothesis}

In accordance with the notation introduced in \eqref{ALh4.2}--\eqref{ALh4.6} we denote the bounded difference operator defined on 
$\ell^2(\bbZ)$, generated by the finite difference expression $P_{\ul p}$ in 
\eqref{ALP_p}, by the symbol $\breve P_{\ul p}$. Similarly, the bounded finite difference operator in $\ell^2(\bbZ)$ generated by $P^\top_{\ul p}$ in \eqref{ALP_ptop} is then denoted by $\breve P^\top_{\ul p}$. 

We start with the following existence result.

\begin{theorem} \lb{tALh5.2} 
Assume Hypothesis \ref{hALh5.1} and suppose $\alpha,\beta$ satisfy 
$\AL_{\ul p}(\alpha,\beta)=0$ for some $\ul p\in\bbN_0^2$. In addition,  
let $t_{\ul p}\in\bbR$ and $z \in \bbC\setminus\big(\spec\big(\breve L(t_{\ul p})\big)
\cup\{0\}\big)$. Then there exist Weyl--Titchmarsh-type solutions 
$u_\pm=u_\pm(z,n,t_{\ul p})$ and $v_\pm=v_\pm(z,n,t_{\ul p})$ such that for all 
$n_0\in\bbZ$,
\begin{equation}
\begin{pmatrix} u_\pm (z,\dott,t_{\ul p}) \\  v_\pm (z,\dott,t_{\ul p}) \end{pmatrix} 
\in\ell^2([n_0,\pm\infty)\cap\bbZ)^2,  
\quad u_\pm (z,n,\dott), v_\pm (z,n,\dott) \in C^1(\bbR),   \lb{ALh5.3} 
\end{equation}
and $u_\pm$ and $v_\pm$ simultaneously satisfy the following equations in the weak sense 
\begin{align}
& \breve L(t_{\ul p})u_\pm(z,\dott,t_{\ul p})=z u_\pm(z,\dott,t_{\ul p}),  \lb{ALh5.4}  \\
& u_{\pm,t_{\ul p}}(z,\dott,t_{\ul p})= \breve P_{\ul p}(t_{\ul p}) u_\pm(z,\dott,t_{\ul p}),   
\lb{ALh5.5} 
\end{align}
and
\begin{align}
& \breve L^\top(t_{\ul p}) v_\pm(z,\dott,t_{\ul p})=z v_\pm(z,\dott,t_{\ul p}),  \lb{ALh5.6}  \\
& v_{\pm,t_{\ul p}}(z,\dott,t_{\ul p})= -\breve P^\top_{\ul p}(t_{\ul p}) v_\pm(z,\dott,t_{\ul p}), 
\lb{ALh5.7}
\end{align}
respectively.  
\end{theorem}
\begin{proof}
Applying $\big(\breve L(t_{\ul p})-zI\big)^{-1}$ to $\delta_0$ (cf.\ \eqref{ALh4.21}) yields the existence of Weyl--Titchmarsh-type solutions $\tilde u_\pm$ of $Lu=zu$ satisfying 
\eqref{ALh5.3}. Next, using the Lax commutator equation \eqref{ALLaxtp} one computes 
\begin{align}
\begin{split}
z\tilde u_{\pm,t_{\ul p}}&=(L \tilde u_\pm)_{t_{\ul p}}= L_{t_{\ul p}}\tilde u_\pm
+ L\tilde u_{\pm,t_{\ul p}}
=[P_{\ul p},L]\tilde u_\pm + L \tilde u_{\pm,t_{\ul p}}  \\
&=zP_{\ul p}\tilde u_\pm - L P_{\ul p}\tilde u_\pm
+ L\tilde u_{\pm,t_{\ul p}}  \lb{ALh5.8}
\end{split}
\end{align}
and hence
\begin{equation}
(L-zI)(\tilde u_{\pm,t_{\ul p}} - P_{\ul p} \tilde u_{\pm}) = 0.    \lb{ALh5.9}
\end{equation}
Thus, $\tilde u_\pm$ satisfy
\begin{equation}
\tilde u_{\pm,t_{\ul p}} - P_{\ul p} \tilde u_{\pm} 
= C_\pm \tilde u_\pm + D_\pm \tilde u_\mp.  \lb{ALh5.10}
\end{equation}
Introducing $\tilde u_\pm = c_\pm u_\pm$, and choosing $c_\pm$ such that 
$c_{\pm,t_{\ul p}} = C_\pm c_\pm$, one obtains
\begin{equation}
u_{\pm,t_{\ul p}} - P_{\ul p} u_{\pm} = D_\pm u_\mp.  \lb{ALh5.11}
\end{equation}
Since $u_\pm \in\ell^2([n_0,\pm\infty)\cap\bbZ)$, $n_0\in\bbZ$, and 
$\alpha, \beta$ satisfy Hypothesis \ref{hALh5.1}, \eqref{ALPp} shows that  
$P_{\ul p} u_\pm \in\ell^2([n_0,\pm\infty)\cap\bbZ)$. Moreover, since 
\begin{equation}
u_\pm (z,n,t_{\ul p}) 
= d_\pm (t_{\ul p}) (\breve L(t_{\ul p})-zI)^{-1} \delta_0)(n)  
\end{equation}
for $n\in [\pm 1,\infty)\cap\bbZ$ and some $d_\pm \in C^1(\bbR)$, the calculation
\begin{equation}
u_{\pm,t_{\ul p}}=d_{\pm,t_{\ul p}} \big(\breve L-zI\big)^{-1} \delta_0  
- d_\pm \big(\breve L-zI\big)^{-1} \breve L_{t_{\ul p}} \big(\breve L-zI\big)^{-1} \delta_0  
\lb{ALh5.12}
\end{equation}
also yields $u_{\pm,t_{\ul p}}\in\ell^2([n_0,\pm\infty)\cap\bbZ)$. But then $D_\pm =0$ in \eqref{ALh5.11} since $u_\mp \notin \ell^2([n_0,\pm\infty)\cap\bbZ)$. This proves 
\eqref{ALh5.5}. 

Equations \eqref{ALh5.3}, \eqref{ALh5.6}, and \eqref{ALh5.7}  for $v_\pm$ are proved similarly replacing $L, P_{\ul p}$ by $L^\top, P^\top_{\ul p}$ and observing that 
\eqref{ALLaxtp} implies 
\begin{equation}
L^\top_{t_{\ul p}}(t_{\ul p}) + \big[P_{\ul p}^\top (t_{\ul p}), L^\top (t_{\ul p})\big] = 0, \quad 
t_{\ul p} \in \bbR.
\end{equation} 
\end{proof}

For the remainder of this section we will always refer to the Weyl--Titchmarsh solutions  $u_\pm$, $v_\pm$ introduced in Theorem \ref{tALh5.2}. Given $u_\pm$, $v_\pm$, we now introduce 
\begin{equation} 
\Psi_{\pm}(z,\dott,t_{\ul p})=\begin{pmatrix}\psi_{1,\pm}(z,\dott,t_{\ul p}) 
\\ \psi_{2,\pm}(z,\dott,t_{\ul p}) \end{pmatrix},   \quad  
z \in \bbC\setminus\big(\spec\big(\breve L(t_{\ul p})\big) \cup\{0\}\big), \; t_{\ul p} \in \bbR, 
\lb{ALh5.14a}
\end{equation}
by (cf.\ \eqref{ALhuv})
\begin{align}
\begin{split}
& \begin{pmatrix} \psi_{1,\pm}(z,n,t_{\ul p}) \\ \psi_{2,\pm}(z,n,t_{\ul p}) \end{pmatrix}
=  D(t_{\ul p}) z^{n/2}  \bigg(\prod_{n'=1}^n \rho(n',t_{\ul p})\bigg)
A(z,n)^{-1} \begin{pmatrix} u_{\pm}(z,n,t_{\ul p}) \\ v_{\pm}(z,n,t_{\ul p}) \end{pmatrix} , \\ 
& \hspace*{4.05cm} z \in \bbC\setminus\big(\spec\big(\breve L(t_{\ul p})\big) \cup\{0\}\big), \; (n,t_{\ul p}) \in\bbZ\times\bbR,   \lb{ALhpsi}
\end{split}
\end{align}
with the choice of normalization 
\begin{equation}
D(t_{\ul p}) = \exp\bigg(\f{i}{2}\big(g_{p_+,+}(0) - g_{p_-,-}(0)\big)t_{\ul p}\bigg) D(0), \quad t_{\ul p} \in \bbR,   \lb{ALhD}
\end{equation}
for some constant $D(0) \in\bbC\setminus\{0\}$.

\begin{lemma} \lb{lALh5.3}
Assume Hypothesis \ref{hALh5.1} and suppose $\alpha,\beta$ satisfy 
$\AL_{\ul p}(\alpha,\beta)=0$ for some $\ul p\in\bbN_0^2$. In addition,  
let $t_{\ul p}\in\bbR$ and $z \in \bbC\setminus\big(\spec\big(\breve L(t_{\ul p})\big)
\cup\{0\}\big)$. Then $\Psi_{\pm}(z,\dott,t_{\ul p})$ defined in \eqref{ALhpsi} satisfy 
\begin{align}
& U(z,\dott,t_{\ul p}) \Psi_{\pm}^- (z,\dott,t_{\ul p}) = \Psi_{\pm} (z,\dott,t_{\ul p}),  
\lb{ALh5.15a} \\
& \Psi_{\pm,t_{\ul p}} (z,\dott,t_{\ul p}) = V^+_{\ul p}(z,\dott,t_{\ul p}) 
\Psi_{\pm} (z,\dott,t_{\ul p}).   \lb{ALh5.15b} 
\end{align}
In addition, $\Psi_- (z,\dott,t_{\ul p})$ and $\Psi_+ (z,\dott,t_{\ul p})$ are linearly independent. 
\end{lemma}
\begin{proof}
Equation \eqref{ALh5.15a} is equivalent to
\begin{equation} \lb{ALh5.15c} 
 \begin{pmatrix} \psi_{1,\pm} \\ \psi_{2,\pm} \end{pmatrix}=
 \begin{pmatrix} z\psi_{1,\pm}^- +\alpha\psi_{2,\pm}^-  \\ 
 z\beta\psi_{1,\pm}^- +\psi_{2,\pm} \end{pmatrix}. 
\end{equation}
Using \eqref{ALhAm} and \eqref{ALhpsi} one obtains
\begin{equation} \lb{ALh5.15d} 
\begin{pmatrix} \psi_{1,\pm} \\ \psi_{2,\pm} \end{pmatrix}
=D z^{n/2}\bigg(\prod_{n'=1}^n \rho(n')\bigg)
 \begin{cases}
\begin{pmatrix} z^{-1/2} u_{\pm} \\  z^{1/2}v_{\pm} \end{pmatrix}, & \text{$n$ odd}, \\[5mm] 
\begin{pmatrix}  v_{\pm} \\  u_{\pm} \end{pmatrix}, & \text{$n$ even}.
 \end{cases}
\end{equation}
Inserting \eqref{ALh5.15d} into  \eqref{ALh5.15c}, one finds that \eqref{ALh5.15c} is equivalent to  
\eqref{ALhTuv}, thereby proving \eqref{ALh5.15a}. 

Equation \eqref{ALh5.15b} is equivalent to
\begin{equation}\lb{ALh5.15e}
 \begin{pmatrix} \psi_{1,\pm,t_{\ul p}} \\ \psi_{2,\pm,t_{\ul p}} \end{pmatrix}=
i \begin{pmatrix} G_{\ul p}\psi_{1,\pm}  - F_{\ul p} \psi_{2,\pm}   \\
H_{\ul p}\psi_{1,\pm}- K_{\ul p}\psi_{2,\pm} \end{pmatrix}. 
\end{equation}
We first consider the case when $n$ is odd. Using \eqref{ALh5.15d}, the right-hand side of  \eqref{ALh5.15e} reads, 
\begin{equation}\lb{ALh5.15f}
i \begin{pmatrix} G_{\ul p}\psi_{1,\pm}  - F_{\ul p} \psi_{2,\pm}   \\
H_{\ul p}\psi_{1,\pm}- K_{\ul p}\psi_{2,\pm} \end{pmatrix}=
i D z^{(n-1)/2}\bigg(\prod_{n'=1}^n \rho(n') \bigg) 
\begin{pmatrix}    G_{\ul p} u_{\pm} - z F_{\ul p}v_{\pm}\\
  H_{\ul p} u_{\pm} - z  K_{\ul p}v_{\pm}   \end{pmatrix}.
\end{equation}
Equation \eqref{ALh5.15d} then implies 
\begin{align}\lb{ALh5.15g}
& \begin{pmatrix} \psi_{1,\pm,t_{\ul p}} \\ \psi_{2,\pm,t_{\ul p}} \end{pmatrix} 
= D_{t_{\ul p}} z^{n/2}\bigg(\prod_{n'=1}^n \rho(n')\bigg)
 \begin{pmatrix} z^{-1/2} u_{\pm} \\  z^{1/2} v_{\pm} \end{pmatrix}   
\\ 
& \quad + D z^{n/2}\bigg(\prod_{n'=1}^n \rho(n')\bigg)
 \begin{pmatrix} z^{-1/2} u_{\pm,t_{\ul p}} \\  z^{1/2} v_{\pm,t_{\ul p}} \end{pmatrix}
 + D z^{n/2} \bigg(\partial_{t_{\ul p}} \prod_{n'=1}^n \rho(n')\bigg)
 \begin{pmatrix} z^{-1/2} u_{\pm} \\  z^{1/2} v_{\pm} \end{pmatrix}.   \no 
\end{align}
Next, one observes that
\begin{align}
&\bigg(\partial_{t_{\ul p}}\prod_{n'=1}^n \rho(n')\bigg)
\bigg(\prod_{n'=1}^n \rho(n')\bigg)^{-1}
=\partial_{t_{\ul p}}\ln \bigg(\prod_{n'=1}^n \rho(n')\bigg) 
=\frac12\partial_{t_{\ul p}}\ln \bigg(\prod_{n'=1}^n \rho(n')^2\bigg) \no \\
&\quad =\frac12\partial_{t_{\ul p}}\ln \bigg(\prod_{n'=1}^n \gamma(n')\bigg) 
=\frac12\sum_{n'=1}^n\frac{\gamma_{t_{\ul p}}(n') }{\gamma(n')}. \lb{ALh5.15h} 
\end{align}
Thus, \eqref{ALh5.15g} reads
\begin{align}
& D^{-1} z^{-(n-1)/2}\bigg(\prod_{n'=1}^n \rho(n')\bigg)^{-1}
 \begin{pmatrix} \psi_{1,\pm,t_{\ul p}} \\ \psi_{2,\pm,t_{\ul p}} 
 \end{pmatrix}   \\ 
 & \quad =
\begin{pmatrix}  u_{\pm,t_{\ul p}} \\  z v_{\pm,t_{\ul p}} \end{pmatrix} 
 +\frac12\bigg(\sum_{n'=1}^n\frac{\gamma_{t_{\ul p}}(n') }{\gamma(n')}\bigg)
\begin{pmatrix}  u_{\pm} \\  z v_{\pm} \end{pmatrix} + 
\f{D_{t_{\ul p}}}{D} \begin{pmatrix}  u_{\pm} \\  z v_{\pm} \end{pmatrix}.   \lb{ALh5.15i}
\end{align}
Combining \eqref{ALh5.15f} and \eqref{ALh5.15i}  one finds that \eqref{ALh5.15e} is equivalent to 
\begin{equation}\lb{ALh5.15k}
 \begin{pmatrix}  u_{\pm,t_{\ul p}} \\ v_{\pm,t_{\ul p}} \end{pmatrix} 
 +\frac{1}2\bigg(\sum_{n'=1}^n\frac{\gamma_{t_{\ul p}}(n') }{\gamma(n')}\bigg)
\begin{pmatrix}  u_{\pm} \\   v_{\pm} \end{pmatrix} 
+ \f{D_{t_{\ul p}}}{D} \begin{pmatrix}  u_{\pm} \\  v_{\pm} \end{pmatrix} =
i \begin{pmatrix}    G_{\ul p} u_{\pm} - z  F_{\ul p}v_{\pm}\\
 z^{-1} H_{\ul p} u_{\pm} -  K_{\ul p}v_{\pm}   \end{pmatrix}.
\end{equation}
Using \eqref{AL2.14}, \eqref{ALK_p}, and \eqref{ALhD} we find
\begin{align}
\sum_{n'=1}^n\frac{\gamma_{t_{\ul p}}(n') }{\gamma(n')}
&= i\sum_{n'=1}^n(I-S^{-}) \big(g_{p_+,+} - g_{p_-,-} \big)\no  \\
&= i \big((g_{p_+,+}(n) - g_{p_-,-}(n))- (g_{p_+,+}(0) - g_{p_-,-}(0)) \big)\no \\
&= i\big(G_{\ul p}-K_{\ul p}\big)-2\f{D_{t_{\ul p}}}{D}. \lb{ALh5.15l}
\end{align}

{}From \eqref{ALPp}, \eqref{ALPpTop},  \eqref{ALh5.5}, and  \eqref{ALh5.7} one obtains (we recall that $n$ is assumed to be odd)
\begin{equation} \lb{ALh5.15m}
\begin{pmatrix}u_{\pm,t_{\ul p}}\\  v_{\pm,t_{\ul p}}\end{pmatrix}
=i\begin{pmatrix}-zF_{\ul p}v_{\pm}+\frac12( G_{\ul p}+K_{\ul p})u_{\pm}\\[1mm]
-z^{-1}H_{\ul p}u_{\pm}+\frac12( G_{\ul p}+K_{\ul p})v_{\pm}
\end{pmatrix},
\end{equation}
using \eqref{ALh4.19b}.

Inserting \eqref{ALh5.15l} into \eqref{ALh5.15k}, we see that it reduces to \eqref{ALh5.15m}, thereby proving \eqref{ALh5.15e} in the case when $n$ is odd.
The case with $n$ even follows from analogous computations.

Linear independence of $\Psi_- (z,\dott,t_{\ul p})$ and $\Psi_+ (z,\dott,t_{\ul p})$ follows from 
\begin{align}
\begin{pmatrix} \psi_{1,-}(z,n,t_{\ul p}) & \psi_{1,+}(z,n,t_{\ul p}) \\ 
\psi_{2,-}(z,n,t_{\ul p}) & \psi_{2,+}(z,n,t_{\ul p}) \end{pmatrix}
& =  D(t_{\ul p})  z^{n/2}  \bigg(\prod_{n'=1}^n \rho(n',t_{\ul p})\bigg)
A(z,n)^{-1}    \no \\
& \quad \times \begin{pmatrix} u_{-}(z,n,t_{\ul p}) & u_{+}(z,n,t_{\ul p}) \\ 
v_{-}(z,n,t_{\ul p}) & v_{+}(z,n,t_{\ul p}) \end{pmatrix},     \lb{ALhPsi}
\end{align}
the fact that $\rho(n,t_{\ul p}) \neq 0$,  $\det(A(z,n))=(-1)^{n+1}$, and from  
\begin{equation}
\det\bigg( \begin{pmatrix} u_{-}(z,n,t_{\ul p}) & u_{+}(z,n,t_{\ul p}) \\ 
v_{-}(z,n,t_{\ul p}) & v_{+}(z,n,t_{\ul p}) \end{pmatrix}\bigg) \neq 0,  \quad 
(n,t_{\ul p}) \in\bbZ\times\bbR,
\end{equation}
since by hypothesis $z \in \bbC\setminus\spec\big(\breve L(t_{\ul p})\big)$. 
\end{proof}

In the following we will always refer to the solutions  $\Psi_\pm$ introduced in 
\eqref{ALh5.14a}--\eqref{ALhD}. 

The next result recalls the existence of a propagator $W_{\ul p}$ associated 
with $P_{\ul p}$. (Below we denote by $\calB(\calH)$ the Banach space of all bounded linear operators defined on the Hilbert space $\calH$.) 

\begin{theorem} \lb{tALh5.4}
Assume Hypothesis \ref{hALh5.1} and suppose $\alpha,\beta$ satisfy 
$\AL_{\ul p}(\alpha,\beta)=0$ for some $\ul p\in\bbN_0^2$. Then there is a propagator $W_{\ul p}(s,t) \in \calB(\ell^2(\bbZ))$, $(s,t)\in\bbR^2$, satisfying 
\begin{align}
& (i)\;\;\; W_{\ul p}(t,t)=I, \quad t\in\bbR,  \lb{ALh5.13}  \\
& (ii)\;\; W_{\ul p}(r,s)W_{\ul p}(s,t)=W_{\ul p}(r,t), \quad (r,s,t)\in\bbR^3,   \lb{ALh5.14}  \\
& (iii) \; W_{\ul p}(s,t) \text{ is jointly strongly continuous in $(s,t)\in\bbR^2$,}  
\lb{ALh5.15} 
\end{align}
such that for fixed $t_0\in\bbR$, $f_0\in\ell^2(\bbZ)$, 
\begin{equation}
f(t)=W_{\ul p}(t,t_0)f_0, \quad  t\in\bbR,  \lb{ALh5.16}
\end{equation}
satisfies
\begin{equation}
\f{d}{dt} f(t) = \breve P_{\ul p}(t) f(t), \quad f(t_0)=f_0.  \lb{ALh5.17}
\end{equation}
Moreover, $\breve L(t)$ is similar to $\breve L(s)$ for all $(s,t)\in\bbR^2$,
\begin{equation}
\breve L(s) = W_{\ul p}(s,t) \breve L(t) W_{\ul p}(s,t)^{-1}, \quad (s,t)\in\bbR^2.  \lb{ALh5.18}
\end{equation} 
This extends to appropriate functions of $\breve L(t)$ and so, in particular, to its resolvent 
$\big(\breve L(t)-zI\big)^{-1}$, $z\in \bbC\setminus \sigma\big(\breve L(t)\big)$, and hence also yields
\begin{equation}
\sigma \big(\breve L(s)\big) = \sigma \big(\breve L(t)\big), \quad (s,t)\in\bbR^2.  \lb{ALh5.19}
\end{equation}
Consequently, the spectrum of $\breve L(t)$ is independent of $t\in\bbR$. 
\end{theorem}
\begin{proof}
\eqref{ALh5.13}--\eqref{ALh5.17} are standard results which follow, for instance, from Theorem X.69 of \cite{ReedSimon:1975} under even weaker hypotheses on 
$\alpha, \beta$. In particular, the propagator $W_{\ul p}$ admits the norm convergent Dyson series  
\begin{align}
& W_{\ul p}(s,t)=I+\sum_{k\in\bbN} \int_s^t dt_1 \int_s^{t_1} dt_2 \cdots \int_s^{t_{k-1}} dt_k \, 
\breve P_{\ul p}(t_1) \breve P_{\ul p}(t_2) \cdots \breve P_{\ul p}(t_k),   \lb{ALh5.20} \\
& \hspace*{9cm} (s,t)\in\bbR^2. \no 
\end{align}
Fixing $s\in\bbR$ and introducing the operator-valued function 
\begin{equation}
\breve K(t)= W_{\ul p}(s,t) \breve L(t) W_{\ul p}(s,t)^{-1}, \quad t\in\bbR, \lb{ALh5.21}
\end{equation}
one computes 
\begin{equation}
\breve K' (t)f = W_{\ul p}(s,t) \big(\breve L'(t)-\big[\breve P_{\ul p}(t),\breve L(t)\big]\big) 
W_{\ul p}(s,t)^{-1}f = 0, 
\quad t\in\bbR, \; f \in \ell^2(\bbZ),   \lb{ALh5.22}
\end{equation}
using the Lax commutator equation \eqref{ALLaxtp}. 
Thus, $\breve K$ is independent of $t\in\bbR$ and hence taking $t=s$ in \eqref{ALh5.21} then yields $\breve K=\breve L(s)$ and thus proves \eqref{ALh5.18}.
\end{proof}

Next we briefly recall the Ablowitz--Ladik initial value problem in a setting convenient for our purpose.

\begin{theorem} \lb{tALh5.5} 
Let $t_{0,\ul p}\in\bbR$ and suppose $\alpha^{(0)}, \beta^{(0)} \in \ell^q(\bbZ)$ for some 
$q\in [1,\infty)\cup\{\infty\}$. Then the $\ul p$th Ablowitz--Ladik initial value problem
\begin{equation}
\AL_{\ul p}(\alpha,\beta)=0, \quad (\alpha,\beta)\big|_{t_{\ul p}=t_{0,\ul p}} 
= \big(\alpha^{(0)},\beta^{(0)}\big)  \lb{ALh5.23}
\end{equation}
for some $\ul p\in\bbN_0^2$, has a unique, local, and  smooth solution in time, that is, there exists a $T_0>0$ such that
\begin{equation}
\alpha(\dott), \, \beta(\dott) \in C^\infty((t_{0,\ul p}-T_0,t_{0,\ul p}+T_0),\ell^q(\bbZ)).  
\lb{ALh5.24}
\end{equation}
\end{theorem}
\begin{proof}
This follows from standard results in \cite[Sect.\ 4.1]{AbrahamMarsdenRatiu:1988}. More precisely, local existence and uniqueness as well as smoothness of the solution of the initial value problem \eqref{ALh5.23} (cf.\ \eqref{AL_p}) follows from \cite[Theorem\ 4.1.5]{AbrahamMarsdenRatiu:1988} since $f_{p_{\pm}-1,\pm}$, $g_{p_{\pm},\pm}$, and 
$h_{p_{\pm}-1,\pm}$ depend polynomially on $\alpha, \beta$ and certain of their shifts, and the fact that the Ablowitz--Ladik flows are autonomous. 
\end{proof}

For an analogous result in connection with the Toda hierarchy we refer to 
\cite{GesztesyHoldenTeschl:2007} and \cite[Sect.\ 12.2]{Teschl:2000}.

\begin{remark} \lb{rALh5.6} 
In the special defocusing case, where $\beta = \ol\alpha$ and hence $\breve L(t)$, $t\in\bbR$, is unitary, one obtains 
\begin{equation}
\sup_{(n,t_{\ul p})\in\bbN\times(t_{0,\ul p}-T_0,t_{0,\ul p}+T_0)}
|\alpha(n,t_{\ul p})|\leq1  \lb{ALh5.25}
\end{equation}
using $\gamma=1-|\alpha|^2$ and 
$\gamma_{t_{\ul p}} = i \gamma \big((g_{p_+,+} - g_{p_+,+}^-) - (g_{p_-,-} - g_{p_-,-}^-) \big)$ in \eqref{AL2.14}. 
A further application of \cite[Proposition 4.1.22]{AbrahamMarsdenRatiu:1988} then yields a unique, global, and 
smooth solution of  the $\ul p$th AL initial value problem \eqref{ALh5.23}.  Moreover, the same 
argument shows that if $\alpha$ satisfies Hypothesis \ref{hALh5.1} 
and the $\ul p$th AL equation $\AL_{\ul p}(\alpha, \ol \alpha)=0$, then $\alpha$ is actually smooth with respect to 
$t_{\ul p}\in\bbR$, that is, 
\begin{equation}
\alpha(n,\dott) \in C^\infty(\bbR), \quad  n\in\bbZ.  \lb{ALh5.26}
\end{equation} 
\end{remark}

Equation \eqref{ALh5.15b}, that is, $\Psi_{\pm,t_{\ul p}} = V^+_{\ul p} \Psi_\pm$,  implies that
\begin{align}
\partial_{t_{\ul p}}\ln\bigg(\frac{\psi_{1,\pm}^+}{\psi_{1,\pm}} \bigg)
&= (S^+ -I) \partial_{t_{\ul p}}
\ln(\psi_{1,\pm}) \no \\
&= (S^+ -I)\frac{\partial_{t_{\ul p}}\psi_{1,\pm}}{\psi_{1,\pm}} \no \\
&= i(S^+ -I)(G_{\ul p}-F_{\ul p} \phi_\pm).
\end{align}
On the other hand, equation \eqref{ALh5.15a}, that is, $U\Psi^-_\pm=\Psi_\pm$, yields 
\begin{equation}
\partial_{t_{\ul p}}\ln\bigg(\frac{\psi_{1,\pm}^+}{\psi_{1,\pm}} \bigg)
= \partial_{t_{\ul p}}\ln(z+\alpha^+ \phi_\pm),
\end{equation}
and thus one concludes that
\begin{equation}\lb{ALh_cons}
\partial_{t_{\ul p}}\ln(z+\alpha^+ \phi_\pm)=  i(S^+ -I)(G_{\ul p}-F_{\ul p} \phi_\pm).
\end{equation}
Below we  will refer to (\ref{ALh_cons}$\pm$) according to the upper or lower sign in 
\eqref{ALh_cons}.  
Expanding (\ref{ALh_cons}$\pm$) in powers of $z$ and $1/z$ then yields the following conserved densities:

\begin{theorem}\lb{tALh5.7}
Assume Hypothesis \ref{hALh5.1} and suppose $\alpha,\beta$ satisfy 
$\AL_{\ul p}(\alpha,\beta)=0$ for some $\ul p\in\bbN_0^2$. Then the following infinite sequences of local conservation laws hold:

Expansion of $($\ref{ALh_cons}$+$$)$ at $1/z=0$$:$
\begin{align} \no
\partial_{t_{\ul p}}\rho_{j,+}^\infty &=  i(S^+ -I)\bigg(g_{p_--j,-}
- \sum_{\ell=0}^{j-1}f_{p_--j+\ell,-}\phi_{\ell,+}^\infty  
 -\sum_{\ell=0}^{p_+-1} f_{p_+-1-\ell,+}\phi_{j+\ell,+}^\infty \bigg),\\ 
& \hspace{7.8cm} j=1,\dots,p_-,   \lb{ALh_cons1} \\
\partial_{t_{\ul p}}\rho_{j,+}^\infty &= -i(S^+ -I)\bigg( \sum_{\ell=1}^{p_-}f_{p_--\ell,-}\phi_{j-\ell,+}^\infty  
+\sum_{\ell=0}^{p_+-1} f_{p_+-1-\ell,+}\phi_{j+\ell,+}^\infty \bigg),   \no  \\ 
& \hspace*{7.55cm}  j\geq p_-+1, \lb{ALh_cons2} 
\end{align}
where $\rho_{j,+}^\infty$ and $\phi_{j,+}^\infty$ are given by \eqref{ALhrhoinfp} and \eqref{ALhphiinfp}.

Expansion of $($\ref{ALh_cons}$-$$)$ at $1/z=0$$:$  
\begin{align} \no
\partial_{t_{\ul p}}\rho_{j,-}^\infty &=i(S^+ -I) \bigg(g_{p_--j,-}
- \sum_{\ell=-1}^{j-1}f_{p_-+\ell-j,-} \phi_{\ell,-}^\infty 
- \sum_{\ell=0}^{p_+-1} f_{p_+-1-\ell,+} \phi_{j+\ell,-}^\infty \bigg),\\  
&\hspace{7.8cm} j=1,\dots,p_-,  \lb{ALh_cons3} \\
\partial_{t_{\ul p}}\rho_{j,-}^\infty &=-i(S^+ -I)\bigg(\sum_{\ell=1}^{p_-}f_{p_--\ell,-}\phi_{j-\ell,-}^\infty 
+ \sum_{\ell=0}^{p_+-1} f_{p_+-1-\ell,+} \phi_{j+\ell,-}^\infty  \bigg),  \no\\
& \hspace*{7.55cm} j \geq p_-+1,   \lb{ALh_cons4}
\end{align}
where $\rho_{j,-}^\infty$ and $\phi_{j,-}^\infty$ are given by \eqref{ALhrhoinfm} and 
\eqref{ALhphiinfm}.

Expansion of $($\ref{ALh_cons}$+$$)$ at $z=0$$:$  
\begin{align} \no
\partial_{t_{\ul p}}\rho_{j,+}^0&=  i(S^+ -I)\bigg(g_{p_+-j,+}
- \sum_{\ell=1}^{p_-}\phi_{j+\ell,+}^0f_{p_--\ell,-} -
\sum_{\ell=0}^{j}\phi_{\ell,+}^0f_{p_+-1-j+\ell,+} \bigg), \\ 
&\hspace{7cm} j=1,\dots,p_+-1,  \lb{ALh_cons5}\\
\partial_{t_{\ul p}}\rho_{p_+,+}^0&= i(S^+ -I)\bigg(g_{0,+} 
-  \sum_{\ell=1}^{p_-}\phi_{j+\ell,+}^0f_{p_--\ell,-} -
\sum_{\ell=0}^{p_+-1}\phi_{j+\ell-p_++1,+}^0f_{\ell,+} \bigg), \lb{ALh_cons5a}\\ 
\partial_{t_{\ul p}}\rho_{j,+}^0&= -i(S^+ -I)\bigg(\sum_{\ell=1}^{p_-}\phi_{j+\ell,+}^0f_{p_--\ell,-} +
\sum_{\ell=0}^{p_+-1}\phi_{j+\ell-p_++1,+}^0f_{\ell,+} \bigg), \no \\  
&\hspace{7.6cm} j\geq p_++1,  \lb{ALh_cons6}
\end{align}
where $\rho_{j,+}^0$ and $\phi_{j,+}^0$ are given by \eqref{ALhrhozerop} and \eqref{ALhphizerop}.

Expansion of $($\ref{ALh_cons}$-$$)$ at $z=0$$:$   
\begin{align} \no
\partial_{t_{\ul p}}\rho_{j,-}^0&=i(S^+ -I) \bigg(g_{p_+-j,+}
- \sum_{\ell=1}^{j}\phi_{\ell,-}^0f_{p_+-j+\ell-1,+} 
-\sum_{\ell=1}^{p_-}\phi_{j+\ell,-}^0f_{p_--\ell,-} \bigg), \\ 
&\hspace{8cm} j=1,\dots,p_+, \lb{ALh_cons7}\\ 
\partial_{t_{\ul p}}\rho_{j,-}^0&=-i(S^+ -I)\bigg(\sum_{\ell=j+1-p_+}^{j}\phi_{\ell,-}^0f_{p_+-j+\ell-1,+} +
\sum_{\ell=1}^{p_-}\phi_{j+\ell,-}^0f_{p_--\ell,-} \bigg), \no\\ 
&\hspace{8.25cm} j \geq p_++1, \lb{ALh_cons8} 
\end{align}
where $\rho_{j,-}^0$ and $\phi_{j,-}^0$ are given by \eqref{ALhrhozerom} and \eqref{ALhphizerom}.
\end{theorem}
\begin{proof} 
The proof consists of expanding (\ref{ALh_cons}$\pm$) in powers of $z$ and $1/z$ and applying \eqref{ALh4.13}--\eqref{ALhrhozerom}. 

Expansion of $($\ref{ALh_cons}$+$$)$ at $1/z=0$: For the right-hand side of 
(\ref{ALh_cons}$+$) one finds
\begin{align}
 G_{\ul p}- F_{\ul p} \phi_+&= \sum_{\ell=1}^{p_-} g_{p_--\ell,-}z^{-\ell}  
+ \sum_{\ell=0}^{p_+} g_{p_+-\ell,+}z^\ell \no \\
&\quad -\bigg(\sum_{\ell=1}^{p_-} f_{p_--\ell,-} z^{-\ell} + \sum_{\ell=0}^{p_+-1} f_{p_+-1-\ell,+}z^\ell\bigg)
\sum_{j=0}^\infty \phi_{j,+}^\infty z^{-j} \no \\
&=g_{0,+}z^{p_+}+\sum_{j=0}^{p_+-1}\bigg(g_{p_+-j,+}
-\sum_{\ell=0}^{p_+-j-1} f_{p_+-j-1-\ell,+}\phi_{\ell,+}^\infty \bigg) z^j\no \\
&\quad +\sum_{j=1}^{p_-} \bigg(g_{p_--j,-}- \sum_{\ell=0}^{j-1}f_{p_--j+\ell,-}\phi_{\ell,+}^\infty  
 -\sum_{\ell=0}^{p_+-1} f_{p_+-1-\ell,+}\phi_{j+\ell,+}^\infty \bigg)z^{-j} \no \\
&\quad -\sum_{j=p_-+1}^{\infty}\bigg( \sum_{\ell=1}^{p_-}f_{p_--\ell,-}\phi_{j-\ell,+}^\infty  
+\sum_{\ell=0}^{p_+-1} f_{p_+-1-\ell,+}\phi_{j+\ell,+}^\infty \bigg)z^{-j}. \label{ALh5.44}
\end{align}
Here we used that that all positive powers vanish because of \eqref{ALh_cons}. This yields the following additional formulas:

Conservation laws derived from $\phi_+$ at $1/z=0$:
\begin{align}
&(S^+ -1)\bigg(g_{p_+-j,+}- \sum_{\ell=0}^{p_+-j-1}f_{p_+-j-1-\ell,+}\phi_{\ell,+}^\infty \bigg)=0, 
\quad j=0,\dots,p_+-1, \no \\
&(S^+ -1)g_{0,+}=0.
\end{align}
Expansion of $($\ref{ALh_cons}$-$$)$ at $1/z=0$: The right-hand side of 
(\ref{ALh_cons}$-$) yields
\begin{align}
 G_{\ul p}- F_{\ul p} \phi_-&= \sum_{\ell=1}^{p_-} g_{p_--\ell,-}z^{-\ell}  
+ \sum_{\ell=0}^{p_+} g_{p_+-\ell,+}z^\ell \no \\
&\quad -\bigg(\sum_{\ell=1}^{p_-} f_{p_--\ell,-} z^{-\ell} + \sum_{\ell=0}^{p_+-1} f_{p_+-1-\ell,+}z^\ell\bigg)
\sum_{j=-1}^\infty \phi_{j,-}^\infty z^{-j} \no \\
&=\sum_{j=1}^{p_+} \bigg(g_{p_+-j,+}
- \sum_{\ell=-1}^{p_+-j-1}f_{p_+-j-1-\ell,+} \phi_{\ell,-}^\infty \bigg)z^j\no \\
&\quad+\bigg(g_{p_+,+}- \sum_{\ell=0}^{p_+-1}f_{p_+-1-\ell,+} \phi_{\ell,-}^\infty 
- f_{p_--1,-} \phi_{-1,-}^\infty \bigg)\no \\
&\quad+\sum_{j=1}^{p_-} \bigg(g_{p_--j,-}- \sum_{\ell=-1}^{j-1}f_{p_-+\ell-j,-} \phi_{\ell,-}^\infty 
- \sum_{\ell=0}^{p_+-1} f_{p_+-1-\ell,+} \phi_{j+\ell,-}^\infty \bigg)z^{-j}\no \\
&\quad-\sum_{j=p_-+1}^{\infty} \bigg(\sum_{\ell=1}^{p_-}f_{p_--\ell,-}\phi_{j-\ell,-}^\infty 
+ \sum_{\ell=0}^{p_+-1} f_{p_+-1-\ell,+} \phi_{j+\ell,-}^\infty \bigg) z^{-j}.   
\label{ALh5.46}
\end{align}
Conservation laws derived from $\phi_+$ at $1/z=0$:
\begin{align}
& (S^+ -I) \bigg(g_{p_+-j,+}- \sum_{\ell=-1}^{p_+-j-1}f_{p_+-j-1-\ell,+} 
\phi_{\ell,-}^\infty \bigg)=0, \quad j=1,\dots,p_+,\\
& i(S^+ -I) \bigg(g_{p_+,+}- \sum_{\ell=0}^{p_+-1}f_{p_+-1-\ell,+} 
\phi_{\ell,-}^\infty - f_{p_--1,-} \phi_{-1,-}^\infty \bigg)  \no \\ \lb{ALh5.52}
& \quad =
\partial_{t_{\ul p}}\ln\bigg( \frac{\alpha^{++}}{\alpha^+}\bigg) 
+ \partial_{t_{\ul p}}\ln (\gamma^+).
\end{align}
Expansion of $($\ref{ALh_cons}$+$$)$ at $z=0$: For the right-hand side of 
(\ref{ALh_cons}$+$) one finds
\begin{align}
G_{\ul p} - F_{\ul p} \phi_+ &=
\sum_{j=1}^{p_-} \bigg(g_{p_--j,-} - \sum_{\ell=0}^{p_--j} \phi_{\ell,+}f_{p_--j-\ell,-}\bigg)z^{-j}  \\
&\quad +\sum_{j=0}^{p_+-1} \bigg(g_{p_+-j,+}- \sum_{\ell=1}^{p_-}\phi_{j+\ell,+}^0f_{p_--\ell,-} -
\sum_{\ell=0}^{j}\phi_{\ell,+}^0f_{p_+-1-j+\ell,+} \bigg)z^{j}\no \\ \no
&\quad +\sum_{j=p_+}^\infty\bigg(g_{0,+}\chi_{jp_+} -  \sum_{\ell=1}^{p_-}\phi_{j+\ell,+}^0f_{p_--\ell,-} -
\sum_{\ell=0}^{p_+-1}\phi_{j+\ell-p_++1,+}^0f_{\ell,+} \bigg)z^{j}.
\end{align}
Conservation laws derived from $\phi_+$ at $z=0$: 
\begin{align}
& (S^+ -I)\bigg(g_{j,-} - \sum_{\ell=0}^{j} \phi_{\ell,+}^0f_{j-\ell,-}\bigg)=0, 
\quad j=1, \dots, p_-,\\
& (S^+ -I) \bigg(g_{p_+,+}-\phi_{0,+}^0f_{p_+-1,+} 
- \sum_{\ell=1}^{p_-}\phi_{\ell,+}^0f_{p_--\ell,-} \bigg)=\partial_{t_{\ul p}}
\ln \bigg(\frac{\alpha}{\alpha^+}\bigg).
\end{align}
Expansion of $($\ref{ALh_cons}$-$$)$ at $z=0$: For the right-hand side of 
(\ref{ALh_cons}$-$) one finds
\begin{align}
G_{\ul p} - F_{\ul p} \phi_- &= g_{0,-}z^{-p_-} + 
\sum_{j=1}^{p_--1} \bigg(g_{p_--j,-} - \sum_{\ell=1}^{p_--j} \phi_{\ell,-}^0f_{p_--j-\ell,-}\bigg)z^{-j}  \\
&\quad + g_{p_+,+} - \sum_{\ell=1}^{p_-} \phi_{\ell,-}^0f_{p_--\ell,-}\no \\
&\quad +\sum_{j=1}^{p_+} \bigg(g_{p_+-j,+}- \sum_{\ell=1}^{j}\phi_{\ell,-}^0f_{p_+-j+\ell-1,+} -
\sum_{\ell=1}^{p_-}\phi_{j+\ell,-}^0f_{p_--\ell,-} \bigg)z^{j}\no \\ \no
&\quad -\sum_{j=p_++1}^\infty\bigg(\sum_{\ell=j+1-p_+}^{j}\phi_{\ell,-}^0f_{p_+-j+\ell-1,+} +
\sum_{\ell=1}^{p_-}\phi_{j+\ell,-}^0f_{p_--\ell,-} \bigg)z^{j}.
\end{align}
Conservation laws derived from $\phi_-$ at $z=0$:  
\begin{align}
& (S^+ -I) g_{0,-}=0, \\
& (S^+ -I)\bigg(g_{j,-} - \sum_{\ell=1}^{j} \phi_{\ell,-}^0f_{j-\ell,-}\bigg)=0, \quad j=1, \dots, p_--1,\\
& (S^+ -I)\bigg( g_{p_+,+} - \sum_{\ell=1}^{p_-} \phi_{\ell,-}^0f_{p_--\ell,-}\bigg)
=\partial_{t_{\ul p}}\ln (\gamma^+).
\end{align}
Combining these expansions with \eqref{ALh4.13}--\eqref{ALh4.19} finishes the proof.
\end{proof}
\begin{remark} \lb{rALh5.8}
$(i)$ There is a certain redundancy in the conservation laws \eqref{ALh_cons1}--\eqref{ALh_cons8} as can be observed from Lemma \ref{lALh4.4}. Equations \eqref{ALh4.23}--\eqref{ALh4.24} imply
\begin{align}
\rho_{j,+}^\infty&= -\rho_{j,-}^\infty +\frac1j(S^+ - I)(d_{j,+}-e_{j,+}),  
\quad j\in\bbN,   \lb{ALh5.11A} \\
\rho_{j,+}^0 &= -\rho_{j,-}^0+\frac1j(S^+ - I)(d_{j,-}-e_{j,-}), \quad j\in\bbN. \lb{ALh5.11B} 
\end{align}
Thus one can, for instance, transfer \eqref{ALh_cons1}--\eqref{ALh_cons2} into
\eqref{ALh_cons3}--\eqref{ALh_cons4}. \\
$(ii)$ In addition to the conservation laws listed in Theorem \ref{tALh5.7}, we recover the familiar conservation law $($cf.\ \eqref{AL2.14}$)$
\begin{equation}\lb{ALh5.11C} 
\partial_{t_{\ul p}}\ln (\gamma) = i  (I-S^{-}) (g_{p_+,+} - g_{p_-,-}), \quad 
 \ul p \in\bbN_0^2.
\end{equation}
$(iii)$  Another consequence of Theorem \ref{tALh5.7} and Lemma \ref{lALh4.4} is that for $\alpha$, $\beta$ satisfying Hypothesis \ref{hALh5.1} and $\alpha, \beta\in C^1(\bbR, \ell^2(\bbZ))$, one has 
\begin{equation}
\frac{d}{d t_{\ul p}}\sum_{n\in\bbZ} \ln (\gamma(n,t_{\ul p})) = 0, \quad 
\frac{d}{d t_{\ul p}}\sum_{n\in\bbZ} \hat g_{j,\pm}(n,t_{\ul p}) = 0,  
 \quad  j\in\bbN, \; \ul p \in\bbN_0^2. \lb{ALh5.11D}  
 \end{equation}
\end{remark}

\begin{remark} \lb{rALh5.8a}
The infinite sequence of conservation laws has been studied in the literature; we refer to  
 \cite{AblowitzLadik:1976}, \cite[Ch.\ 3]{AblowitzPrinariTrubatch:2004},  
\cite{DingSunXu:2006},  \cite{ZhangChen:2002}, and  \cite{ZhangNingBiChen:2006}. In particular, Zhang and Chen \cite{ZhangChen:2002} study local conservation laws for the full $4\times 4$ Ablowitz--Ladik  system in a similar way to the one employed here. However, they only expand their equation around  a point that corresponds to $1/z=0$. 
The systematic derivation of infinite sequences of conserved densities and currents 
$($cf.\ the corresponding discussion in the introduction$)$ as presented in 
Theorem \ref{tALh5.7} appears to be new. 

The two local conservation laws coming from expansions around $z=0$ are essentially the same since the two conserved densities, $\rho_{j,+}^0$ and $\rho_{j,+}^0$, differ by a first-order difference expression $($cf.\ Remark  \ref{rALh5.8}$)$.  A similar argument applies to the expansions around $1/z=0$.   That there are two independent sequences of conservation laws is also clear from \eqref{ALh5.11D} which yields that 
$\sum_{n\in\bbZ} \hat g_{j,\pm}(n,t_{\ul p})$ are time-independent. One observes that the quantities  $\hat g_{j,+}$, $j\in\bbN$, are related to the expansions around $1/z=0$, that is, to $\rho_{j,\pm}^\infty$, while  $\hat g_{j,-}$, $j\in\bbN$, are related to  $\rho_{j,\pm}^0$ $($cf.\ Lemma \ref{lALh4.4}$)$. In addition to the two infinite sequences of polynomial conservation laws, there is a logarithmic conservation law 
$($cf.\ \eqref{ALh5.11C} and \eqref{ALh5.11D}$)$.
\end{remark}

The first conservation laws explicitly read as follows:\\
$p_+=p_-=1$:
\begin{align}
\partial_{t_{(1,1)}}\rho_{j,\pm}^\infty
&= -i (S^+ - I)(f_{0,-}\phi_{j-1,\pm}^\infty + f_{0,+}\phi_{j,\pm}^\infty), \quad j\geq 1,\\
\partial_{t_{(1,1)}}\rho_{j,\pm}^0
&= -i (S^+ - I)(f_{0,-}\phi_{j+1,\pm}^0 + f_{0,+}\phi_{j,\pm}^0), \quad j\geq 1.
\end{align}
For $j=1$ this yields using \eqref{ALh4.55}
\begin{align}
\partial_{t_{(1,1)}}\rho_{1,+}^\infty&=\partial_{t_{(1,1)}} \alpha^+\beta =i (S^+ - I)(-c_{0,-}\alpha\beta +c_{0,+}\alpha^+\beta^-\gamma),  \no \\
\partial_{t_{(1,1)}}\rho_{1,-}^\infty&=\partial_{t_{(1,1)}}\Big(-\alpha^{+++}\beta^{++} + (S^+-I) \frac{\alpha^{++}}{\alpha^+}\Big)  \no \\
&=i (S^+ - I)\Big(c_{0,+}\frac{\alpha^{+++}}{\alpha^+}\gamma^+\gamma^{++}
-c_{0,-}\frac{\alpha \alpha^{++}}{(\alpha^+)^2}\gamma^+
-c_{0,+}\Big(\frac{\alpha \alpha^{++}}{\alpha^+}\Big)^2\gamma^+ \Big),  \no \\
\partial_{t_{(1,1)}}\rho_{1,+}^0
&=\partial_{t_{(1,1)}}\Big(\alpha^-\beta + (S^+-I) \frac{\alpha^-}{\alpha}\Big)\\
&=i (S^+ - I)\Big(c_{0,-}\frac{\alpha^{--}}{\alpha}\gamma^-\gamma-c_{0,+}\frac{\alpha^- \alpha^{+}}{\alpha^2}\gamma
-c_{0,-}\Big(\frac{\alpha^-}{\alpha}\Big)^2\gamma \Big),  \no \\
\partial_{t_{(1,1)}}\rho_{1,-}^0&=\partial_{t_{(1,1)}} \alpha^+\beta^{++} =i (S^+ - I)(c_{0,+}\alpha^+\beta^+ -
c_{0,-}\alpha\beta^{++}\gamma^+).  \no
\end{align}
This shows in particular that we obtain two sets of conservation laws (one from expanding near $\infty$ and the other from
expanding near $0$), where the first few equations of each set explicitly read ($p_+=p_-=1$):    
\begin{align}
\begin{split}
&j=1: \quad \partial_{t_{(1,1)}} \alpha^+\beta =i (S^+ - I)(-c_{0,-}\alpha\beta +c_{0,+}\alpha^+\beta^-\gamma),\\
& \hspace*{1.45cm} \partial_{t_{(1,1)}} \alpha\beta^+ =i (S^+ - I)(c_{0,+}\alpha \beta -
c_{0,-}\alpha^-\beta^+\gamma), 
\end{split} \\
&j=2: \quad \partial_{t_{(1,1)}} \big( - \tfrac{1}{2}(\alpha^+\beta)^2+\gamma\alpha^+\beta^- \big)  \no \\
& \hspace*{1.8cm} = i (S^+ - I)\gamma\big(-c_{0,-}\alpha\beta^- -c_{0,+}\alpha \alpha^+(\beta^-)^2 
+c_{0,+}\gamma^-\alpha^+\beta^{--} \big),   \no \\
& \hspace*{1.45cm} 
\partial_{t_{(1,1)}} \big( \tfrac{1}{2}(\alpha\beta^{+})^2-\gamma^{+}\alpha\beta^{++}\big)\\
&\hspace*{1.8cm} =  i (S^+ - I)\gamma\big(-c_{0,+}\alpha\beta^+ -c_{0,-}\alpha^- \alpha(\beta^+)^2 
+c_{0,-}\gamma^+\alpha^-\beta^{++} \big).   \no 
\end{align}

Using Lemma~\ref{lALh4.4}, one observes that one can replace $\rho_{j,\pm}^{\infty, 0}$ in Theorem~\ref{tALh5.7}
by $\hat g_{j,\pm}$ by suitably adjusting the right-hand sides in 
\eqref{ALh_cons1}--\eqref{ALh_cons8}.

\section{Hamiltonian formalism, variational derivatives} \lb{sALh6}

We start this section by a short review of variational derivatives for discrete systems. 
Consider the functional
\begin{align}
\begin{split}
& \calG\colon \ell^1(\bbZ)^\kappa\to \bbC,   \lb{ALh6.1} \\
& \calG(u)=\sum_{n\in\bbZ} 
G\big(u(n), u^{(+1)}(n), u^{(-1)}(n), \dots, u^{(k)}(n), u^{(-k)}(n)\big)
\end{split}
\end{align}
for some $\kappa\in\bbN$ and $k\in\bbN_0$, where $G\colon \bbZ\times \bbC^{2r\kappa} \to \bbC$ is $C^1$ with respect to the $2r\kappa$ complex-valued entries and where
\begin{equation}
u^{(s)}= S^{(s)} u, \quad
 S^{(s)}= \begin{cases} (S^+)^s u & \text{if $s\ge0$}, \\
(S^{-})^{-s} u & \text{if $s<0$}, \end{cases} \quad u \in \ell^\infty(\bbZ)^\kappa.  
\lb{ALh6.2}
\end{equation}
For brevity we  write 
\begin{equation}
G(u(n))=G\big(u(n), u^{(+1)}(n), u^{(-1)}(n), \dots, u^{(k)}(n), u^{(-k)}(n)\big).
\lb{ALh6.3}
\end{equation}

The functional $\calG$ is Frechet-differentiable and one computes
for any $v\in \ell^1(\bbZ)^\kappa$ for the differential $d\calG$
\begin{align}
(d\calG)_u (v)&=\frac{d}{d\epsilon}\calG(u+\epsilon v)\big|_{\epsilon=0} \no \\
&=\sum_{n\in\bbZ}\bigg(\frac{\partial G(u(n))}{\partial u} v(n)+ \frac{\partial G(u(n))}{\partial u^{(+1)}} v^{(+1)}(n)
+\frac{\partial G(u(n))}{\partial u^{(-1)}} v^{(-1)}(n) \no \\
&\hspace*{1.3cm}  +\cdots+ \frac{\partial G(u(n))}{\partial u^{(k)}} v^{(k)}(n)
+\frac{\partial G(u(n))}{\partial u^{(-k)}} v^{(-k)}(n) \bigg)\no \\
&= \sum_{n\in\bbZ}\bigg(\frac{\partial G(u(n))}{\partial u}+ S^{(-1)}\frac{\partial G(u(n))}{\partial u^{(+1)}} + 
S^{(+1)}\frac{\partial G(u(n))}{\partial u^{(-1)}}\no \\
&\hspace*{1.3cm}  +\cdots+S^{(-k)}\frac{\partial G(u(n))}{\partial u^{(k)}} 
+S^{(k)}\frac{\partial G(u(n))}{\partial u^{(-k)}} \bigg)v(n)\no \\
& = \sum_{n\in\bbZ}\frac{\delta G}{\delta u}(n)  v(n) ,   \lb{ALh6.4}
\end{align}
where we introduce the gradient and the 
variational derivative of $\calG$ by
\begin{align}
(\nabla\calG)_u&=\frac{\delta G}{\delta u}   \lb{ALh6.6}  \\
&= \frac{\partial G}{\partial u}+ S^{(-1)}\frac{\partial G}{\partial u^{(+1)}} + 
S^{(+1)}\frac{\partial G}{\partial u^{(-1)}}+\cdots
+S^{(-k)}\frac{\partial G}{\partial u^{(k)}} 
+S^{(k)}\frac{\partial G}{\partial u^{(-k)}}, \no
\end{align} 
assuming
\begin{align}
\begin{split}
& \{G(u(n))\}_{n\in\bbZ}, \, 
\bigg\{\frac{\partial G(u(n))}{\partial u^{(\pm j)}}\bigg\}_{n\in\bbZ} \in \ell^1(\bbZ),  
\quad j=1,\dots,k.   \lb{ALh6.5}
\end{split}
\end{align}

To establish the connection with the Ablowitz--Ladik hierarchy we make the following assumption for the remainder of this section.

\begin{hypothesis} \lb{hALh6.1}
Suppose
\begin{equation}
\alpha, \beta\in \ell^1(\bbZ), \quad 
\alpha(n)\beta(n)\notin \{0,1\}, \; n\in\bbZ. \lb{ALh6.6a}
\end{equation}
\end{hypothesis}

Next, let $\calG$ be a functional of the type 
\begin{align}
&\calG\colon \ell^\infty(\bbZ)^2 \to\bbC,   \no \\
& \calG(\alpha,\beta) = \sum_{n\in\bbZ} G(\alpha(n),\beta(n), \dots, 
\alpha(n+k), \beta(n+k),\alpha(n-k), \beta(n-k))  \no \\
& \qquad\quad  = \sum_{n\in\bbZ} G(\alpha(n),\beta(n)),   \lb{ALh6.8} 
\end{align}
where $G(\alpha,\beta)$ is polynomial in $\alpha, \beta$ and some of their shifts.  
The gradient $\nabla \calG$ and symplectic gradient $\nabla_s \calG$ of $\calG$ are then defined by
\begin{equation}
(\nabla \calG)_{\alpha,\beta}=\begin{pmatrix} (\nabla \calG)_\alpha \\ (\nabla \calG)_\beta \end{pmatrix} 
= \begin{pmatrix}  \f{\delta \calG}{\delta \alpha} \\[1.5mm]  \f{\delta \calG}{\delta \beta} 
\end{pmatrix}
\end{equation}
and
\begin{equation}
(\nabla_s \calG)_{\alpha,\beta}= \calD(\nabla\calG)_{\alpha,\beta}
= \calD\begin{pmatrix} 
(\nabla\calG)_{\alpha}\\ (\nabla\calG)_{\beta}\end{pmatrix},   \lb{ALh6.11}
\end{equation}
respectively.   Here $\calD$ is defined by
\begin{equation}
\calD=\gamma\begin{pmatrix}0& 1\\ -1& 0 \end{pmatrix}, \quad 
\gamma=1 - \alpha\beta.
\end{equation}
In addition, we introduce the bilinear form 
\begin{align}
&\Omega\colon \ell^1(\bbZ)^2\times \ell^1(\bbZ)^2\to \bbC, \no \\
&\Omega(u,v)=\sum_{n\in\bbZ} (\calD^{-1}u)(n) \cdot v(n). \lb{ALh6.11a}
\end{align}
One then concludes that
\begin{align}
\begin{split}
\Omega(\calD u,v)&=\sum_{n\in\bbZ} u(n) \cdot v(n)
=\sum_{n\in\bbZ} \big(u_1(n)v_1(n)+u_2(n)v_2(n)\big) \\
&=\langle u, v \rangle_{\ell^2(\bbZ)^2}, \quad u, v \in \ell^1(\bbZ)^2,  \lb{ALh6.11b}
\end{split}
\end{align}
where $\langle \dott,\dott \rangle_{\ell^2(\bbZ)^2}$ denotes the ``real'' inner product in $ \ell^2(\bbZ)^2$, that is,
\begin{align}
\begin{split}
& \langle \dott,\dott \rangle_{\ell^2(\bbZ)^2} \colon \ell^2(\bbZ)^2\times\ell^2(\bbZ)^2 \to \bbC, \\
&\langle u, v \rangle_{\ell^2(\bbZ)^2} =\sum_{n\in\bbZ} u(n)\cdot v(n) 
=\sum_{n\in\bbZ} \big(u_1(n)v_1(n)+u_2(n)v_2(n)\big).
\end{split}
\end{align}
In addition, one obtains 
\begin{equation}
(d\calG)_{\alpha,\beta}(v)=\langle (\nabla\calG)_{\alpha,\beta},v\rangle_{\ell^2(\bbZ)^2}
=\Omega(\calD(\nabla \calG)_{\alpha,\beta},v) =\Omega((\nabla_s \calG)_{\alpha,\beta},v). \lb{ALh6.16}
\end{equation}
Given two functionals $\calG_1,\calG_2$ we define their Poisson bracket by
\begin{align}
\{\calG_1,\calG_2\}&=d\calG_1(\nabla_s\calG_2)= \Omega(\nabla_s\calG_1,\nabla_s\calG_2)\no \\
&= \Omega(\calD\nabla\calG_1,\calD\nabla\calG_2) =\langle \nabla\calG_1,\calD\nabla\calG_2\rangle_{\ell^2(\bbZ)^2} \no \\
&=\sum_{n\in\bbZ}
\begin{pmatrix}\frac{\delta G_1}{\delta \alpha}(n)\\[1.5mm] \frac{\delta G_1}{\delta \beta}(n)\end{pmatrix} \cdot 
\calD\begin{pmatrix}\frac{\delta G_2}{\delta \alpha}(n)\\[1.5mm] \frac{\delta G_2}{\delta \beta}(n)\end{pmatrix}.  
\lb{ALh6.16a}
\end{align}
Since $\Omega(\dott,\dott)$ is a weakly non-degenerate closed $2$-form, both the Jacobi identity 
\begin{equation}
\{\{\calG_1,\calG_2\},\calG_3\}+ \{\{\calG_2,\calG_3\},\calG_1\}+\{\{\calG_3,\calG_1\},\calG_2\}=0, \lb{ALh6.16b}
\end{equation}
as well as the Leibniz rule
\begin{equation}
\{\calG_1,\calG_2\calG_3\}= \{\calG_1,\calG_2\}\calG_3+ \calG_2\{\calG_1,\calG_3\},  
\lb{ALh6.16c}
\end{equation}
hold as discussed in \cite[Theorem 48.8]{KrieglMichor:1997}. 

If $\calG$ is a smooth functional and $(\alpha,\beta)$ develops according to a Hamiltonian 
flow  with Hamiltonian $\calH$, that is,
\begin{equation}
\begin{pmatrix} \alpha\\ \beta\end{pmatrix}_t=(\nabla_s\calH)_{\alpha,\beta}=\calD(\nabla\calH)_{\alpha,\beta}
=\calD\begin{pmatrix} \frac{\delta H}{\delta \alpha}\\[1mm] \frac{\delta H}{\delta \beta}\end{pmatrix},\lb{ALh6.16d}
\end{equation}
then
\begin{align}
\frac{d\calG}{dt}&=\frac{d}{dt}\sum_{n\in\bbZ} G(\alpha(n),\beta(n)) \no \\
&= \sum_{n\in\bbZ}  \begin{pmatrix} \frac{\delta G}{\delta \alpha}(n)\\[1mm] \frac{\delta G}{\delta \beta}(n)\end{pmatrix}\cdot \begin{pmatrix} \alpha(n)\\ \beta(n)\end{pmatrix}_t 
=  \sum_{n\in\bbZ}  \begin{pmatrix} \frac{\delta G}{\delta \alpha}(n)\\[1mm] 
\frac{\delta G}{\delta \beta}(n)\end{pmatrix} \cdot 
\calD \begin{pmatrix} \frac{\delta H}{\delta \alpha}(n)\\[1mm] 
\frac{\delta H}{\delta \beta}(n)\end{pmatrix} \no \\
&= \{\calG,\calH\}. \lb{ALh6.16e}
\end{align}
Here, and in the remainder of this section, time-dependent equations  such as \eqref{ALh6.16e} 
are viewed locally in time, that is, assumed to hold on some open 
$t$-interval $\bbI\subseteq\bbR$.
 
If a functional $\calG$ is in involution with the Hamiltonian $\calH$, that is,
\begin{equation}
\{\calG,\calH\}=0, \lb{ALh6.16f}
\end{equation}
then it is conserved in the sense that 
\begin{equation}
\frac{d\calG}{dt}=0. \lb{ALh6.16g}
\end{equation}

Next, we turn to the specifics of the AL hierarchy. We define
\begin{equation} \lb{ALh6.13}
\hatt\calG_{\ell,\pm}=\sum_{n\in\bbZ} \hat g_{\ell,\pm}(n).
\end{equation}

\begin{lemma} \lb{lALh6.2}
Assume Hypothesis \ref{hALh6.1} and $v\in\ell^1(\bbZ)$. Then, 
\begin{align}\lb{ALh6.14}
(d\hatt\calG_{\ell,\pm})_{\beta}(v)
&=\sum_{n\in\bbZ} \frac{\delta\hat g_{\ell,\pm}(n)}{\delta\beta}v(n)=
 \pm \ell \sum_{n\in\bbZ} (\delta_n, L^{\pm\ell-1}M_\beta(v)\delta_n),  \quad \ell\in \bbN,  \\
\lb{ALh6.15}
(d\hatt\calG_{\ell,\pm})_{\alpha}(v)
&=\sum_{n\in\bbZ} \frac{\delta\hat g_{\ell,\pm}(n)}{\delta\alpha}v(n)=
 \pm \ell \sum_{n\in\bbZ} (\delta_n, L^{\pm\ell-1}M_\alpha(v)\delta_n),  \quad \ell\in \bbN,  
\end{align}
where
\begin{align} \no
M_\beta(v)&=-v\alpha^+ 
+\Big(\big(v^-\rho-\beta^-\frac{v\alpha}{2\rho}\big) \deven+ \alpha^+\frac{v\alpha}{2\rho}\dodd \Big) S^- \\
&\quad +\Big(\big(v\rho^+-\beta\frac{v^+\alpha^+}{2\rho^+}\big)\deven 
+\alpha^{++}\frac{v^+\alpha^+}{2\rho^+}\dodd \Big)S^+ \\ \no
&\quad-\Big(\rho\frac{v^-\alpha^-}{2\rho^-}+\rho^-\frac{v\alpha}{2\rho} \Big)\deven S^{--}
 -\Big(\rho^+\frac{v^{++}\alpha^{++}}{2\rho^{++}}+\rho^{++}\frac{v^+\alpha^+}{2\rho^+}\Big) 
\dodd S^{++}, \\ \no
M_\alpha(v)&= -v^+\beta 
-\Big(\big(v^+\rho-\alpha^+\frac{v\beta}{2\rho}\big) \dodd+ \beta^-\frac{v\beta}{2\rho}\deven \Big) S^- \\
&\quad -\Big(\big(v^{++}\rho^+-\alpha^{++}\frac{v^+\beta^+}{2\rho^+}\big)\dodd 
-\beta\frac{v^+\beta^+}{2\rho^+}\deven \Big)S^+ \\ \no
&\quad-\Big(\rho\frac{v^-\beta^-}{2\rho^-}+\rho^-\frac{v\beta}{2\rho} \Big)\deven S^{--}
 -\Big(\rho^+\frac{v^{++}\beta^{++}}{2\rho^{++}}+\rho^{++}\frac{v^+\beta^+}{2\rho^+}\Big) 
\dodd S^{++}.
\end{align}
\end{lemma}
\begin{proof}  
We first consider the derivative with respect to $\beta$. 
By a slight abuse of notation we write $L=L(\beta)$. Using \eqref{ALh6.13} and 
\eqref{ALh3.5} one finds
\begin{align} \lb{ALh6.18}
(d\hatt\calG_{\ell,\pm})_{\beta}v=\frac{d}{d\epsilon}\calG(\beta+\epsilon v)\big|_{\epsilon=0}
&=\sum_{n\in\bbZ} \frac{d}{d\epsilon}\hat g_{\ell,\pm}(\beta+\epsilon v)(n)\big|_{\epsilon=0}\no \\
&=\sum_{n\in\bbZ} (\delta_{n},\frac{d}{d\epsilon} L(\beta+\epsilon v)^{\pm\ell} \delta_{n}) |_{\epsilon=0}.
\end{align}
Next, one considers 
\begin{align}
\frac{d}{d\epsilon}L( \beta+\epsilon v)^\ell |_{\epsilon=0}&= 
\lim_{\epsilon\to 0}\frac{1}{\epsilon}\big(L( \beta+\epsilon v)^\ell-L(\beta)^\ell\big) \no \\
&= \lim_{\epsilon\to 0}\frac{1}{\epsilon}\Big((L( \beta+\epsilon v)-L(\beta)) L(\beta)^{\ell-1}\no \\
&\qquad\qquad+L(\beta) (L( \beta+\epsilon v)-L(\beta)) L(\beta)^{\ell-2}\no \\
&\qquad\qquad+\cdots+
L(\beta)^{\ell-1}(L( \beta+\epsilon v)-L(\beta)) \Big)\no\\
&= M_\beta L(\beta)^{\ell-1}+ L(\beta) M_\beta L(\beta)^{\ell-2}+\cdots+ L(\beta)^{\ell-1}M_\beta,
  \lb{ALh6.19}
\end{align}
where 
\begin{align}  \lb{ALh6.20} \no
M_\beta&= \lim_{\epsilon\to 0}\frac{1}{\epsilon}\big(L( \beta+\epsilon v)-L(\beta)\big) \\
&= \bigg( -v(n)\alpha(n+1)\delta_{m,n}
+ \Big(\big(v(n-1)\rho(n) - \beta(n-1)\frac{v(n)\alpha(n)}{2\rho(n)}\big)\dodd(n)  \no \\ 
&\quad+\alpha(n+1)\frac{v(n)\alpha(n)}{2\rho(n)}\deven(n) \Big)\delta_{m,n-1} \no \\ 
&\quad+\Big(\big(v(n)\rho(n+1)-\beta(n)\frac{v(n+1)\alpha(n+1)}{2\rho(n+1)}\big)\dodd(n)  \\
&\quad+\alpha(n+2)\frac{v(n+1)\alpha(n+1)}{2\rho(n+1)}\deven(n) \Big)\delta_{m,n+1} \no \\
&\quad-\Big(\rho(n+1)\frac{v(n+2)\alpha(n+2)}{2\rho(n+2)}+\rho(n+2)\frac{v(n+1)\alpha(n+1)}{2\rho(n+1)}\Big)  \deven(n)\delta_{m,n+2}  \no \\
&\quad-\Big(\rho(n)\frac{v(n-1)\alpha(n-1)}{2\rho(n-1)}+\rho(n-1)\frac{v(n)\alpha(n)}{2\rho(n)} \Big)\dodd(n)\delta_{m,n-2}
\bigg)_{m,n\in\bbZ}.\no
\end{align}
Similarly one obtains
\begin{align} \lb{ALh6.21}
& \frac{d}{d\epsilon}L( \beta+\epsilon v)^{-\ell}|_{\epsilon=0}  \\
& \quad =- \Big(L(\beta)^{-1} M_\beta L(\beta)^{-\ell}+ L(\beta)^{-2} 
M_\beta L(\beta)^{-\ell+1}
+\cdots+ L(\beta)^{-\ell}M_\beta L(\beta)^{-1} \Big).   \no 
\end{align}
Inserting the expression \eqref{ALh6.19} into \eqref{ALh6.18} one finds
\begin{align}
(d\hatt\calG_{\ell,+})_{\beta}v &=
\sum_{n\in\bbZ}(\delta_{n},\frac{d}{d\epsilon} L(\beta+\epsilon v)^{ \ell} \delta_{n}) |_{\epsilon=0} \no \\
&= \sum_{n\in\bbZ}(\delta_{n}, \sum_{k=0}^{\ell-1}L^{k}M_\beta L^{\ell-1-k}\delta_{n})\no \\
&=\sum_{k=0}^{\ell-1}  \sum_{n\in\bbZ}(\delta_{n}, L^{k}M_\beta L^{\ell-1-k}\delta_{n})\no \\
&= \sum_{k=0}^{\ell-1}  \sum_{n\in\bbZ}(\delta_{n}, 
(L^{\ell-1}M_\beta+[L^{k}M_\beta, L^{\ell-1-k}])\delta_{n})\no \\
&= \sum_{k=0}^{\ell-1}  \sum_{n\in\bbZ}\Big( (\delta_{n}, L^{\ell-1}M_\beta\delta_{n})
+(\delta_{n}, [L^{k}M_\beta, L^{\ell-1-k}]\delta_{n})\Big)\no \\
&= \ell \sum_{n\in\bbZ}(\delta_{n}, L^{\ell-1}M_\beta\delta_{n})
+ \sum_{k=0}^{\ell-1}  \tr\big([L^{k}M_\beta, L^{\ell-1-k}]\big)\no \\
&= \ell \sum_{n\in\bbZ}(\delta_{n}, L^{\ell-1}M_\beta\delta_{n}).
\end{align}
Similarly, using \eqref{ALh6.18} and  \eqref{ALh6.21}, one concludes that
\begin{equation}
(d\hatt\calG_{\ell,-})_{\beta}v
= -\ell \sum_{n\in\bbZ}(\delta_{n}, L^{-\ell-1}M_\beta\delta_{n}).
\end{equation}
For the derivative with respect to $\alpha$ we set $L=L(\alpha)$ and replace $M_\beta$ by
\begin{align} \lb{ALh5.2b}
M_\alpha&= \lim_{\epsilon\to 0}\frac{1}{\epsilon}\big(L( \alpha+\epsilon v)-L(\alpha)\big)\no \\
&= \bigg( -v(n+1)\beta(n)\delta_{m,n}
+ \Big(- \beta(n-1)\frac{v(n)\beta(n)}{2\rho(n)}\dodd(n) \no\\ 
&\quad - \big(v(n+1)\rho(n) -\alpha(n+1)\frac{v(n)\beta(n)}{2\rho(n)}\big)\deven(n) \Big)\delta_{m,n-1} \no \\ 
&\quad-\Big(\beta(n)\frac{v(n+1)\beta(n+1)}{2\rho(n+1)}\dodd(n) + \big(v(n+2)\rho(n+1) \\
&\quad + \alpha(n+2)\frac{v(n+1)\beta(n+1)}{2\rho(n+1)}\big)\deven(n) \Big)\delta_{m,n+1} \no \\
&\quad-\Big(\rho(n+1)\frac{v(n+2)\beta(n+2)}{2\rho(n+2)}+\rho(n+2)\frac{v(n+1)\beta(n+1)}{2\rho(n+1)}\Big)  \deven(n)\delta_{m,n+2}  \no \\
&\quad-\Big(\rho(n)\frac{v(n-1)\beta(n-1)}{2\rho(n-1)}+\rho(n-1)\frac{v(n)\beta(n)}{2\rho(n)} \Big)\dodd(n)\delta_{m,n-2}
\bigg)_{m,n\in\bbZ}. \no 
\end{align}
\end{proof}

\begin{lemma} \label{lALh6.3} 
Assume Hypothesis \ref{hALh6.1}. Then the following relations hold:
\begin{align}
\frac{\delta \hat g_{\ell,+}}{\delta \beta}&
= \frac{\ell}{\gamma}\big(\hat f_{\ell-1,+}-\alpha\hat g_{\ell,+} \big),  \quad \ell\in \bbN, 
\lb{ALh6.25}\\
\frac{\delta \hat g_{\ell,-}}{\delta \beta}&
=- \frac{\ell}{\gamma}\big(\hat f_{\ell-1,-}^- +\alpha\hat g_{\ell,-}^- \big), \quad \ell\in \bbN.  \lb{ALh6.26}
\end{align} 
\end{lemma}
\begin{proof} We consider \eqref{ALh6.25} first. By \eqref{ALh3.5} one concludes that
\begin{align}
\begin{split}
& \hat f_{\ell-1,+}(n)-\alpha(n)\hat g_{\ell,+}(n) \\
& \quad =(\delta_n,EL^{\ell-1} \delta_{n})\deven(n)
 +(\delta_{n},L^{ \ell-1}D \delta_{n})\dodd(n) - \alpha(n) (\delta_{n},L^{ \ell} \delta_{n}). 
 \end{split}
\end{align}
Thus one has to show that
\begin{equation}
 \sum_{n\in\bbZ} (\delta_n, L^{\ell-1}M_\beta\delta_n) 
 =  \sum_{n\in\bbZ} \f{v(n)}{\rho(n)^2}\big(\hat f_{\ell-1,+}(n)-\alpha(n)\hat g_{\ell,+}(n) \big), \lb{ALh6.28}
\end{equation}
since this implies  \eqref{ALh6.25}, using \eqref{ALh6.14}. 
By \eqref{ALh6.20}, \eqref{ALD}, and \eqref{ALE}, and assuming $v \in \ell^1(\bbZ)$, one obtains 
\begin{align}
& \sum_{n\in\bbZ}(\delta_n, L^{\ell-1}M_\beta\delta_n)  \no \\
&= \sum_{n\in\bbZ} \Big(-v\alpha^+(\delta_n, L^{\ell-1}\delta_n) + v^-\rho (\delta_n, L^{\ell-1}\delta_{n-1})\dodd
+ v\rho^+ (\delta_n, L^{\ell-1}\delta_{n+1})\dodd \no \\
&\quad - \f{v\alpha}{2\rho}\big(-\alpha^+(\delta_n, L^{\ell-1}\delta_{n-1})\deven
+ \rho^+(\delta_{n+1}, L^{\ell-1}\delta_{n-1})\deven \big) \no \\
&\quad - \f{v\alpha}{2\rho}\big(\beta^-(\delta_n, L^{\ell-1}\delta_{n-1})\dodd
+ \rho^-(\delta_n, L^{\ell-1}\delta_{n-2})\dodd \big) \no \\
&\quad - \f{v^+\alpha^+}{2\rho^+}\big(-\alpha^{++}(\delta_n, L^{\ell-1}\delta_{n+1})\deven
+ \rho^{++}(\delta_{n}, L^{\ell-1}\delta_{n+1})\deven \big) \no \\
&\quad - \f{v^+\alpha^+}{2\rho^+}\big(\beta(\delta_n, L^{\ell-1}\delta_{n+1})\dodd
+ \rho(\delta_{n-1}, L^{\ell-1}\delta_{n+1})\dodd \big)\Big) \no \\
&=  \sum_{n\in\bbZ} \Big(-v\alpha^+(\delta_n, L^{\ell-1}\delta_n) + v\rho^+ (\delta_{n+1}, L^{\ell-1}\delta_n)\deven
+ v\rho^+ (\delta_n, L^{\ell-1}\delta_{n+1})\dodd \no \\
&\quad - \f{v\alpha}{2\rho}\big((\delta_n, EL^{\ell-1}\delta_{n-1})\deven
+ (\delta_n, L^{\ell-1}D\delta_{n-1})\dodd \big) \no \\
&\quad - \f{v^+\alpha^+}{2\rho^+}\big((\delta_n, L^{\ell-1}D\delta_{n+1})\deven
+ (\delta_n, EL^{\ell-1}\delta_{n+1})\dodd \big)\Big) \no \\
&=  \sum_{n\in\bbZ} \Big( v (\delta_n, EL^{\ell-1}\delta_n)\deven + v(\delta_n, L^{\ell-1}D\delta_n)\dodd \no \\
&\quad - \f{v\alpha}{2\rho}\big((\delta_n, EL^{\ell-1}\delta_{n-1})\deven
+ (\delta_n, L^{\ell-1}D\delta_{n-1})\dodd  \no \\
&\quad + (\delta_{n-1}, L^{\ell-1}D\delta_n)\dodd
+ (\delta_{n-1}, EL^{\ell-1}\delta_n)\deven \big)\Big) \no \\
&=  \sum_{n\in\bbZ} \Big(\f{v}{\rho^2}\big(\hat f_{\ell-1,+} - \alpha \hat g_{\ell,+} \big) + \f{v\alpha}{2\rho}\big( 
(\delta_{n-1}, EL^{\ell-1}\delta_n)\deven    \\
&\quad + (\delta_n, L^{\ell-1}D\delta_{n-1})\dodd
- (\delta_n, EL^{\ell-1}\delta_{n-1})\deven - (\delta_{n-1}, L^{\ell-1}D\delta_n)\dodd\big)\Big),   \no 
\end{align}
where we used \eqref{ALh3.5} and
\begin{align}
\hat g_{\ell,+} &= (\delta_n, L^{\ell-1}DE\delta_n)\no \\
&=\beta (\delta_n, L^{\ell-1}D\delta_n)\dodd + \rho(\delta_n, L^{\ell-1}D\delta_{n-1})\dodd \no \\
&\quad+ \beta (\delta_n, EL^{\ell-1}\delta_n)\deven+\rho(\delta_{n-1}, EL^{\ell-1}\delta_n)\deven \no \\
&=\beta \hat f_{\ell-1,+} +  \rho(\delta_n, L^{\ell-1}D\delta_{n-1})\dodd 
+ \rho(\delta_{n-1}, EL^{\ell-1}\delta_n)\deven. 
\end{align}
Hence it remains to show that
\begin{align} \lb{ALh6.31}
\begin{split} 
& (\delta_{n-1}, EL^{\ell-1}\delta_n)\deven + (\delta_n, L^{\ell-1}D\delta_{n-1})\dodd  \\
& \quad - (\delta_n, EL^{\ell-1}\delta_{n-1})\deven - (\delta_{n-1}, L^{\ell-1}D\delta_n)\dodd = 0,
\end{split}
\end{align}
but this follows from $(E L^\ell)^\top= E L^\ell$ respectively
$(L^\ell D)^\top=  L^\ell D$ by \eqref{ALtheta}, \eqref{ALfact}.

In the case \eqref{ALh6.26} one similarly shows that
\begin{align}
\sum_{n\in\bbZ} (\delta_n, L^{-\ell-1}M_\beta\delta_n) 
& =-\sum_{n\in\bbZ} \frac{v(n+1)}{\rho(n)^2}\Big( (\delta_{n},D^{-1}L^{- \ell+1} \delta_{n})\deven(n)    \\
&\quad  +(\delta_{n},L^{- \ell+1}E^{-1} \delta_{n})\dodd(n) + \alpha(n+1)(\delta_{n},L^{- \ell} \delta_{n})\Big). \no
\end{align}
\end{proof}

\begin{lemma}\label{lALh6.4} 
Assume Hypothesis \ref{hALh6.1}. Then the following relations hold:
\begin{align}
\frac{\delta \hat g_{\ell,+}}{\delta \alpha}&
= - \frac{\ell}{\gamma}\big(\hat h_{\ell-1,+}^- +\beta \hat g_{\ell,+}^- \big),  \quad \ell\in \bbN, \lb{ALh6.34} \\
\frac{\delta \hat g_{\ell,-}}{\delta \alpha}&
= \frac{\ell}{\gamma}\big(\hat h_{\ell-1,-} - \beta\hat g_{\ell,-} \big), \quad \ell\in \bbN.  
\lb{ALh6.34a}
\end{align} 
\end{lemma}
\begin{proof} 
We consider \eqref{ALh6.34} first. Using \eqref{ALh6.15}, \eqref{ALh5.2b}, \eqref{ALD}, and \eqref{ALE}, and assuming $v\in\ell^1(\bbZ)$ one obtains 
\begin{align}
& \sum_{n\in\bbZ}(\delta_n, L^{\ell-1}M_\alpha\delta_n)  \no \\
&= \sum_{n\in\bbZ} \Big(-v^+\beta(\delta_n, L^{\ell-1}\delta_n) - v\rho (\delta_n, L^{\ell-1}\delta_{n-1})\deven
- v^{++}\rho^+ (\delta_n, L^{\ell-1}\delta_{n+1})\deven \no \\
&\quad - \f{v\beta}{2\rho}\big(\beta^-(\delta_n, L^{\ell-1}\delta_{n-1})\dodd
+ \rho^-(\delta_{n}, L^{\ell-1}\delta_{n-2})\dodd \big) \no \\
&\quad - \f{v\beta}{2\rho}\big(-\alpha^+(\delta_n, L^{\ell-1}\delta_{n-1})\deven
+ \rho^+(\delta_{n+1}, L^{\ell-1}\delta_{n-1})\deven \big) \no \\
&\quad - \f{v^+\beta^+}{2\rho^+}\big(-\alpha^{++}(\delta_n, L^{\ell-1}\delta_{n+1})\deven
+ \rho^{++}(\delta_{n}, L^{\ell-1}\delta_{n+1})\deven \big) \no \\
&\quad - \f{v^+\alpha^+}{2\rho^+}\big(\beta(\delta_n, L^{\ell-1}\delta_{n+1})\dodd
+ \rho(\delta_{n-1}, L^{\ell-1}\delta_{n+1})\dodd \big)\Big) \no \\
&=  \sum_{n\in\bbZ} \Big(-v^+\big((\delta_n, L^{\ell-1}D\delta_n)\deven + (\delta_{n}, EL^{\ell-1}\delta_n)\dodd\big)\no \\
&\quad - \f{v\beta}{2\rho}\big((\delta_n, EL^{\ell-1}\delta_{n-1})\deven
+ (\delta_n, L^{\ell-1}D\delta_{n-1})\dodd \big) \no \\
&\quad - \f{v^+\alpha^+}{2\rho^+}\big((\delta_n, L^{\ell-1}D\delta_{n+1})\deven
+ (\delta_n, EL^{\ell-1}\delta_{n+1})\dodd \big)\Big) \no \\ 
&= - \sum_{n\in\bbZ} \f{v}{\rho^2}\big(\hat h_{\ell-1,+}^- + \beta \hat g_{\ell,+}^- \big),
\end{align}
since by \eqref{ALh3.5} and \eqref{ALh5.10},
\begin{align}
\begin{split}
2 \alpha \hat h_{\ell-1,+}^- + 2 \hat g_{\ell,+}^- &=
\rho\big((\delta_n, L^{\ell-1}D\delta_{n-1})\dodd + (\delta_n, EL^{\ell-1}\delta_{n-1})\deven  \\ 
&\quad+ (\delta_{n-1}, L^{\ell-1}D\delta_{n})\dodd + (\delta_{n-1}, EL^{\ell-1}D\delta_{n})\deven\big).
\end{split}
\end{align}
The result \eqref{ALh6.34a} follows similarly.
\end{proof}

Next, we introduce the Hamiltonians
\begin{align} \label{ALh6.35}
\widehat \calH_0 &= \sum_{n\in\bbZ} \ln(\gam(n)), \qquad
\widehat \calH_{p_\pm,\pm} = \frac{1}{p_\pm} \sum\limits_{n\in\bbZ}\hat g_{p_\pm,\pm}(n), \quad p_\pm \in \bbN,\\  \label{ALh6.36}
\calH_{\ul p} &=
\sum_{\ell=1}^{p_+} c_{p_+-\ell,+} \widehat \calH_{\ell,+} +
\sum_{\ell=1}^{p_-} c_{p_--\ell,-} \widehat \calH_{\ell,-} + c_{\ul{p}} \widehat \calH_0, 
\quad \ul p=(p_-,p_+) \in\bbN_0^2.
\end{align}
(We recall that $c_{\ul p} = (c_{p,-} + c_{p,+})/2$.)
 
\begin{theorem} \lb{tALh6.5}
Assume Hypothesis \ref{hALh6.1}. Then the following relations hold: 
\begin{align}  \label{ALh6.37}
\AL_{\ul p}(\alpha,\beta)&= 
\begin{pmatrix}-i\alpha_{t_{\ul p}} \\ -i \beta_{t_{\ul p}} \end{pmatrix}
+\calD\nabla \calH_{\ul p}=0, \quad \ul p \in \bbN_0^2.
\end{align}
\end{theorem}
\begin{proof}
This follows directly from Lemmas \ref{lALh6.3} and \ref{lALh6.4}, 
\begin{align}  
(\nabla \widehat \calH_{\ell,+})_\alpha&=
\frac1\gamma\big(-\beta\hat g_{\ell,+}^- - \hat h_{\ell-1,+}^-\big),\quad
(\nabla \widehat \calH_{\ell,+})_\beta=\frac1\gamma\big(-\alpha\hat g_{\ell,+}+ \hat f_{\ell-1,+}\big),   \no \\ 
(\nabla \widehat \calH_{\ell,-})_\alpha&=
\frac1\gamma\big(-\beta\hat g_{\ell,-}+ \hat h_{\ell-1,-} \big),\quad
(\nabla \widehat \calH_{\ell,-})_\beta=\frac1\gamma\big(-\alpha\hat g_{\ell,-}^-  - \hat f_{\ell-1,-}^-\big),   \lb{ALh6.49a}  \\
& \hspace*{8.55cm}  \ell \in \bbN,  \no
\end{align}
together with \eqref{ALhat f}.
\end{proof}

\begin{theorem} \lb{tALh6.6}
Assume Hypothesis~\ref{hALh5.1} and suppose that $\alpha, \beta$ 
satisfy $\AL_{\ul p}(\alpha,\beta)=0$ for some 
$\ul p \in\bbN_{0}^2$. Then, 
\begin{equation}
\frac{d \calH_{\ul r}}{dt_{\ul p}} =0, \quad \ul r\in\bbN_{0}^2.   \lb{ALh6.41}
\end{equation}
\end{theorem}
\begin{proof} From Lemma \ref{lALh4.4} and Theorem \ref{tALh5.7} one obtains 
\begin{equation}
\frac{d\hat g_{r_\pm, \pm}}{dt_{\ul p}}=(S^+ -I) J_{r_\pm,\pm}, \quad r_\pm \in\bbN_{0}, 
\end{equation}
for some $J_{r_\pm,\pm}$, $r_\pm\in\bbN_{0}$, which are polynomials in $\alpha$ and $\beta$ and certain shifts thereof. 
Using definition \eqref{ALh6.36} of $\calH_{\ul r}$,  the result \eqref{ALh6.41} follows in the homogeneous case and 
then by linearity in the general case. 
\end{proof}

\begin{theorem} \lb{tALh6.7}
Assume Hypothesis~\ref{hALh6.1} and let $\ul p, \ul r\in\bbN_0^2$. Then, 
\begin{equation}
\{ \calH_{\ul p},  \calH_{\ul r}\}=0,   \lb{ALh6.46}
\end{equation}
that is, $\calH_{\ul p}$ and $\calH_{\ul r}$ are in involution for all $\ul p, \ul r\in\bbN_0^2$. 
\end{theorem}
\begin{proof}  By Theorem~\ref{tALh5.5}, there exists $T>0$ such that the initial value problem 
\begin{equation}
\AL_{\ul p}(\alpha,\beta)=0, \quad (\alpha,\beta)\big|_{t_{\ul p}=0}=\big(\alpha^{(0)},\beta^{(0)}\big),
\end{equation}
where $\alpha^{(0)},\beta^{(0)}$ satisfy Hypothesis~\ref{hALh6.1}, has unique, local, and smooth solutions 
$\alpha(t), \beta(t)$ satisfying Hypothesis~\ref{hALh6.1} for each $t\in[0,T)$. For this solution we know that
\begin{equation}
\frac{d}{dt_{\ul p}} \calH_{\ul p}(t)=\{\calH_{\ul r}(t), \calH_{\ul p}(t)\}=0.
\end{equation}
Next, let $t\downarrow 0$. Then
\begin{equation}
0=\{\calH_{\ul r}(t), \calH_{\ul p}(t)\}\underset{t\downarrow 0}{\to}\{\calH_{\ul r}(0), \calH_{\ul p}(0)\}
=\{\calH_{\ul r}, \calH_{\ul p}\}\big|_{(\alpha,\beta)=(\alpha^{(0)},\beta^{(0)})}. 
\end{equation}
Since $\alpha^{(0)},\beta^{(0)}$ are arbitrary coefficients satisfying Hypothesis~\ref{hALh6.1} one
concludes \eqref{ALh6.46}.
\end{proof}

\bigskip
\noindent {\bf Acknowledgments.}
Fritz Gesztesy, Johanna Michor, and Gerald Teschl gratefully acknowledge the extraordinary
hospitality of the Department of Mathematical Sciences of
the Norwegian University of Science and Technology, Trondheim, during 
extended stays in the summers of 2006 and 2007, where parts of this paper
were written. 


\end{document}